\documentclass[aps,notitlepage,nofootinbib,longbibliography,twocolumn,superscriptaddress,10pt]{revtex4-2}
\usepackage{graphicx,amssymb,amsmath,amsfonts,empheq}
\usepackage[dvipsnames]{xcolor}
\usepackage{color}
\usepackage{braket}
\usepackage{latexsym}
\usepackage{upgreek}
\usepackage{dsfont}
\usepackage{bm}
\usepackage{multirow}
\usepackage{enumitem}
\usepackage[normalem]{ulem}
\usepackage{url}
\usepackage{hyperref}
\hypersetup{
colorlinks = true,
linkcolor=blue,
citecolor=[rgb]{0.15,0.65,0.13},
urlcolor = [rgb]{0.25, 0.41, 0.88}}
\usepackage{bbm}

\usepackage{bm}
\renewcommand{\vec}[1]{\boldsymbol{\mathbf{#1}}}

\graphicspath{{./}{./images/}}

\usepackage{subfigure}

\newcommand{\bit}{\begin{itemize}}
\newcommand{\eit}{\end{itemize}}

\renewcommand{\>}{\right\rangle}
\newcommand{\<}{\left\langle}
\newcommand{\ba}{\begin{align}}
\newcommand{\ea}{\end{align}}
\newcommand{\be}{\begin{equation}}
\newcommand{\ee}{\end{equation}}
\newcommand{\bi}{\begin{itemize}}
\newcommand{\ei}{\end{itemize}}

\newcommand{\fineq}[4][1=-.8ex,2=1,3=1]{
  \begin{tikzpicture}[baseline={([yshift=#1]current  bounding  box.center)}, scale = #2, every node/.style={scale = #3}]
    #4
  \end{tikzpicture}
}

\newcommand{\wickpair}[5][1=0,2=0,3=0.3,4=0.2,5=0.8]{
  \begin{scope}[shift={(#1,#2)}]
    \draw[line width = 0.5pt] (0,0) to[out=90,in=180] (#3/2,#4) to[out=0,in=90] (#3,0);
    \begin{scope}[shift={(0,#5)}]
      \draw[line width = 0.5pt] (0,0) to[out=-90,in=180] (#3/2,-#4) to[out=0,in=-90] (#3,0);
    \end{scope}
  \end{scope}
}

\newcommand{\wickswitch}[4][1=0,2=0,3=0.3,4=0.8]{
  \begin{scope}[shift={(#1,#2)}]
    \pgfmathsetmacro{\w}{#3};
    \pgfmathsetmacro{\h}{#4};
    \draw[line width = 0.5pt] (0,0) to[out=90,in=-135] (\w/2,\h/2) to[out=45,in=-90] (\w,\h);
    \draw[line width = 0.5pt] (\w,0) to[out=90,in=-45] (\w/2,\h/2) to[out=135,in=-90] (0,\h);
  \end{scope}
}

\newcommand{\bcid}[4][1=0,2=0,3=0.3,4=0.8]
{
  \begin{scope}[shift={(#1,#2)}]
    \draw (0,0)--++(0,#4);
    \draw (#3,0)--++(0,#4);
  \end{scope}
}

\newcommand{\bcwick}[1][1=0]
{
  \fineq[-0.4ex]{
    \pgfmathsetmacro{\h}{0.8};
    \pgfmathsetmacro{\w}{0.3};
    \ifthenelse{\equal{#1}{0}}{
      \wickpair[0][0];
    }{}
    \ifthenelse{\equal{#1}{1}}{
      \wickpair[0][0][\w*2];
    }{}
    \ifthenelse{\equal{#1}{2}}{
      \wickpair[0][0][\w*3];
    }{}
    \node () at (0,\h/2) {};
    \node () at (\w,\h/2) {};
  }
}

\newcommand{\idw}[0]
{
  \draw[line width = 0.5pt] (0,0) to[out=90,in=180] (0.15,0.2) to[out=0,in=90] (0.3,0);
  \draw[line width = 0.5pt] (0.6,0) to[out=90,in=180] (0.6+0.15,0.2) to[out=0,in=90] (0.9,0);
}

\newcommand{\swapw}[0]
{
  \draw[line width = 0.5pt] (0,0) to[out=90,in=180] (0.45,0.25) to[out=0,in=90] (0.9,0);
  \draw[line width = 0.5pt] (0.3,0) to[out=90,in=180] (0.45,0.15) to[out=0,in=90] (0.6,0);
}

\newcommand{\topcont}[3][1=0,2=0,3=0]
{
  \begin{scope}[shift={(#1,#2)}]
    \ifthenelse{\equal{#3}{0}}{
      \idw;
    }{}
    \ifthenelse{\equal{#3}{1}}{
      \swapw;
    }{}
  \end{scope}
}

\newcommand{\bcwickfouru}[2][1=0,2=]
{
  \fineq[-0.6ex]{
    \pgfmathsetmacro{\h}{0.8};
    \pgfmathsetmacro{\w}{0.3};
    \ifthenelse{\equal{#1}{0}}{
      \wickpair[0][0];
      \bcid[2*\w][0];
    }{}
    \ifthenelse{\equal{#1}{1}}{
      \wickpair[\w][0];
      \bcid[0][0][3*\w];
    }{}
    \ifthenelse{\equal{#1}{2}}{
      \wickpair[0][0][\w*3];
      \bcid[\w][0]
    }{}
    \ifthenelse{\equal{#1}{3}}{
      \wickpair[2*\w][0];
      \bcid[0][0]
    }{}
    \node () at (0,\h/2) {};
    \node () at (3*\w,\h/2) {};
    \ifthenelse{\equal{#2}{}}{}{
      \topcont[0][\h+0.1][#2];
    }
  }
}

\newcommand{\bcidfouru}[2][1=,2=]
{
  \fineq[-0.6ex]{
    \pgfmathsetmacro{\h}{0.8};
    \pgfmathsetmacro{\w}{0.3};
    \bcid;
    \bcid[2*\w];
    \node () at (0,\h/2) {};
    \node () at (3*\w,\h/2) {};
    \ifthenelse{\equal{#2}{}}{}{
      \topcont[0][\h+0.1][#2];
    }
  }
}

\newcommand{\bcswitchfouru}[2][1=0,2=]
{
  \fineq[-0.6ex]{
    \pgfmathsetmacro{\h}{0.8};
    \pgfmathsetmacro{\w}{0.3};
    \ifthenelse{\equal{#1}{0}}{
      \wickswitch[0][0][2*\w][\h]
      \bcid[\w][0][2*\w];
    }{}
    \ifthenelse{\equal{#1}{1}}{
      \wickswitch[\w][0][2*\w][\h]
      \bcid[0][0][2*\w];
    }{}
    \node () at (0,\h/2) {};
    \node () at (3*\w,\h/2) {};
    \ifthenelse{\equal{#2}{}}{}{
      \topcont[0][\h+0.1][#2];
    }
  }
}

\newcommand{\linearrow}[4][1=0,2=0,3=0.8,4=0]
{
  \pgfmathsetmacro{\h}{#3};
  \begin{scope}[shift={(#1,#2)}]
    \ifthenelse{\equal{#4}{0}}{
      \draw[-stealth] (0,0)--(0,\h*0.6);
      \draw (0,\h*0.6)--(0,\h);
    }{}
    \ifthenelse{\equal{#4}{1}}{
      \draw[-stealth] (0,\h)--(0,\h*0.45);
      \draw (0,\h*0.45)--(0,0);
    }{}
  \end{scope}
 
}

\begin{document}
\date{\today}

\renewcommand{\bra}[1]{\< #1 \right|}
\renewcommand{\ket}[1]{\left| #1 \>}
\newcommand{\bbra}[1]{\<\< #1 \right|\right|}
\newcommand{\kket}[1]{\left|\left| #1 \>\>}

\title{Universality of the cross entropy in $\mathbb{Z}_2$ symmetric monitored quantum circuits}

\author{Maria Tikhanovskaya}
\affiliation{Department of Physics, Harvard University, Cambridge, MA 02138}
\affiliation{Kavli Institute for Theoretical Physics, University of California, Santa Barbara, CA 93106 }
\author{Ali Lavasani}
\affiliation{Kavli Institute for Theoretical Physics, University of California, Santa Barbara, CA 93106 }
\author{Matthew P. A. Fisher}
\affiliation{Department of Physics, University of California, Santa Barbara, CA 93106}
\author{Sagar Vijay}
\affiliation{Department of Physics, University of California, Santa Barbara, CA 93106}

\definecolor{ali}{RGB}{	0, 117, 117}
\newcommand{\ali}[1]{{\color{ali}\footnotesize{(AL)[#1]}}}

\definecolor{maria}{RGB}{84,39,143}
\newcommand{\maria}[1]{{\color{maria}\footnotesize{(MT)[#1]}}}

\begin{abstract}
The linear cross-entropy (LXE) has been recently proposed as a {scalable probe of the measurement-driven phase transition between volume- and area-law-entangled phases of pure-state trajectories in certain monitored quantum circuits. Here, we demonstrate that the LXE can distinguish distinct area-law-entangled phases of monitored circuits with symmetries, and extract universal behavior at the critical points separating these phases. We focus on (1+1)-dimensional monitored circuits with an on-site $\mathbb{Z}_{2}$ symmetry.  For an appropriate choice of initial states, the LXE distinguishes the area-law-entangled spin glass and paramagnetic phases of the monitored trajectories.  At the critical point, described by two-dimensional percolation, the LXE exhibits universal behavior which depends sensitively on boundary conditions, and the choice of initial state.  With open boundary conditions, we show that the LXE relates to crossing probabilities in critical percolation, and is thus given by a {known} universal function of the aspect ratio of the dynamics, which quantitatively agrees with numerical studies of the LXE at criticality.  The LXE probes correlations of other operators in percolation with periodic boundary conditions.  We show that the LXE is sensitive to the richer phase diagram of the circuit model in the presence of symmmetric unitary gates. Lastly, we consider the effect of noise during the circuit evolution, and propose potential solutions to counter it.} 
\end{abstract}

\maketitle

\tableofcontents

\section{Introduction}

 Recent studies of quantum many-body systems which are being measured frequently by an external observer, have uncovered new phases of quantum matter which are manifest in the pure-state trajectories of the quantum dynamics \cite{fisher2023random, potter2022entanglement}.  Novel phase transitions are now known to occur in these ``monitored" quantum dynamics due to the competition between the entangling nature of chaotic unitary evolution, and the disentangling action of projective measurements. {Quantum circuits with random, local unitary gates and frequent measurements exhibit a measurement-induced phase transition (MIPT) as the rate of projective measurements is tuned}. As was first observed numerically and followed later by theoretical justifications, for {a small enough measurement rate, the late-time pure-state trajectories sustain a volume-law scaling of the entanglement entropy \cite{li2018quantum, skinner2019measurement,chan2019unitary,choi2020quantum,bao2020theory,jian2020measurement,li2023entanglement}. As this rate is increased, however, there will be a MIPT to a phase characterized by an area-law scaling within these pure-states.} Many aspects of these transitions were extensively studied previously \cite{nahum2021measurement,turkeshi2020measurement,vijay2020measurement,hsieh2021,lavasani2021topological,sharma2022measurement,lunt2021measurement,bao2021symmetry,weinstein2022measurement,alberton2021entanglement,PhysRevX.11.041004,theoryOftransitionsBao,PhysRevLett.128.010605,PhysRevB.101.104302,PhysRevB.105.064305,Sierant_2022,Turkeshi_2021,Turkeshi_2022,ChargeSharp2022PRL,lavasani_monitored_2022,zabalo_infinite-randomness_2022,zabalo_operator_2022,ippoliti_postselection-free_2021,ippoliti_dynamical_2022,szyniszewski_disordered_2022,arxiv.2207.07096,arxiv.2201.12704,PhysRevB.106.024304,Sierant2022dissipativefloquet,PhysRevLett.127.140601}. 

Experimental efforts to observe MIPT have been limited to relatively small system sizes \cite{koh2022experimental}, which are classically simulable \cite{noel2022, hoke2023quantum}, mainly due to the so-called ``post-selection'' problem{: the entanglement structure of the late-time state of the circuit must be accessed by preparing many identical copies of the final state, which requires repeating the experiment exponentially many times to obtain the same measurement outcomes in each realization of the evolution.}  As the number of qubits increases, the post-selection problem becomes more challenging and eventually impossible to overcome. Therefore, to experimentally observe MIPT, it is necessary to find efficient probes that do not rely on post-selection of measurement outcomes \cite{gullans2020scalable,li2023cross}. Alternatively, one may study other types of non-equlibrium phase transitions that share some essential features of the MIPT, but do not suffer from the post-selection problem \cite{weinstein2022scrambling}.

{Post-processing the measurement record is required to overcome the post-selection problem.  Recently, 
it has been proposed \cite{li2023cross, bao2021} that the circuit averaged linear cross-entropy (LXE, denoted by $\overline{\chi}$) can detect the MIPT in random circuits with projective measurements.}
Starting with two different initial states and running them through identical quantum circuits, {the LXE measures how the distribution of measurement outcomes correlates with the initial states.}  
It was shown in Ref.~\cite{li2023cross} that when one starts with two random initial states,  in the area-law phase, measurement outcomes can be used to distinguish the 
initial states with finite probability, and the LXE takes values $\overline{\chi}<1$. 
On the other hand, in the volume law phase, measurement outcomes cannot distinguish the initial states, and  $\overline{\chi}=1$. 
For a circuit with {Clifford unitary gates and Pauli measurements, one can choose to run the evolution with a non-stabilizer initial state on a quantum simulator and a stabilizer initial state on a classical computer, and compute the LXE efficiently. In this setting, the post-selection problem is  partially mitigated, allowing one to detect MIPT in quantum dynamics that are not classically simulable \cite{li2023cross}.} Nonetheless, it is worth noting that this approach does not completely solve the post-selection problem; for a  completely generic quantum circuit, computing the LXE still requires exponentially large classical resources. 

In this paper, we study the utility of LXE beyond the typical volume law to area law phase transition by focusing on MIPT's between different area-law-entangled phases that arise in monitored dynamics with symmetries. More specifically, {we study monitored dynamics in $(1+1)$D with an on-site $\mathbb{Z}_{2}$ symmetry.} These circuits can host two area-law-entangled phases: {a ``spin glass" phase which spontaneously breaks the $\mathbb{Z}_{2}$ symmetry and a paramagnet in which this symmetry is restored \cite{hsieh2021, bao2021}.} We show that by 
choosing appropriate initial states, the LXE can distinguish these phases.

{The critical point separating these two phases is in the same universality class as two-dimensional critical bond percolation \cite{nahum2020entanglement,skinner2019measurement}. We show that depending on the boundary conditions, and the choice of initial states, the LXE is sensitive to different operators in the percolation conformal field theory (CFT), and thus encodes universal information about the critical point separating these monitored quantum phases. In particular, the LXE between two particular initial states in a monitored quantum dynamics proceeding for a time $T$ with open boundary conditions in a system of size $L$ is described by a \emph{universal} function of the ``aspect ratio" $T/L$ which is related to the celebrated Cardy formula for crossing probabilities in critical percolation \cite{cardy1992critical}.  With periodic boundary conditions, the LXE is related to the correlation function of other operators in the percolation CFT, which we demonstrate by taking advantage of well-understood description of percolation as a two-dimensional Coulomb gas \cite{nienhuis1984critical,saleur1987exact}. Our analytic predictions here are confirmed by large-scale numerical simulations. Our results demonstrate that the LXE can also be used to understand critical phenomena in monitored quantum dynamics.} 

Furthermore, we study the effect of adding  $\mathbb{Z}_2$ symmetric unitary operators to the circuit and show that, by carefully choosing the initial states for different parts of the phase diagram, one can use LXE to reproduce the phase diagram studied in Ref.~\cite{hsieh2021}. Our results demonstrate that LXE could be an effective tool to analyze and study more general MIPT. 

Lastly, we study the effect of noise when it only affects one of the circuits \cite{li2023cross}, say the one with the initial state $\rho$, and show that in this case, $\overline{\chi}(\rho,\sigma)$ vanishes in the thermodynamic limit even when the noise is $\mathbb{Z}_2$-symmetric. We briefly discuss possible workarounds, while leaving a more detailed study to a future work. 


The rest of this paper is organized as follows. In Section II we define LXE in detail, describe the main circuit model and show how LXE can be used to detect the MIPT in these circuits. In Section III we focus on the critical point and analytically map LXE to a four-point correlation function in percolation CFT for open and closed boundary conditions and show that the results agree with the numerical data. In Section IV we introduce $\mathbb{Z}_2$-symmetric unitary gates to the circuit and obtain the phase diagram. In Section V we discuss the effect of symmetric noise on the general behavior of LXE in random quantum circuits.

\section{Measurement-only $\mathbb{Z}_2$ symmetric circuit}
\label{sec:Z2}
\subsection{Normalized Linear Cross Entropy (LXE)}

In this work, we use the normalized linear cross entropy (LXE) as defined in Ref.~\cite{li2023cross} as an order parameter to study the MIPT between ordered area law phases. Let $M$ denote the total number of measurements in a monitored circuit $C$, and let ${\bf m}=(m_1,m_2,\dots,m_M)$ denote a particular sequence of measurement outcomes. The LXE for circuit $C$ and for initial states $\rho$ and $\sigma$ is defined as
\begin{align}\label{eq:chi_definition}
\chi(\rho,\sigma) = \frac{\sum_{\bf m} p_{\bf m} ^\sigma p_{\bf m} ^\rho}{\sum_{\bf m}(p_{\bf m}^\sigma)^2},
\end{align}
where $p_{\bf m}^\rho$ and $p_{\bf m}^\sigma$ are the probabilities of observing the sequence of measurements $\bf m$ when the input state of the circuit is $\rho$ and $\sigma$, respectively. The summations are over all possible measurement outcomes ${\bf m}\in \mathbb{Z}_2^M$. We use $\overline{\chi}$ to denote the LXE averaged over different circuit realizations $C$. If there are more than one type of measurements in the circuit, e.g. $X$ and $ZZ$ measurements, one can use the same expression in Eq.\eqref{eq:chi_definition} to define the LXE for specific kinds of measurements, by including only the corresponding outcomes in $\bf{m}$. 

To gain some intuition about $\overline{\chi}$, it is helpful to consider two extreme cases. First, note that if the probability distributions of the measurement outcomes for the $\rho$ and $\sigma$ circuits are exactly the same, i.e. $p_{\bf m}^\rho=p_{\bf m}^\sigma$ for all $\bf{m}$, we have $\chi(\rho,\sigma)=1$. In this case, the measurement outcomes carry zero information about whether the initial state of the circuit has been $\rho$ or $\sigma$. On the other hand, if $p_{\bf m}^\rho$ is non-zero only for $\bf{m}$ with $p^\sigma_{\bf m}=0$ and vice versa, then the measurement outcomes in principle uniquely determine  which state has been used as the initial state, and we have $\chi=0$. Therefore, one can view $1-\chi$ as a measure of the information leaked to the environment about the initial states.

\subsection{Circuit model and initial states} 
\begin{figure}
\center{\includegraphics[width=3.5in]{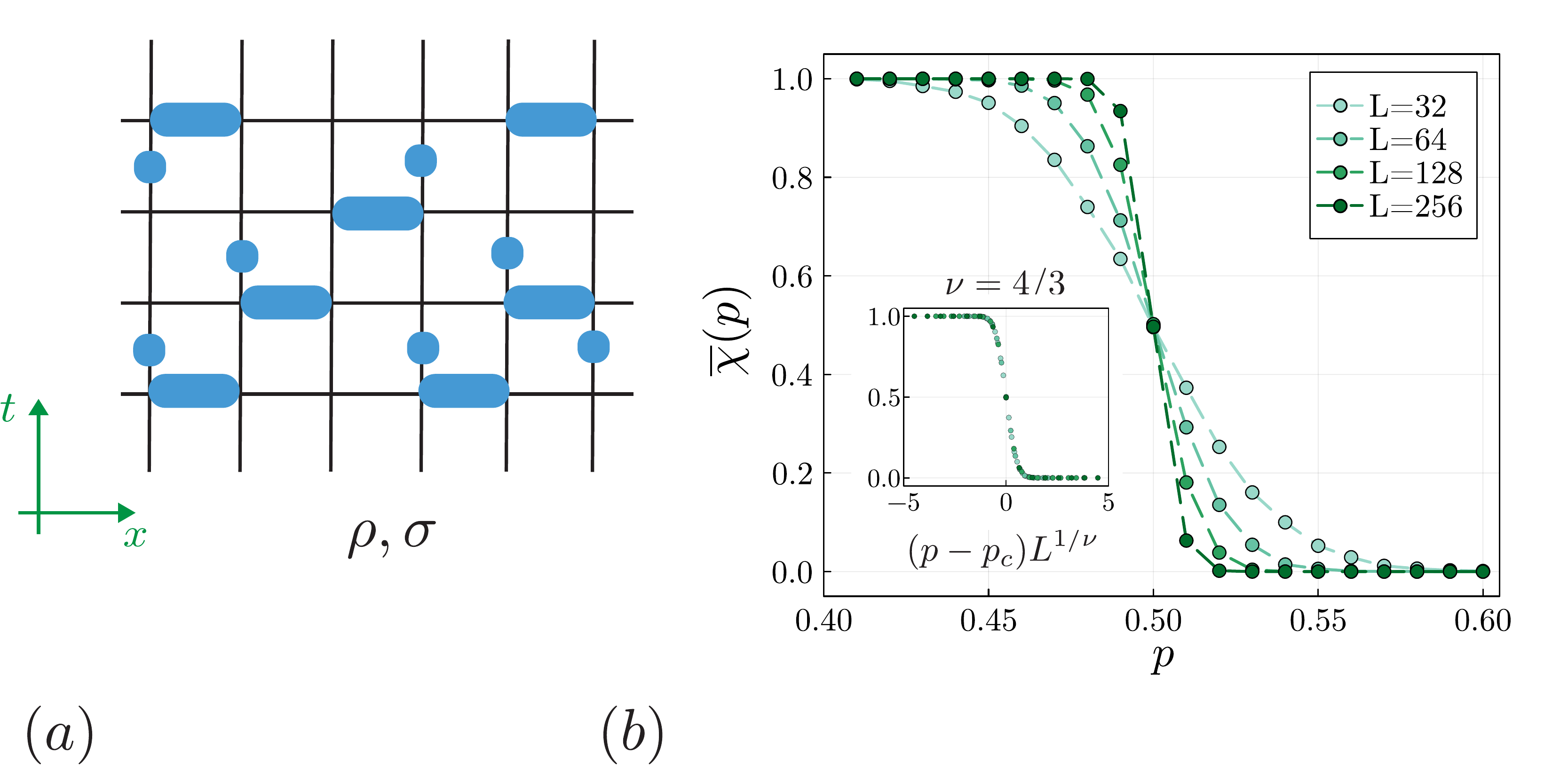}}
\caption{(a)  Typical architecture of the measurement-only $Z_2$ symmetric circuit with open boundary conditions. ${ZZ}$ measurement (operation acting on two qubits) is applied with probability $1-p$ and $X$ measurement (single qubit operation) with probability $p$. (b) Linear cross entropy in the circuit with circuit depth $T=L$. 
The crossing point (at $p=0.5$) is the location of the phase transition from spin glass ($p<0.5$) to paramagnet ($p>0.5$). Inset: scaling collapse with the critical exponent $\nu=4/3$. We average over $N_{iter}=1000$ iterations of the random circuit.} 
\label{Fig:MO_circuit}
\end{figure}

We start by showing that the linear cross entropy can detect the MIPT in a measurement-only $Z_2$-symmetric circuit studied in Ref.~\cite{hsieh2021,nahum2020entanglement}. Consider a $1D$ arrangement of $L$ qubits. The measurement-only dynamics consists of two types of measurements: two-qubit nearest neighbor $Z_iZ_{i+1}$ measurements and single-qubit $X_i$ measurements. Both measurements respect the global $\mathbb{Z}_2$ symmetry generated by
\begin{align}\label{eq:global_symmetry_gen}
    G=\prod_{i=1}^L X_i.
\end{align}
Each time step of the circuit is comprised of a layer of $ZZ$ measurements which is followed by a layer of $X$ measurements. The measurements are performed randomly with probability $p$ for $X$ measurements and probability $1-p$ for $ZZ$ measurements. 
We assume open boundary conditions unless explicitly stated otherwise.
A typical realization of the circuit is shown in Fig.~\ref{Fig:MO_circuit}(a).  The depth of the circuit which we denote by $T$, will be set to be $T=L$ throughout this Section. 

As is shown in Ref.~\cite{hsieh2021}, there is a MIPT between two different area law phases at the critical probability $p_c=1/2$. The phase for $p<p_c$ is characterized by a non-vanishing spin glass order parameter \cite{hsieh2021} in the late time states of the circuit  and accordingly is called the spin glass phase. The other area law phase for $p>p_c$ is called the paramagnetic phase, in which the spin glass order parameter vanishes.

Equivalently, one can view the $\mathbb{Z}_2$ symmetric circuit as a faulty implementation of the active error correction scheme for the quantum repetition code. The code space of the quantum repetition code is specified by the set of $Z_i Z_{i+1}$ stabilizers, and encodes one logical qubit. The symmetry generator $G$ in Eq.\eqref{eq:global_symmetry_gen} is the logical $X$ and the logical $Z$ can be taken to be $Z_1$. Accordingly, one can interpret the $ZZ$ measurements in the random circuit as the syndrome measurements of the quantum repetition code and the $X$ measurements as errors caused by the environment. As was shown in Refs.~\cite{lang2020entanglement,Fisher2021}, if the initial state of the circuit is in the code space of the quantum repetition code, within the spin-glass phase one can recover the initial state from the final state of the circuit after time $T=O(L)$, whereas within the paramagnetic phase the initial logical information encoded in the initial state would be lost within time $T=O(1)$. As such, the entanglement phase transition at $p=p_c$ can be thought of as a recoverability phase transition in the context of quantum error correction.This observation is a special case of the broader viewpoint that MIPT generally can be viewed as a phase transition in the ability of the random quantum circuit to hide information from the environment \cite{choi2020quantum,gullans2020dynamical,gullans2020scalable,kelly2022coherence}.


The error correction viewpoint can be used as a guide to choose the initial states $\rho$ and $\sigma$ such that $\overline{\chi}(\rho,\sigma)$ would be an order parameter for the phase transition. According to this view, if the initial state is in the code space of the quantum repetition code, the subsequent measurements in the circuit should not leak any information about the encoded state to the environment when $p<p_c$. Hence, if $\rho$ and $\sigma$ are two code states of the quantum repetition code, one would have $\overline{\chi}(\rho,\sigma)=1$ throughout the spin glass phase. To use $\overline{\chi}$ as an order parameter, then we need to find two code states $\rho$ and $\sigma$ such that in the paramagnetic phase $\overline{\chi}(\rho,\sigma)<1$. Given the prevalence of $X$ measurements in  the paramagnetic phase, one would expect that if $\rho$ and $\sigma$ were the eigenstates of the logical $X$ operator $G=\prod_{i=1}^L X_i$ with opposite signs, the measurements in the circuit could tell the difference within the paramagnetic phase $p>p_c$ and hence $\overline{\chi}(\rho,\sigma)$ would be less than $1$.  
Therefore, a potentially suitable choice  to detect the transition between two area law phases in the $\mathbb{Z}_2$ symmetric circuit would be the Greenberger-Horne-Zeilinger (GHZ) states, defined as,
\begin{align}
    \ket{\text{GHZ}^{(\pm)}}=\frac{1}{\sqrt{2}}\Big(|\uparrow\dots\uparrow\rangle\pm|\downarrow\dots\downarrow\rangle\Big),
\end{align}
which satisfy, $G\ket{\text{GHZ}^{(\pm)}}=\pm\ket{\text{GHZ}^{(\pm)}}$. 
In the next section, based on the mapping of the circuit to a 2D loop model, we show that the choice of such GHZ-type states gives a natural interpretation of LXE in terms of correlation functions of 2D percolation. 

Since the GHZ states are stabilizer states and the circuit consists of Pauli measurements, LXE can be computed efficiently through Clifford simulation (see Appendix \ref{app:XEB_clifford} for details on computing LXE in Clifford circuits). In Fig.~\ref{Fig:MO_circuit}(b) we plot the numerically obtained LXE for the two GHZ initial states. There are two phases: the spin glass phase ($p<0.5$) where LXE is reaching $\overline{\chi}=1$ in the thermodynamic limit as expected, and the paramagnet phase $p>0.5$ where $\overline{\chi}=0$ for large system sizes. The latter shows that the circuit always measures $G$ in the paramagnetic phase which is consistent with viewing the phase transition as a charge-sharpening phase transition \cite{agrawal2022entanglement,barratt2022field}. At $p=0.5$ there is a clear crossing which indicates the phase transition. 

At the critical point ($p=0.5$), LXE does not depend on system size but does depend on the aspect ratio of the circuit, as we discuss in detail in Section \ref{sec:CP}, where we consider the detailed properties of the critical point and obtain analytic results using percolation theory.
Data collapse of our numerical data for different system sizes (up to $L=256$) shows clearly that the critical exponent is in the percolation class $\nu=4/3$, as expected. 

In Appendix \ref{ZIZ_XX_model}, we study a modified circuit model in which in addition to $ZZ$ and $X$ measurement, with certain probabilities we also measure $ZIZ$ and $XX$ operators. While the critical point of this modified circuit model can still be fixed by  $X\longleftrightarrow ZZ$ duality \cite{hsieh2021,lin2023probing}, its dynamics no longer maps exactly to the classical percolation. As such, it allows us to test generality of our results away from the percolation fixed point. 

\subsection{Scrambling of initial states}
\label{sec:UE}

\begin{figure}
\center{\includegraphics[width=3.4in]{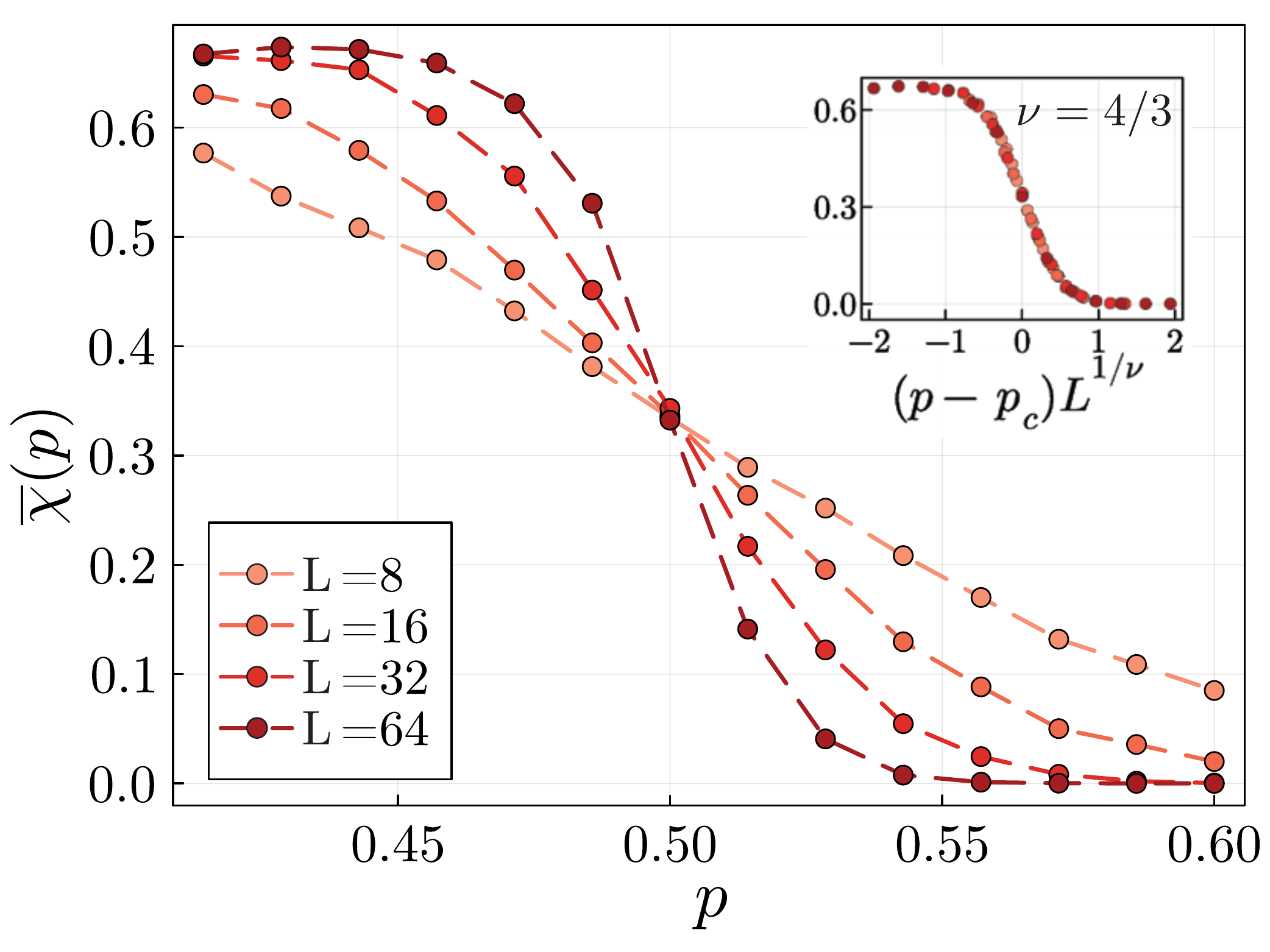}}
\caption{LXE in $ZZ-X$ circuit followed by the scrambling step for initial states $|\text{GHZ}_\pm\rangle$. In thermodynamic limit the cross entropy reaches $\approx2/3$ in spin glass phase. Inset: scaling collapse with the critical exponent $\nu=4/3$. }
\label{fig:chi_GHZ_SU}
\end{figure}

It is interesting to see how the cross entropy changes when the GHZ states are subject to a $\mathbb{Z}_2$ symmetry preserving  initial scrambling stage consisting of random unitaries, which we denote as $U$.  Here, we take the scrambling stage to consist of a brick-work arrangement of random 2-qubit unitary gates, with a depth equal to the system size.
As shown in Fig.~\ref{fig:chi_GHZ_SU},
in this protocol the cross entropy can still detect the two phases and the phase transition but its behavior in the spin glass phase is different compared to the circuit without the encoding step. In particular, in the thermodynamic limit the cross entropy approaches a constant value close to $2/3$ instead of $1$
 throughout the spin glass phase, and again vanishes in the paramagnet.
 
As we argue in Appendix \ref{app:scrambling}, the initial scrambling unitary might--rather counterintuitively--expose the initially non-local difference between $\ket{\text{GHZ}^{(+)}}$ and $\ket{\text{GHZ}^{(-)}}$ to local $ZZ$ measurements, resulting $\overline{\chi}$ to be less than one in the spin glass phase. $\overline{\chi}$ still vanishes for $p>0.5$ since the symmetric scrambling conserves the parity charges of $\ket{\textrm{GHZ}\pm}$, which then gets measured in the paramagnet phase. 
In Fig.~\ref{fig:chi_GHZ_SU} we show the numerical result and scaling collapse near the critical point which gives the same bulk critical exponent $\nu=4/3$, expected since only the initial states have changed due to the scrambling stage. 

\section{Critical point}
\label{sec:CP}

\subsection{Mapping to Two-Dimensional Percolation}
\begin{figure}[t]
$\begin{array}{cc}
              \includegraphics[width=.23\textwidth]{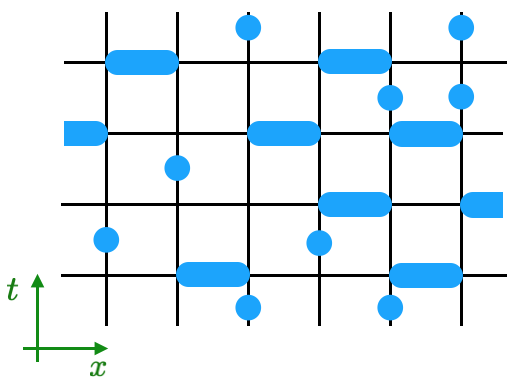}&
              \includegraphics[width=.23\textwidth]{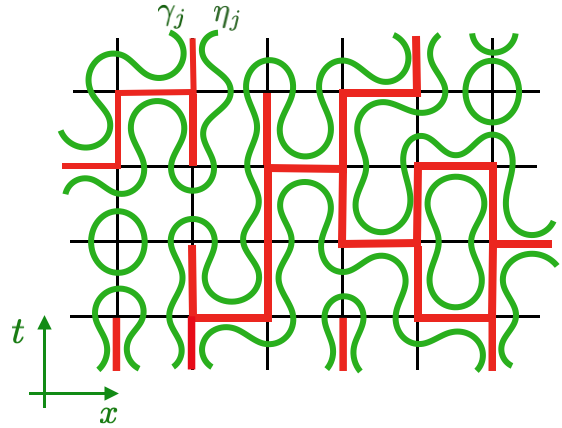}\\
              \text{(a)} & \text{(b)}
\end{array}$
         \caption{The $ZZ$-$X$ measurement-only circuit $(a)$.  The evolving stabilizers after a Jordan-Wigner transformation may be thought of as the hulls of percolating clusters on the square lattice, as in (b). }
         \label{fig:Percolation}
\end{figure}

Consider a one-dimensional array of $L$ qubits, which are acted upon by projective measurements of $Z_{j}Z_{j+1}$ and $X_{j}$ with probability $1-p$ and $p$, respectively.  A spacetime representation of this quantum circuit, for a given sequence of projective measurements, is shown in Fig.~\ref{fig:Percolation}a; the``boxes" on horizontal bonds indicate a measurement of the pair of Pauli $Z$ operators on adjacent sites, while the ``circles" on vertical bonds indicate a measurement of a single-site Pauli $X$ operator on that qubit. 

Starting with a stabilizer initial state $\ket{\psi_{0}}$ which is invariant under the symmetry transformation $G\equiv\prod_{j}X_{j}$, the evolving state of the qubits may be represented by an evolving pairing of Majorana fermions after a Jordan-Wigner (JW) transformation.  In a system with open boundary conditions, the JW transformation takes $X_{j}\rightarrow i\gamma_{j}\eta_{j}$ and $Z_{j}Z_{j+1}\rightarrow i\eta_{j}\gamma_{j+1}$.   The dynamics of these Majorana pairs may be understood as follows \cite{skinner2019measurement}.  First, we may color the bonds of the lattice where ($i$) a $ZZ$ measurement is performed and where ($ii$) \emph{no} $X$ measurement is performed.  The resulting square lattice with shaded bonds (in red) is shown in Fig.~\ref{fig:Percolation}b.   Because of the measurement probabilities, each bond of the square lattice is shaded with probability $1-p$.  When $p\le p_{c}$ ($p_{c} = 1/2$ for the square lattice) the shaded bonds form a percolating cluster spanning a finite fraction of the lattice.  We may now draw the \emph{hulls} of the regions of percolating clusters (in green).  A closed circle is drawn around lattice sites from which no shaded bonds emerge.  The resulting fully-packed loop configuration (FPLC) on the square lattice describes the spacetime trajectories of the Majorana pairs which stabilize the state, as repeated projective measurements are performed.

\subsection{LXE with Open Boundary Conditions and Percolation Crossing Probabilities}
We now run these dynamics, starting with either of the following initial states
\begin{align}\label{eq:state}
\ket{\Psi^{(\pm)}_{r} }\equiv \ket{\rightarrow\cdots}\otimes\ket{\mathrm{GHZ}^{(\pm)}_{r}}\otimes\ket{\rightarrow\cdots}
\end{align}
in which an $r$-qubit GHZ state $\ket{\mathrm{GHZ}^{(\pm)}_{r}} \sim \ket{\uparrow\cdots} \pm \ket{\downarrow\cdots}$ is surrounded by spins which are aligned in the $+x$ direction.  Without loss of generality, we take $L$ and $r$ to be even.  We initially consider dynamics with open boundary conditions, starting from initial  states $\ket{\Psi^{(\pm)}_{r} }$ in which the $r$-qubit GHZ state is centered in the middle of the $L$-site system.  For convenience of presentation, we number the lattice sites $j\in[-L/2,L/2]$.  

\begin{figure}
$\begin{array}{c}
              \includegraphics[width=.34\textwidth]{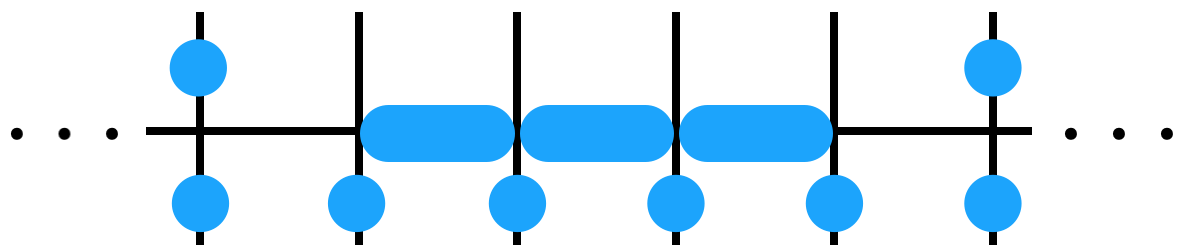} \\
              \text{(a)}\\
              \includegraphics[width=.34\textwidth]{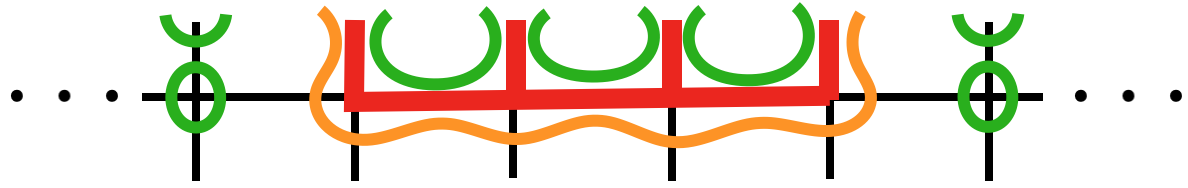} \\
              \text{(b)}
\end{array}$
                       \caption{The initial states $\ket{\Psi^{(\pm)}_{r=4}}$ may be prepared using the circuit in (a).  The observable distinguishing these states may be drawn as the orange strand in (b).  }
         \label{fig:GHZ_r}
\end{figure}

We may prepare the initial states $\ket{\Psi^{(\pm)}_{r}}$ by measuring each of the qubits in the Pauli $X$ basis, and then performing measurements of the two-qubit $ZZ$ stabilizers on adjacent pairs of the $r$ qubits in the center of the system. Adaptive feedback may then be used to deterministically prepare $\ket{\Psi^{(\pm)}_{r}}$ based on the measurement outcomes.  The composition of these two operations can be thought of as a sequence of ``forced" measurements of $X$ and $ZZ$, as represented in the circuit in Fig.~\ref{fig:GHZ_r}a to prepare $\ket{\Psi^{(\pm)}_{r=4} }$. A representation of this circuit as a colored configuration of bonds, following  in Fig.~\ref{fig:GHZ_r}a.

We may study the LXE between these two states, and averaged over realizations of the dynamics $\overline{\chi}$ as various parameters are tuned. In a given realization of the dynamics where projective measurements of $Z_{j}Z_{j+1}$ and $X_{j}$ are independently performed with probability $1-p$ and $p$, respectively, $\chi$ is zero if an operator that distinguishes the two initial states is measured during the dynamics, and is constant otherwise.  The generators of the stabilizer groups for the two states $\ket{\Psi_r^{(\pm)}}$ can be chosen to be identical, with the exception of one generator which distinguishes these states.  This choice for the stabilizer generators can be made at each time in the dynamics.  At the initial time, we may choose the stabilizer generator that distinguishes $\ket{\Psi_r^{(\pm)}}$ to be $\displaystyle Y_{1}\Big(\prod_{1< j < r}X_{j}\Big)Y_{r}$, which becomes the operator $i\gamma_{1}\eta_{r}$ after a Jordan-Wigner transformation, as indicated by the orange strand in Fig.~\ref{fig:GHZ_r}b.  The spacetime evolution of this operator as described by the evolving endpoints of the orange strand, describes a particular choice of stabilizer generator which distinguishes the two evolving states at any time.  A measurement of this operator corresponds to the endpoints of this strand connecting (forming a closed loop) during the evolution.

\begin{figure}
$\begin{array}{cc}
              \includegraphics[width=.2\textwidth]{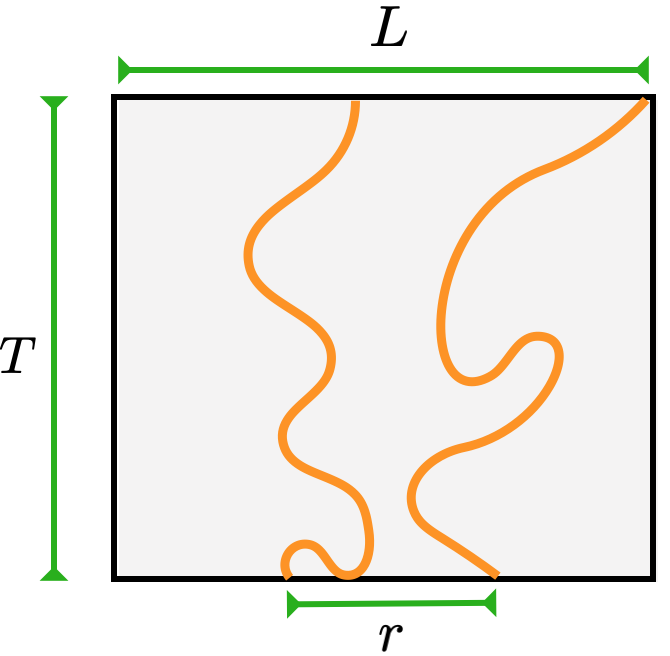} &
              \includegraphics[width=.17\textwidth]{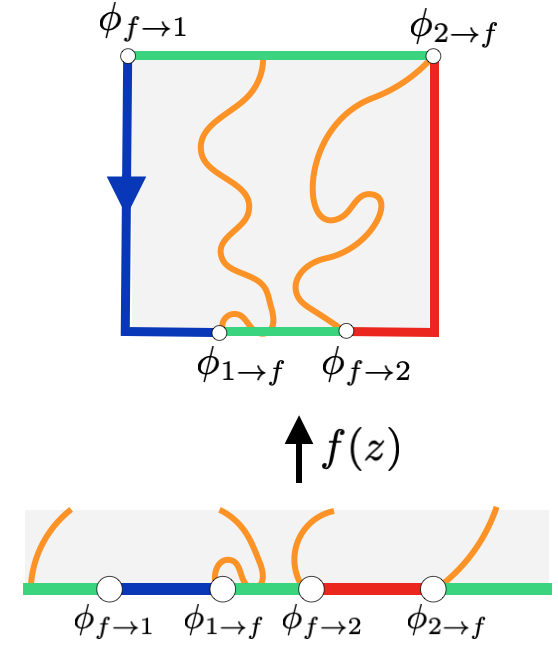} \\
              \text{(a)}&\text{(b)}
\end{array}$
                       \caption{With open boundary conditions, the LXE between the states in Eq. (\ref{eq:state}) is related to the \emph{crossing probability} in a rectangle of dimension $L\times T$ of a percolating hull that stretches between a region of width $r$ at one boundary and any part of the boundary a distance $T$ away, as shown in ($a$). This quantity is related to the four-point correlation function of boundary condition changing operators in the $q\rightarrow 1$ limit of a $q$-state Potts model, as shown in ($b$). }
         \label{fig:XEB_Potts}
\end{figure}

The averaged LXE $\overline{\chi}$ is then proportional to the fraction of trajectories of the orange hull which remain open until time $T$. A graphical depiction of such a trajectory is shown in Fig.~\ref{fig:XEB_Potts}a.  Calculation of similar quantities, related to the probability that a single cluster percolates across a finite geometry, have been obtained \cite{cardy1992critical,cardy1998number,smirnov2001critical} by considering bond percolation as the $q\rightarrow1$ limit of a $q$-state Potts model \cite{francesco2012conformal}.  Here, we review this relation, which we then use to determine the universal behavior of the linear cross-entropy at the phase transition.

In its simplest form, the two-dimensional Potts model describes the interaction of $q$-state degrees of freedom $\boldsymbol{\sigma}_{r}\in\{1,\ldots,q\}$ at each site $r$ on a square lattice via a Hamiltonian $\beta H = -\beta J\sum_{\langle r,s\rangle}\delta_{\mathbf{\sigma}_{r},\mathbf{\sigma}_{s}}$.  The Potts partition function may be written as  
\begin{align}
Z_{q}= \sum_{\Gamma}\left(e^{\beta J} - 1\right)^{B(\Gamma)}q^{C(\Gamma)}
\end{align}
where the sum is over configurations of colored bonds $\Gamma$ on the square lattice.  Here, $B(\Gamma)$ is the number of colored bonds, while $C(\Gamma)$ is the number of connected clusters.  Each connected cluster corresponds to a domain of aligned Potts spins.  In the $q\rightarrow 1$ limit, this partition sum manifestly describes percolation on the square lattice, and the boundaries of the Potts domains become the hulls of percolating clusters.

We now return to the LXE between the states in Eq. (\ref{eq:state}) in a system with $L$ qubits, in which the measurement-only dynamics have been run for a time $T$.  Let $Z_{q}$ now describe the Potts model on an $L\times T$ lattice with free boundary conditions.  It is natural to write the linear cross-entropy as $\overline{\chi} = \lim_{q\rightarrow 1}\left[ {Z'_{q}}/{Z_{q}}\right]$ where the partition sum $Z'_{q}$ is only over configurations  in which a single Potts domain connects any portion of the boundary at the final time $T$ with the strip of length $r$ at the initial time, as shown in Fig.~\ref{fig:XEB_Potts}a.  The counting of these configurations may be performed by fixing the Potts spins at the boundaries of the $L\times T$ system, and as such, the ratio $Z'_{q}/Z_{q}$ may be regarded as the four-point correlation function of appropriately-chosen \emph{boundary-condition-changing} (bcc) operators in the Potts model \cite{cardy1992critical}.  Specifically, we may insert bcc operators at the points $\boldsymbol{x}_{1} = (-L/2,T)$, $\boldsymbol{x}_{2} = (-r/2,0)$, $\boldsymbol{x}_{3} = (r/2,0)$, and $\boldsymbol{x}_{4} = (L/2,T)$, and formally write that
\begin{align}
\overline{\chi} &= \underset{q\rightarrow 1}{\lim}\Big[\langle\phi_{f\rightarrow1}\left(\boldsymbol{x}_{1}\right)\phi_{1\rightarrow f}\left(\boldsymbol{x}_{2}\right)\phi_{f\rightarrow2}\left(\boldsymbol{x}_{3}\right)\phi_{2\rightarrow f}\left(\boldsymbol{x}_{4}\right)\rangle\Big]\nonumber
\end{align}
where $\phi_{i\rightarrow f}(\boldsymbol{x})$ changes the boundary conditions from a region where the Potts spins are pinned to be in state $i\in\{1,\ldots,q\}$ to a region where the boundary conditions are free.  The insertion of these operators manifestly forces a Potts domain to connect the initial and final final-time boundaries as in Fig.~\ref{fig:XEB_Potts}a.  The scaling dimension of the operator $\phi_{i\rightarrow f}(\boldsymbol{x})$ is known as an analytic function of $q$, and vanishes in  the $q\rightarrow 1$ limit \cite{cardy1992critical}; this is sensible, given that $\overline{\chi}$ is identified with a percolation probability which is invariant under a uniform scaling of $L$ and $T$.  As a result, the linear cross entropy is described by a universal function near criticality 
\begin{align}
\overline{\chi}(p) = F\left(L^{1/\nu}(p-p_{c}), r/L,T/L\right)
\end{align}
with the correlation length exponent $\nu = 4/3$.

\begin{figure}
    $\begin{array}{c}
    \includegraphics[width=2.5in]{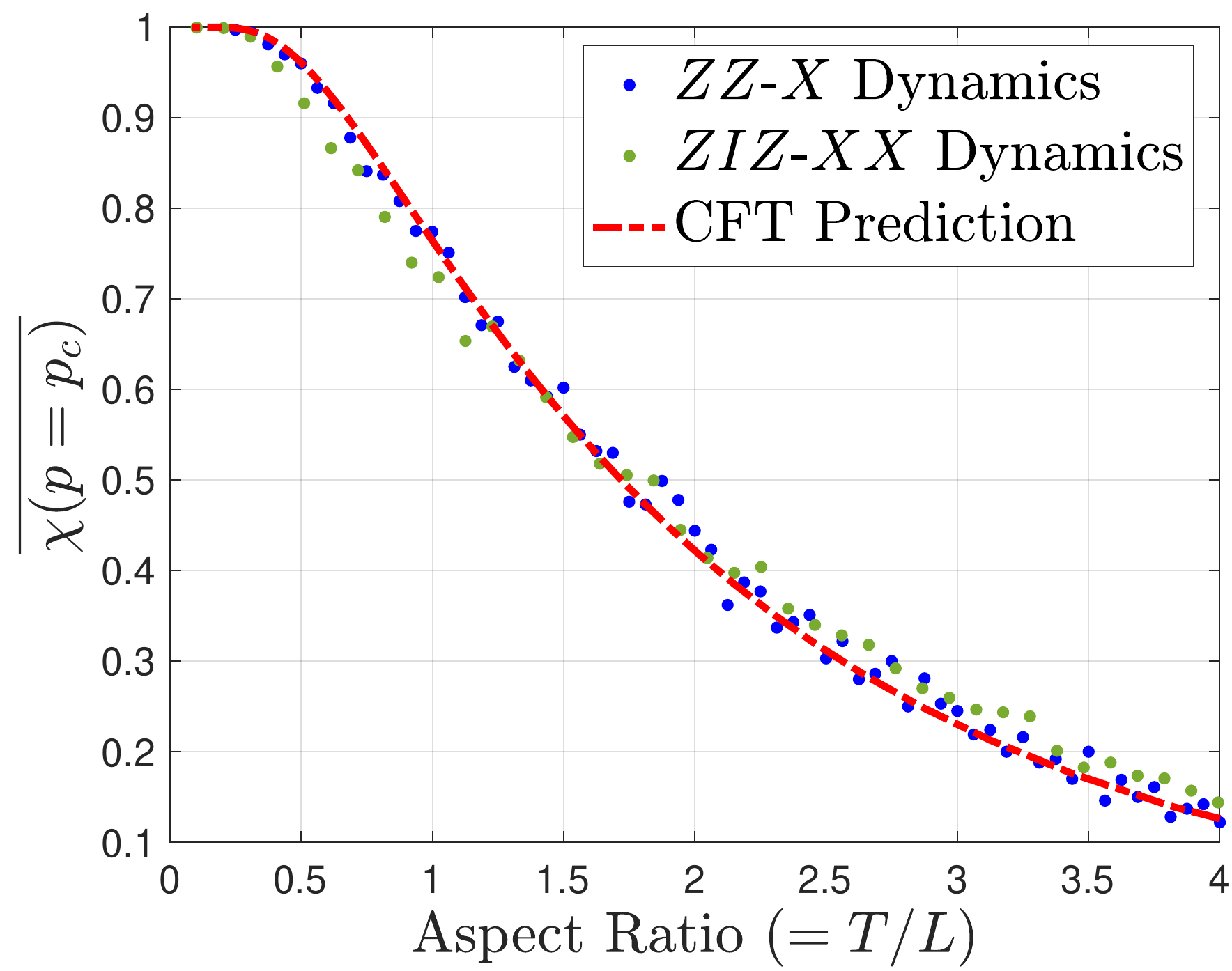}\\
    \text{(a)}\\
\includegraphics[width=2.7in]{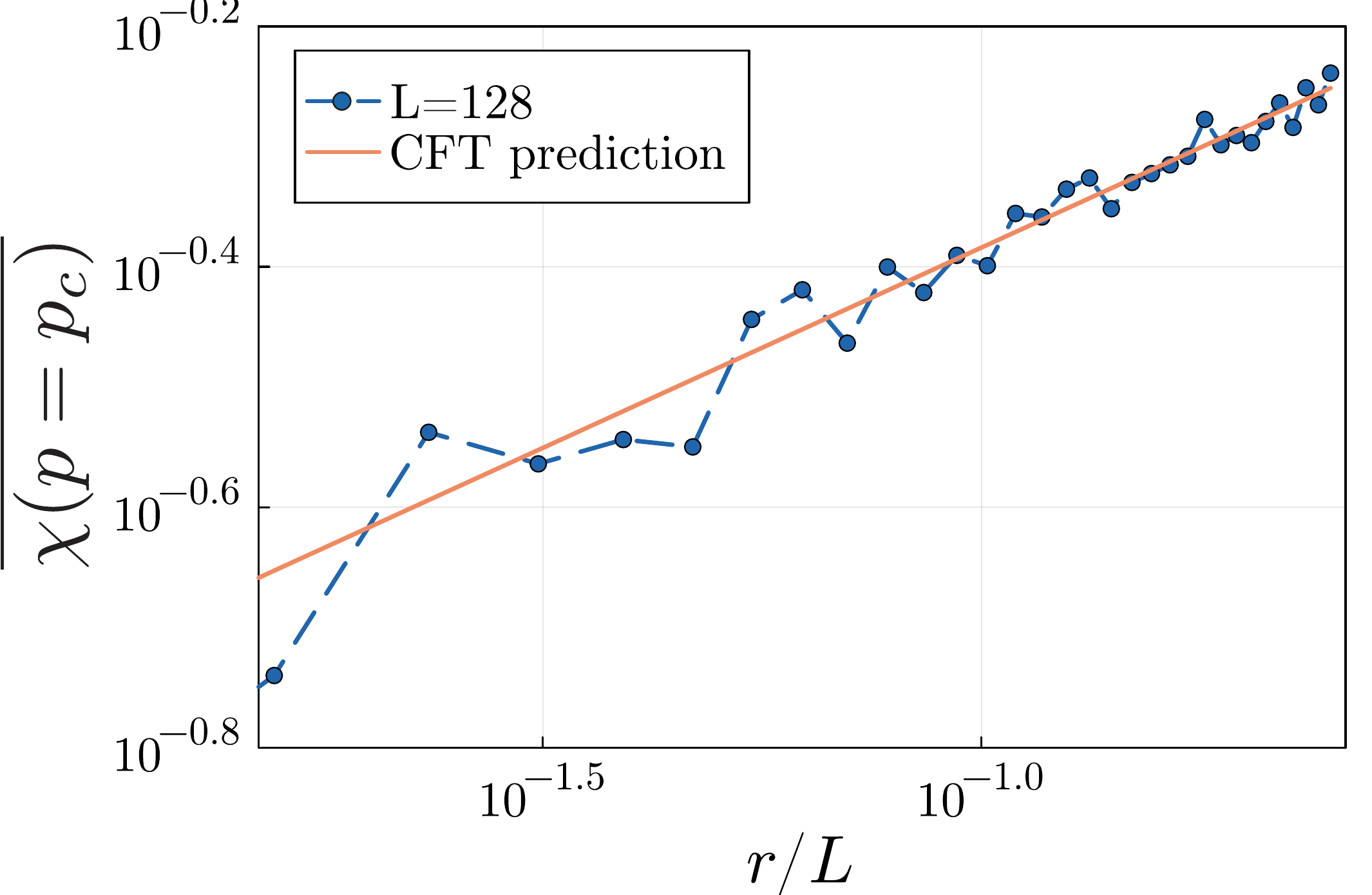}\\
\text{(b)}
\end{array}$
    \caption{LXE at the critical point in two different measurement-only $\mathbb{Z}_{2}$-symmetric circuits with open boundary conditions is shown in (a), and as a function of aspect ratio $T/L$. The theoretical curve is derived from Eq. (\ref{eq:four_pt_fn}) as explained in the text. In (b), we show the LXE at criticality, revealing the universal, power-law decay of $\overline{\chi}$ at criticality, as discussed in Eq. (\ref{eq:small_r_L}).}
    \label{fig:chi_GHZ_AR}
\end{figure}

The four-point correlation function of the boundary-condition-changing operators on the \emph{half-plane}, and with the points ${x}_{1}<x_{2}<x_{3}<x_{4}$ on the real axis is known \cite{cardy1992critical,francesco2012conformal} to be given by
\begin{align}
C(\{x_{i}\}) = \frac{3\Gamma\left(\frac{2}{3}\right)}{\Gamma\left(\frac{1}{3}\right)^{2}}\,\,(1-\eta)^{1/3}\,\,_{2}F_{1}\left(\frac{1}{3},\frac{2}{3},\frac{4}{3};1-\eta\right)\label{eq:four_pt_fn}
\end{align}
where the cross-ratio $\eta \equiv (x_{12}x_{34})/(x_{14}x_{23})$ with $x_{ij} \equiv |{x}_{i} -{x}_{j}|$. In Appendix \ref{app:conf_transf}, we perform a conformal transformation of the half-plane to a rectangle to relate this four-point function to the desired correlation function in which the bcc operators are inserted as shown in Fig.~\ref{fig:XEB_Potts}b. {The resulting function agrees quantitatively with the numerically-obtained linear cross-entropy; as an example, starting with the initial states $\ket{\Psi^{(\pm)}_{L}}$, the averaged linear cross-entropy as a function of the aspect ratio $T/L$ is shown in Fig.~\ref{fig:chi_GHZ_AR}a for two different measurement-only dynamics\footnote{The data points for $ZIZ-XX$ model correspond to the $p=q=1/2$ critical point in Fig.~\ref{fig:phase_diagram_LR}} which are tuned to the critical point separating the $\mathbb{Z}_{2}$ spin-glass and paramagnetic phases, along with the predicted value from the conformal transformation of Eq. (\ref{eq:four_pt_fn}).  Since the definition of a timestep in the numerical simulations is arbitrary, an overal re-scaling of the time $T$ has been performed in the numerical data to obtain agreement with the CFT prediction.}

Additionally, the operator-product-expansion 
\begin{align}
\lim_{\boldsymbol{x}\rightarrow\boldsymbol{y}}\phi_{1\rightarrow f}(\boldsymbol{x})\phi_{f\rightarrow 2}(\boldsymbol{y}) \sim \phi_{1\rightarrow 2}(\boldsymbol{y})
\end{align}
where $\phi_{1\rightarrow 2}(\boldsymbol{y})$ -- the bcc operator which changes the Potts spins from state $1$ to $2$, taken in the limit $q\rightarrow 1$ --   has scaling dimension $\Delta = 1/3$ \cite{cardy1992critical} implies that as $r/L \rightarrow 0$, the cross-entropy vanishes as \cite{francesco2012conformal} 
\begin{align}\label{eq:small_r_L}
\overline{\chi} \,\overset{p=p_{c}}{\sim} \,\left(\frac{r}{L}\right)^{\Delta}
\end{align}
Agreement between this prediction and numerical data is shown in Fig.~\ref{fig:chi_GHZ_AR}b.

\subsection{LXE with Periodic Boundary Conditions} 
\begin{figure}
              \includegraphics[width=.15\textwidth]{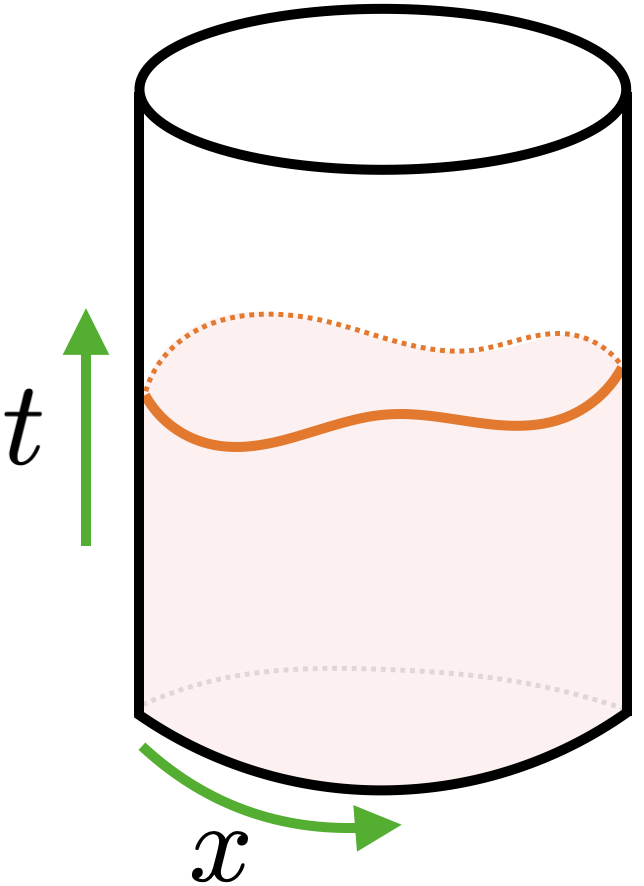}
               \caption{With periodic boundary conditions, the LXE between states $\ket{\Psi^{(\pm)}_{L}}$ probes the probability that on a cylinder with compact direction of length $L$ and height $T$, that there  is no percolating hull that wraps around the compact direction of the cylinder.  }
         \label{fig:PBC}
\end{figure}

With periodic boundary conditions, the linear cross-entropy probes other universal properties of the percolation critical point.  In the following discussion, we restrict our attention to the cross-entropy between the states $\ket{\Psi^{(\pm)}_{L}}$.  For a given realization of the measurement-only dynamics, we may again color the bonds of the quantum circuit depending on whether a $ZZ$ or $X$ measurement has been applied, as described previously.  After preparing the initial states $\ket{\Psi^{(\pm)}_{L}}$, all of the bonds corresponding to the initial time of the quantum circuit are colored.  The information about the observable $\prod_{j=1}^{L}X_{j}$ which distinguish the two initial states, is extracted in a given realization of the dynamics, if this region does not percolate to the final-time boundary of the quantum circuit. This would require that a percolating hull wraps around the compact direction of the cylinder, as shown in Fig.~\ref{fig:PBC}. As a result, the LXE in this setting is proportional to the probability that there is no non-contractible, percolating hull on the cylinder.

This probability may be computed by observing that percolation on the cylinder of height $\tau\in[0,T]$ and compact direction $x$ is described by the following continuum field theory for a bosonic field $\varphi(x,\tau)$ \cite{nienhuis1984critical,kondev1997liouville,saleur1987exact}:
\begin{align}\label{eq:CFT_percolation_cylinder}
    S[\varphi] = \frac{g}{4\pi}\int dx\,d\tau (\nabla\varphi)^{2} + i\delta\left[\varphi(0,T) - \varphi(0,0)\right]
\end{align}
where 
\begin{align}
g = \frac{2}{3},\hspace{.3in}\delta = \frac{1}{3}.    
\end{align}
While a careful derivation of this result has been presented in the literature (see e.g. \cite{kondev1997liouville}), we provide a self-contained, heuristic discussion of this result in Appendix \ref{app:field_theory_percolation}.  The action in Eq. (\ref{eq:CFT_percolation_cylinder}) derives from a continuum description of each connected cluster in percolation as a region of constant ``height".  The action (\ref{eq:CFT_percolation_cylinder}) describes the coarse-grained fluctuations of this continuum height-field $\varphi$, which is defined to jump $\varphi\rightarrow\varphi + \pi$ across a percolation hull.  The second term in the action is required on the cylinder, so that non-contractible percolation hulls appear with the same weight as contractible clusters in the path integral. 

As shown in Appendix \ref{app:field_theory_percolation}, the weights of non-contractible percolation hulls are affected by the insertion of operators $e^{i\alpha \varphi}$ at the ends of the cylinder, and as a result the LXE, which is given by the probability that there is no non-contractible percolating hull as shown in Fig.~\ref{fig:PBC}, can be written as a two-point correlation function 
\begin{align}
\overline{\chi(p_{c})} = \langle e^{i\varphi(0,T)/6}e^{-i\varphi(0,0)/6}\rangle
\end{align}
where the expectation value is taken with respect to the path integral with the action given in Eq. (\ref{eq:CFT_percolation_cylinder}).  Evaluating this correlation function on the cylinder gives
\begin{align}\label{eq:pbc_XEB_prediction}
    \overline{\chi(p_{c})} \sim \left[2\cosh\left(\frac{2\pi T}{L}\right) - 2\right]^{-\Delta}
\end{align}
where
\begin{align}
    \Delta = \frac{5}{48}
\end{align}
is twice the scaling dimension of the operator $\exp(i\varphi/6)$. 

\begin{figure}
              \includegraphics[width=.33\textwidth]{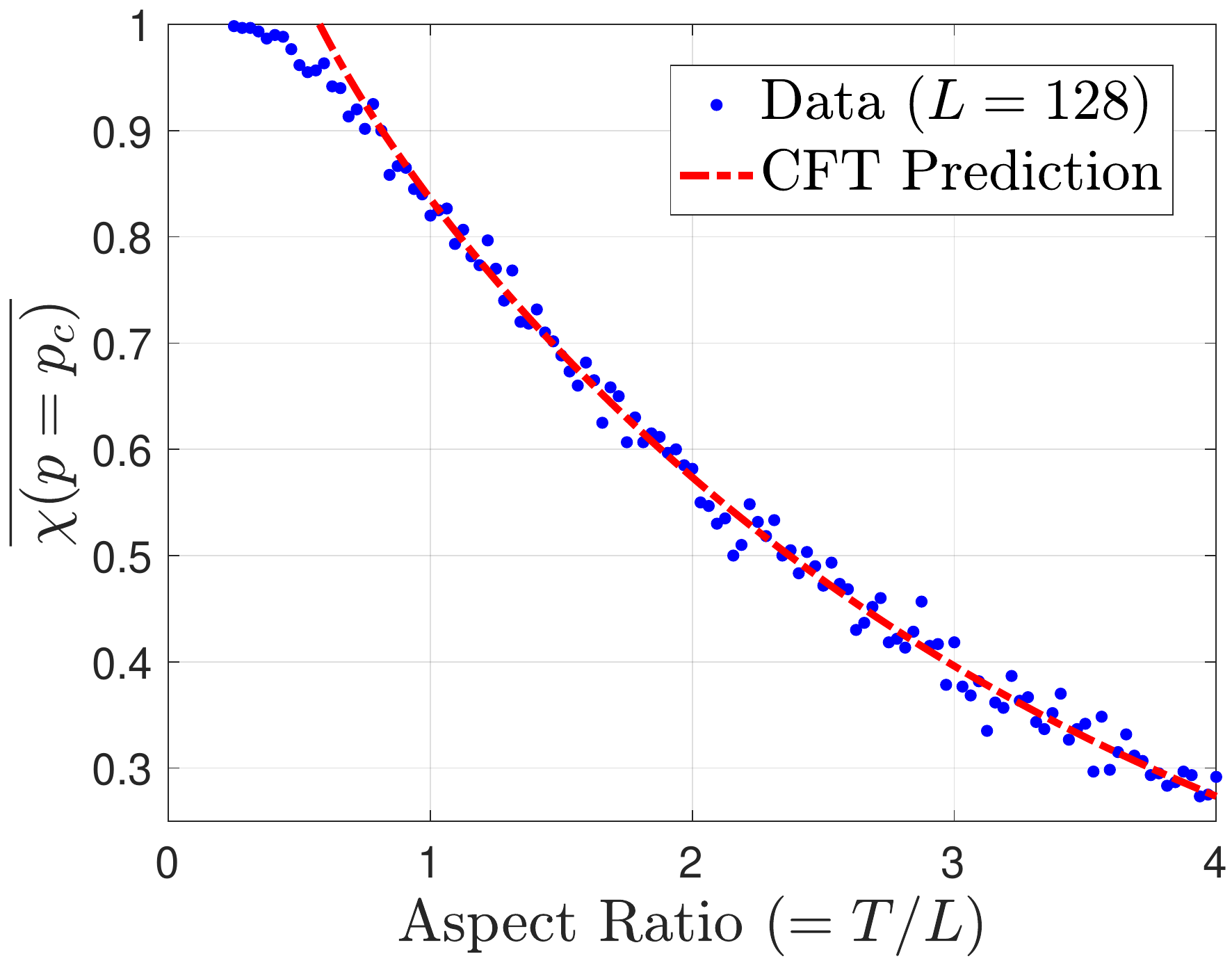}
               \caption{The LXE with periodic boundary conditions, between states $\ket{\Psi_{L}^{(\pm)}}$ as a function of the aspect ratio $T/L$, showing agreement with the predicted behavior in Eq. (\ref{eq:pbc_XEB_prediction}).}
         \label{fig:XEB_PBC}
\end{figure}

We compare this prediction for the scaling of the LXE with aspect ratio with numerical studies of the LXE in dynamics with two-qubit $ZZ$ measurements and single-qubit $X$ measurements.  A comparison is made by re-scaling time $T$ in the function (\ref{eq:pbc_XEB_prediction}) by the same factor used when studying the LXE with open boundary conditions.  One free parameter, given by an overall constant prefactor, is then used to re-scale the resulting function to fit the numerical data.  The resulting quantitative agreement is shown in Fig.~\ref{fig:XEB_PBC}.

\section{Hybrid circuit}
\label{sec:HC}

\begin{figure*}
\center{\includegraphics[width=\textwidth]{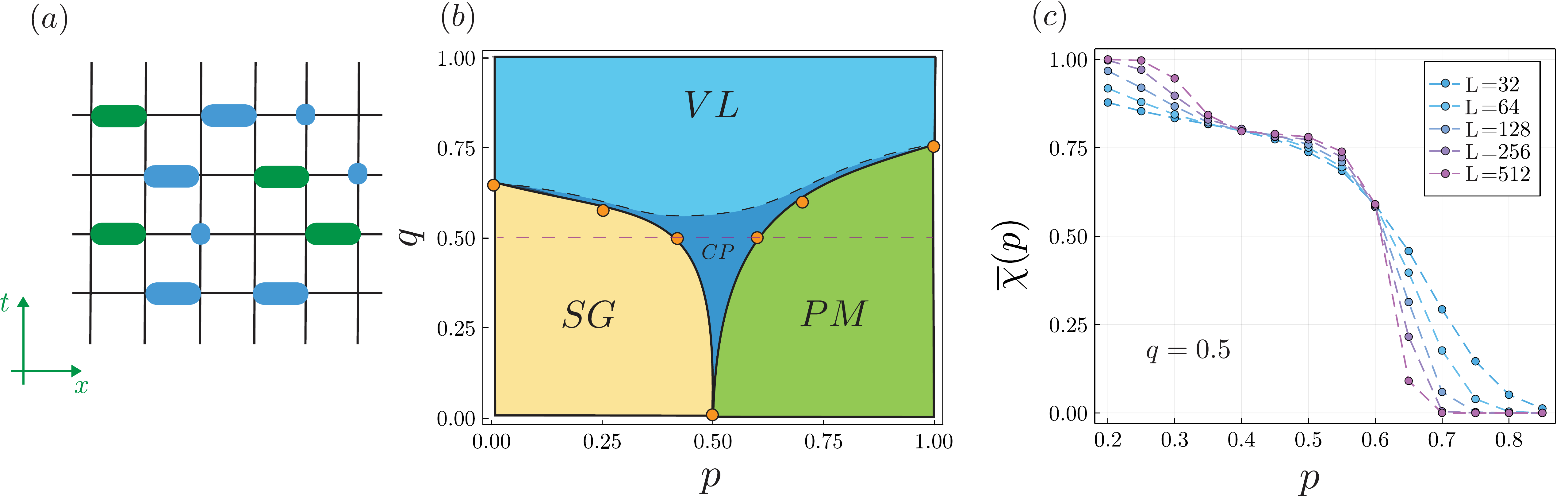}}
\caption{Phase diagram of $\mathbb{Z}_2$-symmetric quantum circuit. (a) Brickwork architecture of hybrid circuit with open boundary conditions. Blue operators correspond to measurements (${ZZ}$ for two qubits, and $X$ for single qubit measurements) and green operators correspond to $\mathbb{Z}_2$-symmetric unitaries. (b) Phase diagram of the model. Orange dots are numerical data computed for up to $L=512$ system size. Number of performed interations is $5000$.  For $|GHZ\pm\rangle$ initial states and in the thermodynamic limit, LXE can be characterized as follows. In the volume law (VL) phase as well as in the spin glass (SG) phase, LXE is $\overline{\chi}=1$ and in paramagnetic phase (PM), LXE becomes $\overline{\chi}=0$. The black dashed line between VL and CP marks the possible boundary of a critical region~\cite{hsieh2021}. (c) LXE along the horizontal dashed line in panel (b) at fixed rate of unitaries $q=0.5$.}
\label{fig:phase_diagram}
\end{figure*}
We now study LXE in the phase diagram of the hybrid circuit model studied in Ref.~\cite{hsieh2021}, where in addition to random ${ZZ}$ and $X$ measurements, the circuit includes random 2-qubit Clifford unitary gates $U$ that respect the $\mathbb{Z}_2$ symmetry. To satisfy this constraint, the unitary gates are required to have a property $U^\dagger XXU = XX$.  In this section we choose the brickwork architecture, similar to the architecture used in Ref.
~\cite{hsieh2021}, and impose open boundary conditions. We choose each brick to be either  $XI$ or ${ZZ}$ measurement or a random symmetric unitary $U$ with probabilities $p(1-q)$, $(1-p)(1-q)$ and $q$ respectively\footnote{We include single qubit $X$ measurements on the rightmost qubit in every other layer with the same probability as $XI$ measurements in the bulk of the circuit, to allow for $X$ measurement on the rightmost qubit in the brickwork architecture with open boundaries.}. A typical circuit realization is shown in Fig.~\ref{fig:phase_diagram}(a). 
The number of time steps in the circuit is equal to the system size $T=L$.

With the presence of the random unitary gates, in addition to the spin glass (SG) and paramagnetic (PM) phases, the system can support a volume law phase (VL), and possibly a critical phase\cite{hsieh2021} (see Fig.~\ref{fig:phase_diagram}(b)).  Besides detecting the phase transition between the two area law phases, SG and PM, (as discussed in Section \ref{sec:Z2}) with an appropriate choice of the initial states the LXE can detect the phase transitions between SG and VL, and between PM and VL.
For the phase transition between PM and VL we can again choose GHZ states since we expect LXE to be $\overline{\chi}=0$ in PM and $\overline{\chi}=1$ in VL in the thermodynamic limit. Therefore, the same initial states can be used to detect the phase transition using LXE. However, for the transition between SG and VL phases, a choice of GHZ initial states is rather inconvenient,
since we expect $\overline{\chi}=1$ in both phases and to be a constant value at the critical point. For this reason, we choose scrambled GHZ states as described in Section \ref{sec:UE} to detect the SG to VL phase boundary. We prepare the scrambled GHZ states by again introducing the ``scrambling" step in the circuit that consists of $\mathbb{Z}_2$ symmetric unitaries for time $T=L/2$, before running the main circuit for time $T=L/2$. 
As discussed in Section \ref{sec:Z2}, we expect LXE to be some non-zero constant in the SG phase in the thermodynamic limit (see Fig.~\ref{fig:chi_GHZ_SU}). Therefore, the phase transition becomes apparent as LXE takes different values in SG and VL phases
(see Fig.~\ref{fig:add_plots1} in Appendix \ref{apx:additional_plots}).

By changing the values of $q$ and $p$, we thereby obtain a phase diagram, as shown in Fig.~\ref{fig:phase_diagram}(b), which is consistent with the phase diagram found before \cite{hsieh2021, Fisher2021, bao2021}. First, we observe that upon increasing the rate of the unitaries $q$, the spin glass and paramagnetic phases eventually give way to a volume law phase.
 We note, however, that the critical phase (CP in the phase diagram Fig.~\ref{fig:phase_diagram}(b)) that was found in Ref.~\cite{hsieh2021} is not exactly apparent for the system sizes that we can reach, and for the number of circuit samples we can simulate. In Fig.~\ref{fig:phase_diagram}(c) we show the LXE for a horizontal cut through the phase diagram for $q=0.5$,
 with no scrambling step and with GHZ initial states. We observe two crossing points at $p\approx 0.4$ and $p\approx 0.6$. In the nominal critical phase, however, LXE increases slowly with system size while we expect LXE to be system size independent in a critical phase. It might be due to finite size effects and its resolution requires further numerical simulations which is beyond the scope of this work (see also Fig.~\ref{fig:add_plots2} in Appendix \ref{apx:additional_plots}).  


\section{LXE with symmetric noise}
\label{sec:Noise}
 Lastly, we study the effect of noise on LXE in the $\mathbb{Z}_2$ symmetric circuit. We focus on the $\mathbb{Z}_2$-symmetric single-qubit bit flip $X$ noise, 
 \begin{align}
     \mathcal{E}_X(\rho)=\frac{1}{2} \rho + \frac{1}{2} X\rho X,
 \end{align}
 where after each layer of measurements in the circuit, the channel $\mathcal{E}_X$ is applied to each qubit with some small probability $q$. 
 Furthermore, following Ref.~\cite{li2023cross}, here we consider the case when only the $\rho$ circuit is affected by the noise, imagining that the quantum simulation is performed on a noisy computer, while the classical simulation is noiseless.
 The initial states $\rho$ and $\sigma$ are chosen to be $\ket{\text{GHZ}\pm}$.
 
 As we shall explain below, the $X$ noise would have drastically different effects depending on whether the cross entropy is evaluated using only the $X$ measurement outcomes or all the measurement outcomes:  while the LXE would not be affected  at all by the presence of noise in the former case, in the latter it will completely vanish in the large system size limit. 

\subsection{Keeping track of the outcomes of $X$ measurements only}
First, we consider LXE when only $X$ measurement outcomes are included.
Note that in the measurement-only circuit with $\ket{GHZ\pm}$ initial states, the stabilizer group which describes the evolving states, $G_S$, retains a special structure.  In particular, the generators of $G_S$ can always be taken to be either purely a Pauli-$X$ string operator or purely a Pauli-$Z$ string operator. As such, $G_S$ may be expressed in the following form,
\begin{align}
    G_S=G_X\times G_Z,
\end{align}
where $G_X$ contains only Pauli-$X$ strings and $G_Z$ only contains Pauli-$Z$ strings. Furthermore, due to the specific choice of gates in the circuit, this structure persists throughout the circuit. Moreover, whether an $X_i$ measurement is random or deterministic -- and the outcome in case the measurement is deterministic -- can be determined completely from $G_X$ alone; If $ X_i$ is in $G_X$, then the measurement is deterministic with an outcome $+1$, if $-X_i \in G_X$ then the measurement is deterministic with the outcome $-1$, and if neither $X_i$ nor $-X_i$ is in $G_X$ then the measurement outcome is random. Importantly, the noise channel $\mathcal{E}_X$ does not alter $G_X$ and only changes $G_Z$. Moreover, the effect of $X$ and $ZZ$ measurements on $G_X$ is completely independent of $G_Z$. Therefore, as far as the probability distribution of $X$ measurement outcomes is concerned, the presence of $\mathcal{E}_X$ has no effect. As such, the LXE between the probability distribution of $X$ measurement outcomes in $\rho$ and $\sigma$ circuit does not change at all by the presence of the $\mathcal{E}_X$ noise channel.

\subsection{Keeping track of the outcomes of all measurements}
\begin{figure}
\center{\includegraphics[width=2.5in]{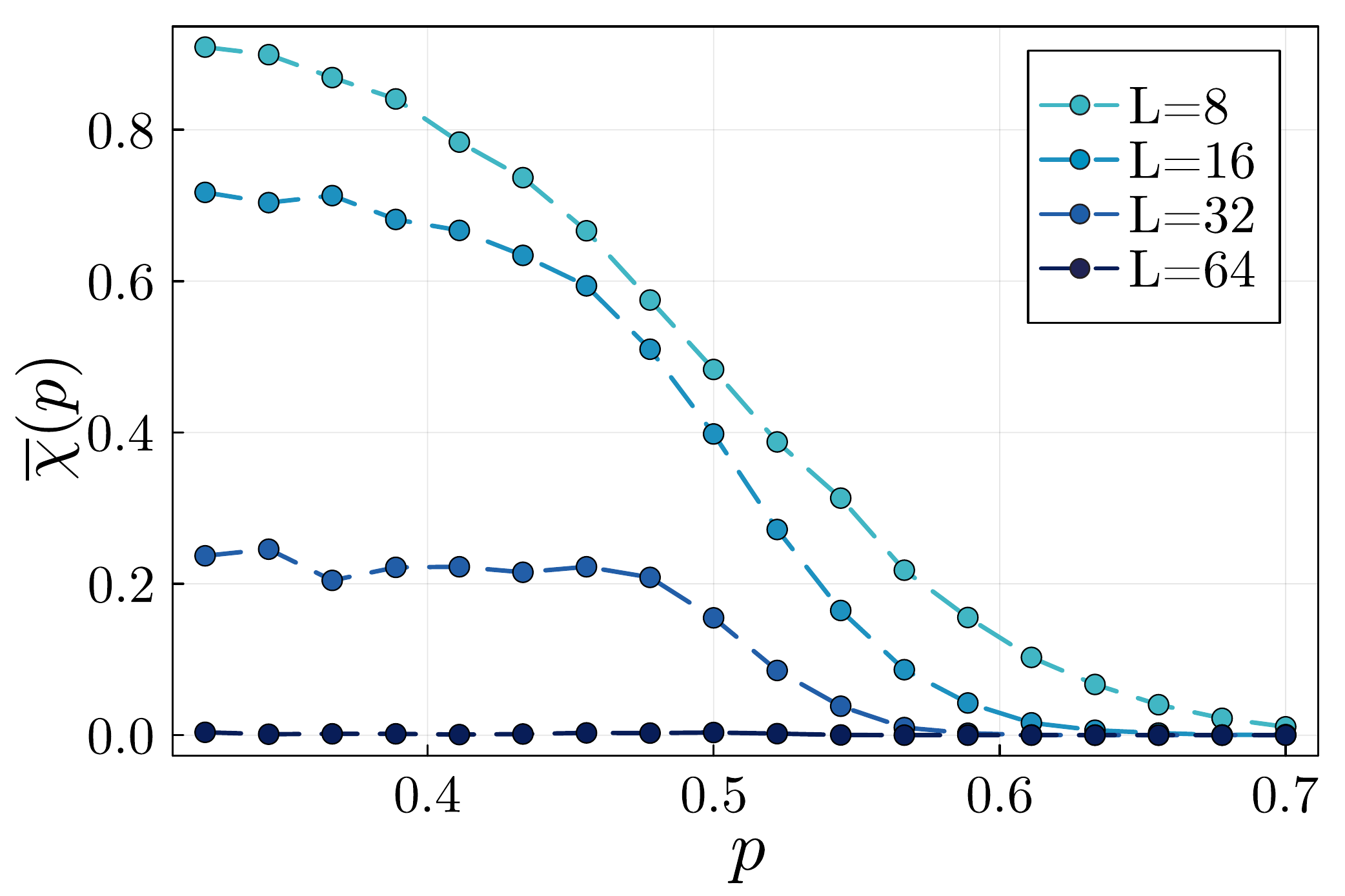}}
\caption{LXE as a function of $p$ at $q=0.01$ of $X$-noise. We keep track of all measurement outcomes. LXE goes to zero with increasing the system size. }
\label{fig:noise}
\end{figure}

We now consider the case where the LXE is evaluated based on all measurement outcomes in the circuit. Fig.~\ref{fig:noise} shows LXE as a function of $p$, where the noise rate is fixed at $q=0.01$ and suggests that in the thermodynamic limit the LXE will be $0$ throughout either of the two phases. Below, we show that this is indeed the case, and argue that this is the generic behavior in the presence of noise. 

\begin{figure}
    \centering
    \subfigure[]{{\label{fig:rhocircuit}\includegraphics[width=0.25\linewidth]{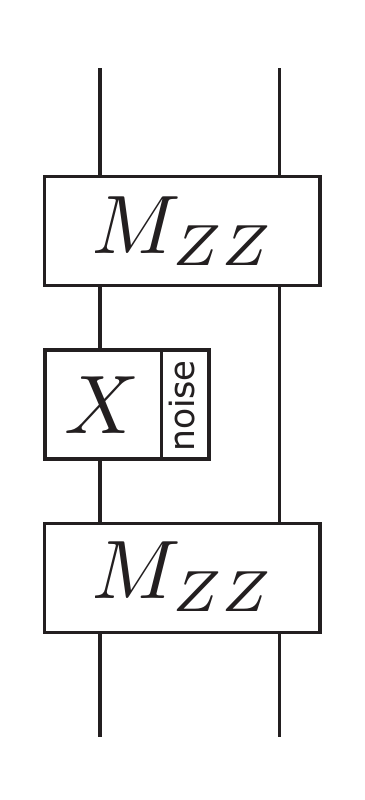}}}
    \subfigure[]{{\label{fig:sigmacircuit}\includegraphics[width=0.25\linewidth]{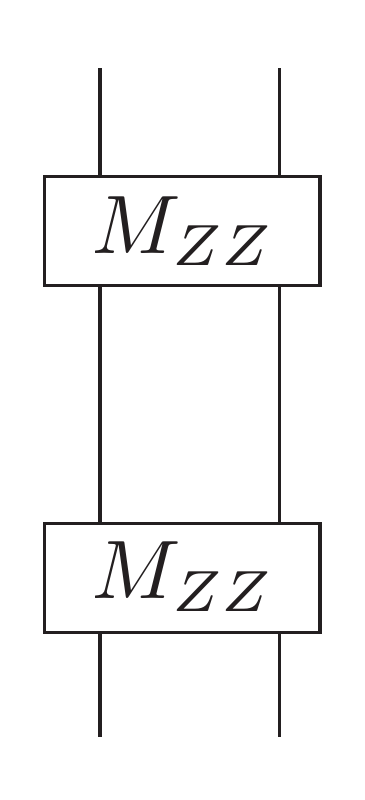}}}
    \caption{a) a part of the $\rho$ circuit, b) the corresponding part in the $\sigma$ circuit.}
    \label{fig:sbar_plots}
\end{figure}


Imagine that the specific sequence of gates shown in Fig.~\ref{fig:rhocircuit} appears somewhere in the $\rho$ circuit. Since the noise only occurs in the $\rho$ circuit, the corresponding part in the $\sigma$ circuit would look like Fig.~\ref{fig:sigmacircuit}. When running the $\rho$ circuit, say the outcome of the first $ZZ$ measurement is $+1$.  Then the $X$ noise decoheres the state and when $ZZ$ is measured for the second time, its outcome could be $-1$ with probability $1/2$. While this sequence of outcomes is compatible with the $\rho$ circuit, they are clearly incompatible with the $\sigma$ circuit because both measurements in Fig.~\ref{fig:sigmacircuit} have to have the same outcome and thus $p_{\vec{m}}^\sigma=0$. Given that this specific  sequence of gates with the aforementioned  outcomes is bound to appear somewhere in the $\rho$ circuit with probability one in the thermodynamic limit, any set of measurement outcomes sampled from the $\rho$ circuit is incompatible with $\sigma$ circuit with probability $1$ and hence $\overline{\chi}(\rho,\sigma)=0$. 

It is worth noting that the argument above is independent of the other details of the random circuit model and holds more generally even in cases with e.g. topological order or unitary scrambling. More precisely, if the noise channel occurs only in one circuit and if it does not commute with all the measurements which are used to compute LXE, then LXE will be zero in the thermodynamic limit. Nonetheless, the reason that LXE is zero, e.g. in the SG phase of the $\mathbb{Z}_2$-symmetric circuit, is not that the measurement outcomes can distinguish between the two initial states $\rho$ and $\sigma$ but rather that they distinguish between the noise-less and noisy circuit. Indeed, LXE would be still $0$ even if $\rho=\sigma$. This in turn suggests that one can first use the measurement outcomes to ``correct" for noise in the $\rho$ circuit, and then compute LXE between the corrected measurement outcome of the $\rho$ circuit and measurement outcomes of the $\sigma$ circuit. A detailed study of this idea is left for a future work. 

\section{Summary and Outlook}
\label{sec:outlook}
The paper presents a study on the utility of linear cross-entropy (LXE) as a probe of measurement-induced phase transitions. Specifically, we focussed on $\mathbb{Z}_2$ symmetric quantum circuits comprised of random $ZZ$ and $X$ measurements, which sustains different entanglement phases depending on the rate of the $X$ measurements. We demonstrated that by using appropriate initial states, LXE can be used as an order parameter to detect the phase transition between the spin glass phase and the paramagnetic phase in this circuit. Furthermore, we showed that at the critical point, LXE corresponds to a four-point correlation function of the underlying conformal field theory (CFT). We studied scaling properties of the correlation function for open and periodic boundary conditions of the circuit analytically.

We also explored the richer phase diagram of the circuit model in the presence of random $\mathbb{Z}_2$-symmetric unitary gates and showed that LXE can probe the phase transitions effectively if proper initial states are chosen. 
Finally, we considered and computed the effect of noise on the measurement-only circuit and proposed potential solutions to counter it.

Looking forward, it will be interesting to explore the LXE for other monitored circuits, for example to probe the MIPT between a topological and trivial phase.  

In the experimental setting one will be comparing the mid-circuit measurement distribution function when running a given circuit on a quantum processor with initial state $\rho$, with a classical computation performed with the same circuit but with a different initial state $\sigma$.  Observing a MIPT will be challenging due to the noise inherent in the quantum processor.  Indeed, for the spin-glass to paramagnetic transition studied in this paper, the presence of a bit flip noise channel drives the LXE to zero for large systems, destroying the crossing point indicative of the transition,
even for a low noise rate, see Fig.~\ref{fig:noise}. It might be possible to partly mitigate the effects of noise by first extracting the LXE when the two initial states are identical, $\rho = \sigma$, which would give $\chi =1$ for a noiseless quantum processor, but will be suppressed with the noise.  This suppression might allow for a baseline estimate of noise, to compare with results obtained for two different initial states.

\begin{acknowledgments}
We thank Ehud Altman, Timothy Hsieh and Yaodong Li for helpful discussions.

M.P.A.F is supported by the Heising-Simons Foundation and the Simons Collaboration on Ultra-Quantum Matter, which is a grant from the Simons Foundation (651457). M.T. is supported by the U.S. Department of Energy under Grant DE-SC0019030. This research was supported in part by the National Science Foundation under Grant No. NSF PHY-1748958, the Heising-Simons Foundation, and the Simons Foundation (216179, LB). The code for this paper was written in part using QuantumClifford.jl package \cite{QuantumClifford}. The computations in this paper were run on the FASRC Cannon cluster supported by the FAS Division of Science Research Computing Group at Harvard University.
\end{acknowledgments}
\bibliography{refs}

\begin{thebibliography}{65}%
\makeatletter
\providecommand \@ifxundefined [1]{%
 \@ifx{#1\undefined}
}%
\providecommand \@ifnum [1]{%
 \ifnum #1\expandafter \@firstoftwo
 \else \expandafter \@secondoftwo
 \fi
}%
\providecommand \@ifx [1]{%
 \ifx #1\expandafter \@firstoftwo
 \else \expandafter \@secondoftwo
 \fi
}%
\providecommand \natexlab [1]{#1}%
\providecommand \enquote  [1]{``#1''}%
\providecommand \bibnamefont  [1]{#1}%
\providecommand \bibfnamefont [1]{#1}%
\providecommand \citenamefont [1]{#1}%
\providecommand \href@noop [0]{\@secondoftwo}%
\providecommand \href [0]{\begingroup \@sanitize@url \@href}%
\providecommand \@href[1]{\@@startlink{#1}\@@href}%
\providecommand \@@href[1]{\endgroup#1\@@endlink}%
\providecommand \@sanitize@url [0]{\catcode `\\12\catcode `\$12\catcode
  `\&12\catcode `\#12\catcode `\^12\catcode `\_12\catcode `\%12\relax}%
\providecommand \@@startlink[1]{}%
\providecommand \@@endlink[0]{}%
\providecommand \url  [0]{\begingroup\@sanitize@url \@url }%
\providecommand \@url [1]{\endgroup\@href {#1}{\urlprefix }}%
\providecommand \urlprefix  [0]{URL }%
\providecommand \Eprint [0]{\href }%
\providecommand \doibase [0]{https://doi.org/}%
\providecommand \selectlanguage [0]{\@gobble}%
\providecommand \bibinfo  [0]{\@secondoftwo}%
\providecommand \bibfield  [0]{\@secondoftwo}%
\providecommand \translation [1]{[#1]}%
\providecommand \BibitemOpen [0]{}%
\providecommand \bibitemStop [0]{}%
\providecommand \bibitemNoStop [0]{.\EOS\space}%
\providecommand \EOS [0]{\spacefactor3000\relax}%
\providecommand \BibitemShut  [1]{\csname bibitem#1\endcsname}%
\let\auto@bib@innerbib\@empty
\bibitem [{\citenamefont {Fisher}\ \emph {et~al.}(2023)\citenamefont {Fisher},
  \citenamefont {Khemani}, \citenamefont {Nahum},\ and\ \citenamefont
  {Vijay}}]{fisher2023random}%
  \BibitemOpen
  \bibfield  {author} {\bibinfo {author} {\bibfnamefont {M.~P.}\ \bibnamefont
  {Fisher}}, \bibinfo {author} {\bibfnamefont {V.}~\bibnamefont {Khemani}},
  \bibinfo {author} {\bibfnamefont {A.}~\bibnamefont {Nahum}},\ and\ \bibinfo
  {author} {\bibfnamefont {S.}~\bibnamefont {Vijay}},\ }\bibfield  {title}
  {\bibinfo {title} {Random quantum circuits},\ }\href
  {https://doi.org/10.1146/annurev-conmatphys-031720-030658} {\bibfield
  {journal} {\bibinfo  {journal} {Annual Review of Condensed Matter Physics}\
  }\textbf {\bibinfo {volume} {14}},\ \bibinfo {pages} {335} (\bibinfo {year}
  {2023})}\BibitemShut {NoStop}%
\bibitem [{\citenamefont {Potter}\ and\ \citenamefont
  {Vasseur}(2022)}]{potter2022entanglement}%
  \BibitemOpen
  \bibfield  {author} {\bibinfo {author} {\bibfnamefont {A.~C.}\ \bibnamefont
  {Potter}}\ and\ \bibinfo {author} {\bibfnamefont {R.}~\bibnamefont
  {Vasseur}},\ }\bibfield  {title} {\bibinfo {title} {{E}ntanglement dynamics
  in hybrid quantum circuits},\ }in\ \href
  {https://link.springer.com/chapter/10.1007/978-3-031-03998-0_9} {\emph
  {\bibinfo {booktitle} {Entanglement in Spin Chains: From Theory to Quantum
  Technology Applications}}}\ (\bibinfo  {publisher} {Springer},\ \bibinfo
  {year} {2022})\ pp.\ \bibinfo {pages} {211--249}\BibitemShut {NoStop}%
\bibitem [{\citenamefont {Li}\ \emph {et~al.}(2018)\citenamefont {Li},
  \citenamefont {Chen},\ and\ \citenamefont {Fisher}}]{li2018quantum}%
  \BibitemOpen
  \bibfield  {author} {\bibinfo {author} {\bibfnamefont {Y.}~\bibnamefont
  {Li}}, \bibinfo {author} {\bibfnamefont {X.}~\bibnamefont {Chen}},\ and\
  \bibinfo {author} {\bibfnamefont {M.~P.~A.}\ \bibnamefont {Fisher}},\
  }\bibfield  {title} {\bibinfo {title} {{Q}uantum {Z}eno effect and the
  many-body entanglement transition},\ }\href
  {https://doi.org/10.1103/PhysRevB.98.205136} {\bibfield  {journal} {\bibinfo
  {journal} {Phys. Rev. B}\ }\textbf {\bibinfo {volume} {98}},\ \bibinfo
  {pages} {205136} (\bibinfo {year} {2018})}\BibitemShut {NoStop}%
\bibitem [{\citenamefont {Skinner}\ \emph {et~al.}(2019)\citenamefont
  {Skinner}, \citenamefont {Ruhman},\ and\ \citenamefont
  {Nahum}}]{skinner2019measurement}%
  \BibitemOpen
  \bibfield  {author} {\bibinfo {author} {\bibfnamefont {B.}~\bibnamefont
  {Skinner}}, \bibinfo {author} {\bibfnamefont {J.}~\bibnamefont {Ruhman}},\
  and\ \bibinfo {author} {\bibfnamefont {A.}~\bibnamefont {Nahum}},\ }\bibfield
   {title} {\bibinfo {title} {{M}easurement-induced phase transitions in the
  dynamics of entanglement},\ }\href
  {https://doi.org/10.1103/PhysRevX.9.031009} {\bibfield  {journal} {\bibinfo
  {journal} {Phys. Rev. X}\ }\textbf {\bibinfo {volume} {9}},\ \bibinfo {pages}
  {031009} (\bibinfo {year} {2019})}\BibitemShut {NoStop}%
\bibitem [{\citenamefont {Chan}\ \emph {et~al.}(2019)\citenamefont {Chan},
  \citenamefont {Nandkishore}, \citenamefont {Pretko},\ and\ \citenamefont
  {Smith}}]{chan2019unitary}%
  \BibitemOpen
  \bibfield  {author} {\bibinfo {author} {\bibfnamefont {A.}~\bibnamefont
  {Chan}}, \bibinfo {author} {\bibfnamefont {R.~M.}\ \bibnamefont
  {Nandkishore}}, \bibinfo {author} {\bibfnamefont {M.}~\bibnamefont
  {Pretko}},\ and\ \bibinfo {author} {\bibfnamefont {G.}~\bibnamefont
  {Smith}},\ }\bibfield  {title} {\bibinfo {title} {Unitary-projective
  entanglement dynamics},\ }\href {https://doi.org/10.1103/PhysRevB.99.224307}
  {\bibfield  {journal} {\bibinfo  {journal} {Phys. Rev. B}\ }\textbf {\bibinfo
  {volume} {99}},\ \bibinfo {pages} {224307} (\bibinfo {year}
  {2019})}\BibitemShut {NoStop}%
\bibitem [{\citenamefont {Choi}\ \emph {et~al.}(2020)\citenamefont {Choi},
  \citenamefont {Bao}, \citenamefont {Qi},\ and\ \citenamefont
  {Altman}}]{choi2020quantum}%
  \BibitemOpen
  \bibfield  {author} {\bibinfo {author} {\bibfnamefont {S.}~\bibnamefont
  {Choi}}, \bibinfo {author} {\bibfnamefont {Y.}~\bibnamefont {Bao}}, \bibinfo
  {author} {\bibfnamefont {X.-L.}\ \bibnamefont {Qi}},\ and\ \bibinfo {author}
  {\bibfnamefont {E.}~\bibnamefont {Altman}},\ }\bibfield  {title} {\bibinfo
  {title} {{Q}uantum error correction in scrambling dynamics and
  measurement-induced phase transition},\ }\href
  {https://doi.org/10.1103/PhysRevLett.125.030505} {\bibfield  {journal}
  {\bibinfo  {journal} {Phys. Rev. Lett.}\ }\textbf {\bibinfo {volume} {125}},\
  \bibinfo {pages} {030505} (\bibinfo {year} {2020})}\BibitemShut {NoStop}%
\bibitem [{\citenamefont {Bao}\ \emph {et~al.}(2020{\natexlab{a}})\citenamefont
  {Bao}, \citenamefont {Choi},\ and\ \citenamefont {Altman}}]{bao2020theory}%
  \BibitemOpen
  \bibfield  {author} {\bibinfo {author} {\bibfnamefont {Y.}~\bibnamefont
  {Bao}}, \bibinfo {author} {\bibfnamefont {S.}~\bibnamefont {Choi}},\ and\
  \bibinfo {author} {\bibfnamefont {E.}~\bibnamefont {Altman}},\ }\bibfield
  {title} {\bibinfo {title} {{T}heory of the phase transition in random unitary
  circuits with measurements},\ }\href
  {https://doi.org/10.1103/PhysRevB.101.104301} {\bibfield  {journal} {\bibinfo
   {journal} {Phys. Rev. B}\ }\textbf {\bibinfo {volume} {101}},\ \bibinfo
  {pages} {104301} (\bibinfo {year} {2020}{\natexlab{a}})}\BibitemShut
  {NoStop}%
\bibitem [{\citenamefont {Jian}\ \emph
  {et~al.}(2020{\natexlab{a}})\citenamefont {Jian}, \citenamefont {You},
  \citenamefont {Vasseur},\ and\ \citenamefont {Ludwig}}]{jian2020measurement}%
  \BibitemOpen
  \bibfield  {author} {\bibinfo {author} {\bibfnamefont {C.-M.}\ \bibnamefont
  {Jian}}, \bibinfo {author} {\bibfnamefont {Y.-Z.}\ \bibnamefont {You}},
  \bibinfo {author} {\bibfnamefont {R.}~\bibnamefont {Vasseur}},\ and\ \bibinfo
  {author} {\bibfnamefont {A.~W.~W.}\ \bibnamefont {Ludwig}},\ }\bibfield
  {title} {\bibinfo {title} {{M}easurement-induced criticality in random
  quantum circuits},\ }\href {https://doi.org/10.1103/PhysRevB.101.104302}
  {\bibfield  {journal} {\bibinfo  {journal} {Phys. Rev. B}\ }\textbf {\bibinfo
  {volume} {101}},\ \bibinfo {pages} {104302} (\bibinfo {year}
  {2020}{\natexlab{a}})}\BibitemShut {NoStop}%
\bibitem [{\citenamefont {Li}\ \emph {et~al.}(2023{\natexlab{a}})\citenamefont
  {Li}, \citenamefont {Vijay},\ and\ \citenamefont
  {Fisher}}]{li2023entanglement}%
  \BibitemOpen
  \bibfield  {author} {\bibinfo {author} {\bibfnamefont {Y.}~\bibnamefont
  {Li}}, \bibinfo {author} {\bibfnamefont {S.}~\bibnamefont {Vijay}},\ and\
  \bibinfo {author} {\bibfnamefont {M.~P.}\ \bibnamefont {Fisher}},\ }\bibfield
   {title} {\bibinfo {title} {Entanglement domain walls in monitored quantum
  circuits and the directed polymer in a random environment},\ }\href
  {https://doi.org/10.1103/PRXQuantum.4.010331} {\bibfield  {journal} {\bibinfo
   {journal} {PRX Quantum}\ }\textbf {\bibinfo {volume} {4}},\ \bibinfo {pages}
  {010331} (\bibinfo {year} {2023}{\natexlab{a}})}\BibitemShut {NoStop}%
\bibitem [{\citenamefont {Nahum}\ \emph {et~al.}(2021)\citenamefont {Nahum},
  \citenamefont {Roy}, \citenamefont {Skinner},\ and\ \citenamefont
  {Ruhman}}]{nahum2021measurement}%
  \BibitemOpen
  \bibfield  {author} {\bibinfo {author} {\bibfnamefont {A.}~\bibnamefont
  {Nahum}}, \bibinfo {author} {\bibfnamefont {S.}~\bibnamefont {Roy}}, \bibinfo
  {author} {\bibfnamefont {B.}~\bibnamefont {Skinner}},\ and\ \bibinfo {author}
  {\bibfnamefont {J.}~\bibnamefont {Ruhman}},\ }\bibfield  {title} {\bibinfo
  {title} {{M}easurement and entanglement phase transitions in all-to-all
  quantum circuits, on quantum trees, and in {L}andau-{G}insburg theory},\
  }\href {https://doi.org/10.1103/PRXQuantum.2.010352} {\bibfield  {journal}
  {\bibinfo  {journal} {PRX Quantum}\ }\textbf {\bibinfo {volume} {2}},\
  \bibinfo {pages} {010352} (\bibinfo {year} {2021})}\BibitemShut {NoStop}%
\bibitem [{\citenamefont {Turkeshi}\ \emph {et~al.}(2020)\citenamefont
  {Turkeshi}, \citenamefont {Fazio},\ and\ \citenamefont
  {Dalmonte}}]{turkeshi2020measurement}%
  \BibitemOpen
  \bibfield  {author} {\bibinfo {author} {\bibfnamefont {X.}~\bibnamefont
  {Turkeshi}}, \bibinfo {author} {\bibfnamefont {R.}~\bibnamefont {Fazio}},\
  and\ \bibinfo {author} {\bibfnamefont {M.}~\bibnamefont {Dalmonte}},\
  }\bibfield  {title} {\bibinfo {title} {{M}easurement-induced criticality in
  $(2+1)$-dimensional hybrid quantum circuits},\ }\href
  {https://doi.org/10.1103/PhysRevB.102.014315} {\bibfield  {journal} {\bibinfo
   {journal} {Phys. Rev. B}\ }\textbf {\bibinfo {volume} {102}},\ \bibinfo
  {pages} {014315} (\bibinfo {year} {2020})}\BibitemShut {NoStop}%
\bibitem [{\citenamefont {Vijay}(2020)}]{vijay2020measurement}%
  \BibitemOpen
  \bibfield  {author} {\bibinfo {author} {\bibfnamefont {S.}~\bibnamefont
  {Vijay}},\ }\href@noop {} {\bibinfo {title} {Measurement-driven phase
  transition within a volume-law entangled phase}} (\bibinfo {year} {2020}),\
  \Eprint {https://arxiv.org/abs/arXiv: 2005.03052} {arXiv: 2005.03052}
  \BibitemShut {NoStop}%
\bibitem [{\citenamefont {Sang}\ and\ \citenamefont {Hsieh}(2021)}]{hsieh2021}%
  \BibitemOpen
  \bibfield  {author} {\bibinfo {author} {\bibfnamefont {S.}~\bibnamefont
  {Sang}}\ and\ \bibinfo {author} {\bibfnamefont {T.~H.}\ \bibnamefont
  {Hsieh}},\ }\bibfield  {title} {\bibinfo {title} {{M}easurement-protected
  quantum phases},\ }\href {https://doi.org/10.1103/PhysRevResearch.3.023200}
  {\bibfield  {journal} {\bibinfo  {journal} {Phys. Rev. Research}\ }\textbf
  {\bibinfo {volume} {3}},\ \bibinfo {pages} {023200} (\bibinfo {year}
  {2021})}\BibitemShut {NoStop}%
\bibitem [{\citenamefont {Lavasani}\ \emph {et~al.}(2021)\citenamefont
  {Lavasani}, \citenamefont {Alavirad},\ and\ \citenamefont
  {Barkeshli}}]{lavasani2021topological}%
  \BibitemOpen
  \bibfield  {author} {\bibinfo {author} {\bibfnamefont {A.}~\bibnamefont
  {Lavasani}}, \bibinfo {author} {\bibfnamefont {Y.}~\bibnamefont {Alavirad}},\
  and\ \bibinfo {author} {\bibfnamefont {M.}~\bibnamefont {Barkeshli}},\
  }\bibfield  {title} {\bibinfo {title} {{T}opological order and criticality in
  $(2+1)${D} monitored random quantum circuits},\ }\href
  {https://doi.org/10.1103/PhysRevLett.127.235701} {\bibfield  {journal}
  {\bibinfo  {journal} {Phys. Rev. Lett.}\ }\textbf {\bibinfo {volume} {127}},\
  \bibinfo {pages} {235701} (\bibinfo {year} {2021})}\BibitemShut {NoStop}%
\bibitem [{\citenamefont {Sharma}\ \emph {et~al.}(2022)\citenamefont {Sharma},
  \citenamefont {Turkeshi}, \citenamefont {Fazio},\ and\ \citenamefont
  {Dalmonte}}]{sharma2022measurement}%
  \BibitemOpen
  \bibfield  {author} {\bibinfo {author} {\bibfnamefont {S.}~\bibnamefont
  {Sharma}}, \bibinfo {author} {\bibfnamefont {X.}~\bibnamefont {Turkeshi}},
  \bibinfo {author} {\bibfnamefont {R.}~\bibnamefont {Fazio}},\ and\ \bibinfo
  {author} {\bibfnamefont {M.}~\bibnamefont {Dalmonte}},\ }\bibfield  {title}
  {\bibinfo {title} {{M}easurement-induced criticality in extended and
  long-range unitary circuits},\ }\href
  {https://scipost.org/SciPostPhysCore.5.2.023/pdf} {\bibfield  {journal}
  {\bibinfo  {journal} {SciPost Physics Core}\ }\textbf {\bibinfo {volume}
  {5}},\ \bibinfo {pages} {023} (\bibinfo {year} {2022})}\BibitemShut {NoStop}%
\bibitem [{\citenamefont {Lunt}\ \emph {et~al.}(2021)\citenamefont {Lunt},
  \citenamefont {Szyniszewski},\ and\ \citenamefont
  {Pal}}]{lunt2021measurement}%
  \BibitemOpen
  \bibfield  {author} {\bibinfo {author} {\bibfnamefont {O.}~\bibnamefont
  {Lunt}}, \bibinfo {author} {\bibfnamefont {M.}~\bibnamefont {Szyniszewski}},\
  and\ \bibinfo {author} {\bibfnamefont {A.}~\bibnamefont {Pal}},\ }\bibfield
  {title} {\bibinfo {title} {{M}easurement-induced criticality and entanglement
  clusters: {A} study of one-dimensional and two-dimensional {C}lifford
  circuits},\ }\href {https://doi.org/10.1103/PhysRevB.104.155111} {\bibfield
  {journal} {\bibinfo  {journal} {Phys. Rev. B}\ }\textbf {\bibinfo {volume}
  {104}},\ \bibinfo {pages} {155111} (\bibinfo {year} {2021})}\BibitemShut
  {NoStop}%
\bibitem [{\citenamefont {Bao}\ \emph {et~al.}(2021{\natexlab{a}})\citenamefont
  {Bao}, \citenamefont {Choi},\ and\ \citenamefont {Altman}}]{bao2021symmetry}%
  \BibitemOpen
  \bibfield  {author} {\bibinfo {author} {\bibfnamefont {Y.}~\bibnamefont
  {Bao}}, \bibinfo {author} {\bibfnamefont {S.}~\bibnamefont {Choi}},\ and\
  \bibinfo {author} {\bibfnamefont {E.}~\bibnamefont {Altman}},\ }\bibfield
  {title} {\bibinfo {title} {{S}ymmetry enriched phases of quantum circuits},\
  }\href {https://doi.org/10.1016/j.aop.2021.168618} {\bibfield  {journal}
  {\bibinfo  {journal} {Annals of Physics}\ }\textbf {\bibinfo {volume}
  {435}},\ \bibinfo {pages} {168618} (\bibinfo {year}
  {2021}{\natexlab{a}})}\BibitemShut {NoStop}%
\bibitem [{\citenamefont {Weinstein}\ \emph
  {et~al.}(2022{\natexlab{a}})\citenamefont {Weinstein}, \citenamefont {Bao},\
  and\ \citenamefont {Altman}}]{weinstein2022measurement}%
  \BibitemOpen
  \bibfield  {author} {\bibinfo {author} {\bibfnamefont {Z.}~\bibnamefont
  {Weinstein}}, \bibinfo {author} {\bibfnamefont {Y.}~\bibnamefont {Bao}},\
  and\ \bibinfo {author} {\bibfnamefont {E.}~\bibnamefont {Altman}},\
  }\bibfield  {title} {\bibinfo {title} {{M}easurement-induced power-law
  negativity in an open monitored quantum circuit},\ }\href
  {https://doi.org/10.1103/PhysRevLett.129.080501} {\bibfield  {journal}
  {\bibinfo  {journal} {Phys. Rev. Lett.}\ }\textbf {\bibinfo {volume} {129}},\
  \bibinfo {pages} {080501} (\bibinfo {year} {2022}{\natexlab{a}})}\BibitemShut
  {NoStop}%
\bibitem [{\citenamefont {Alberton}\ \emph {et~al.}(2021)\citenamefont
  {Alberton}, \citenamefont {Buchhold},\ and\ \citenamefont
  {Diehl}}]{alberton2021entanglement}%
  \BibitemOpen
  \bibfield  {author} {\bibinfo {author} {\bibfnamefont {O.}~\bibnamefont
  {Alberton}}, \bibinfo {author} {\bibfnamefont {M.}~\bibnamefont {Buchhold}},\
  and\ \bibinfo {author} {\bibfnamefont {S.}~\bibnamefont {Diehl}},\ }\bibfield
   {title} {\bibinfo {title} {{E}ntanglement transition in a monitored
  free-fermion chain: {F}rom extended criticality to area law},\ }\href
  {https://doi.org/10.1103/PhysRevLett.126.170602} {\bibfield  {journal}
  {\bibinfo  {journal} {Phys. Rev. Lett.}\ }\textbf {\bibinfo {volume} {126}},\
  \bibinfo {pages} {170602} (\bibinfo {year} {2021})}\BibitemShut {NoStop}%
\bibitem [{\citenamefont {Buchhold}\ \emph {et~al.}(2021)\citenamefont
  {Buchhold}, \citenamefont {Minoguchi}, \citenamefont {Altland},\ and\
  \citenamefont {Diehl}}]{PhysRevX.11.041004}%
  \BibitemOpen
  \bibfield  {author} {\bibinfo {author} {\bibfnamefont {M.}~\bibnamefont
  {Buchhold}}, \bibinfo {author} {\bibfnamefont {Y.}~\bibnamefont {Minoguchi}},
  \bibinfo {author} {\bibfnamefont {A.}~\bibnamefont {Altland}},\ and\ \bibinfo
  {author} {\bibfnamefont {S.}~\bibnamefont {Diehl}},\ }\bibfield  {title}
  {\bibinfo {title} {Effective theory for the measurement-induced phase
  transition of dirac fermions},\ }\href
  {https://doi.org/10.1103/PhysRevX.11.041004} {\bibfield  {journal} {\bibinfo
  {journal} {Phys. Rev. X}\ }\textbf {\bibinfo {volume} {11}},\ \bibinfo
  {pages} {041004} (\bibinfo {year} {2021})}\BibitemShut {NoStop}%
\bibitem [{\citenamefont {Bao}\ \emph {et~al.}(2020{\natexlab{b}})\citenamefont
  {Bao}, \citenamefont {Choi},\ and\ \citenamefont
  {Altman}}]{theoryOftransitionsBao}%
  \BibitemOpen
  \bibfield  {author} {\bibinfo {author} {\bibfnamefont {Y.}~\bibnamefont
  {Bao}}, \bibinfo {author} {\bibfnamefont {S.}~\bibnamefont {Choi}},\ and\
  \bibinfo {author} {\bibfnamefont {E.}~\bibnamefont {Altman}},\ }\bibfield
  {title} {\bibinfo {title} {Theory of the phase transition in random unitary
  circuits with measurements},\ }\href
  {https://doi.org/10.1103/PhysRevB.101.104301} {\bibfield  {journal} {\bibinfo
   {journal} {Phys. Rev. B}\ }\textbf {\bibinfo {volume} {101}},\ \bibinfo
  {pages} {104301} (\bibinfo {year} {2020}{\natexlab{b}})}\BibitemShut
  {NoStop}%
\bibitem [{\citenamefont {M\"uller}\ \emph {et~al.}(2022)\citenamefont
  {M\"uller}, \citenamefont {Diehl},\ and\ \citenamefont
  {Buchhold}}]{PhysRevLett.128.010605}%
  \BibitemOpen
  \bibfield  {author} {\bibinfo {author} {\bibfnamefont {T.}~\bibnamefont
  {M\"uller}}, \bibinfo {author} {\bibfnamefont {S.}~\bibnamefont {Diehl}},\
  and\ \bibinfo {author} {\bibfnamefont {M.}~\bibnamefont {Buchhold}},\
  }\bibfield  {title} {\bibinfo {title} {Measurement-induced dark state phase
  transitions in long-ranged fermion systems},\ }\href
  {https://doi.org/10.1103/PhysRevLett.128.010605} {\bibfield  {journal}
  {\bibinfo  {journal} {Phys. Rev. Lett.}\ }\textbf {\bibinfo {volume} {128}},\
  \bibinfo {pages} {010605} (\bibinfo {year} {2022})}\BibitemShut {NoStop}%
\bibitem [{\citenamefont {Jian}\ \emph
  {et~al.}(2020{\natexlab{b}})\citenamefont {Jian}, \citenamefont {You},
  \citenamefont {Vasseur},\ and\ \citenamefont {Ludwig}}]{PhysRevB.101.104302}%
  \BibitemOpen
  \bibfield  {author} {\bibinfo {author} {\bibfnamefont {C.-M.}\ \bibnamefont
  {Jian}}, \bibinfo {author} {\bibfnamefont {Y.-Z.}\ \bibnamefont {You}},
  \bibinfo {author} {\bibfnamefont {R.}~\bibnamefont {Vasseur}},\ and\ \bibinfo
  {author} {\bibfnamefont {A.~W.~W.}\ \bibnamefont {Ludwig}},\ }\bibfield
  {title} {\bibinfo {title} {Measurement-induced criticality in random quantum
  circuits},\ }\href {https://doi.org/10.1103/PhysRevB.101.104302} {\bibfield
  {journal} {\bibinfo  {journal} {Phys. Rev. B}\ }\textbf {\bibinfo {volume}
  {101}},\ \bibinfo {pages} {104302} (\bibinfo {year}
  {2020}{\natexlab{b}})}\BibitemShut {NoStop}%
\bibitem [{\citenamefont {Piccitto}\ \emph {et~al.}(2022)\citenamefont
  {Piccitto}, \citenamefont {Russomanno},\ and\ \citenamefont
  {Rossini}}]{PhysRevB.105.064305}%
  \BibitemOpen
  \bibfield  {author} {\bibinfo {author} {\bibfnamefont {G.}~\bibnamefont
  {Piccitto}}, \bibinfo {author} {\bibfnamefont {A.}~\bibnamefont
  {Russomanno}},\ and\ \bibinfo {author} {\bibfnamefont {D.}~\bibnamefont
  {Rossini}},\ }\bibfield  {title} {\bibinfo {title} {Entanglement transitions
  in the quantum ising chain: A comparison between different unravelings of the
  same lindbladian},\ }\href {https://doi.org/10.1103/PhysRevB.105.064305}
  {\bibfield  {journal} {\bibinfo  {journal} {Phys. Rev. B}\ }\textbf {\bibinfo
  {volume} {105}},\ \bibinfo {pages} {064305} (\bibinfo {year}
  {2022})}\BibitemShut {NoStop}%
\bibitem [{\citenamefont {Sierant}\ and\ \citenamefont
  {Turkeshi}(2022)}]{Sierant_2022}%
  \BibitemOpen
  \bibfield  {author} {\bibinfo {author} {\bibfnamefont {P.}~\bibnamefont
  {Sierant}}\ and\ \bibinfo {author} {\bibfnamefont {X.}~\bibnamefont
  {Turkeshi}},\ }\bibfield  {title} {\bibinfo {title} {Universal behavior
  beyond multifractality of wave functions at measurement-induced phase
  transitions},\ }\href {https://doi.org/10.1103/PhysRevLett.128.130605}
  {\bibfield  {journal} {\bibinfo  {journal} {Phys. Rev. Lett.}\ }\textbf
  {\bibinfo {volume} {128}},\ \bibinfo {pages} {130605} (\bibinfo {year}
  {2022})}\BibitemShut {NoStop}%
\bibitem [{\citenamefont {Turkeshi}\ \emph {et~al.}(2021)\citenamefont
  {Turkeshi}, \citenamefont {Biella}, \citenamefont {Fazio}, \citenamefont
  {Dalmonte},\ and\ \citenamefont {Schir\'o}}]{Turkeshi_2021}%
  \BibitemOpen
  \bibfield  {author} {\bibinfo {author} {\bibfnamefont {X.}~\bibnamefont
  {Turkeshi}}, \bibinfo {author} {\bibfnamefont {A.}~\bibnamefont {Biella}},
  \bibinfo {author} {\bibfnamefont {R.}~\bibnamefont {Fazio}}, \bibinfo
  {author} {\bibfnamefont {M.}~\bibnamefont {Dalmonte}},\ and\ \bibinfo
  {author} {\bibfnamefont {M.}~\bibnamefont {Schir\'o}},\ }\bibfield  {title}
  {\bibinfo {title} {Measurement-induced entanglement transitions in the
  quantum ising chain: From infinite to zero clicks},\ }\href
  {https://doi.org/10.1103/PhysRevB.103.224210} {\bibfield  {journal} {\bibinfo
   {journal} {Phys. Rev. B}\ }\textbf {\bibinfo {volume} {103}},\ \bibinfo
  {pages} {224210} (\bibinfo {year} {2021})}\BibitemShut {NoStop}%
\bibitem [{\citenamefont {Turkeshi}\ \emph
  {et~al.}(2022{\natexlab{a}})\citenamefont {Turkeshi}, \citenamefont
  {Dalmonte}, \citenamefont {Fazio},\ and\ \citenamefont
  {Schir\`o}}]{Turkeshi_2022}%
  \BibitemOpen
  \bibfield  {author} {\bibinfo {author} {\bibfnamefont {X.}~\bibnamefont
  {Turkeshi}}, \bibinfo {author} {\bibfnamefont {M.}~\bibnamefont {Dalmonte}},
  \bibinfo {author} {\bibfnamefont {R.}~\bibnamefont {Fazio}},\ and\ \bibinfo
  {author} {\bibfnamefont {M.}~\bibnamefont {Schir\`o}},\ }\bibfield  {title}
  {\bibinfo {title} {Entanglement transitions from stochastic resetting of
  non-hermitian quasiparticles},\ }\href
  {https://doi.org/10.1103/PhysRevB.105.L241114} {\bibfield  {journal}
  {\bibinfo  {journal} {Phys. Rev. B}\ }\textbf {\bibinfo {volume} {105}},\
  \bibinfo {pages} {L241114} (\bibinfo {year}
  {2022}{\natexlab{a}})}\BibitemShut {NoStop}%
\bibitem [{\citenamefont {Barratt}\ \emph
  {et~al.}(2022{\natexlab{a}})\citenamefont {Barratt}, \citenamefont {Agrawal},
  \citenamefont {Gopalakrishnan}, \citenamefont {Huse}, \citenamefont
  {Vasseur},\ and\ \citenamefont {Potter}}]{ChargeSharp2022PRL}%
  \BibitemOpen
  \bibfield  {author} {\bibinfo {author} {\bibfnamefont {F.}~\bibnamefont
  {Barratt}}, \bibinfo {author} {\bibfnamefont {U.}~\bibnamefont {Agrawal}},
  \bibinfo {author} {\bibfnamefont {S.}~\bibnamefont {Gopalakrishnan}},
  \bibinfo {author} {\bibfnamefont {D.~A.}\ \bibnamefont {Huse}}, \bibinfo
  {author} {\bibfnamefont {R.}~\bibnamefont {Vasseur}},\ and\ \bibinfo {author}
  {\bibfnamefont {A.~C.}\ \bibnamefont {Potter}},\ }\bibfield  {title}
  {\bibinfo {title} {Field theory of charge sharpening in symmetric monitored
  quantum circuits},\ }\href {https://doi.org/10.1103/PhysRevLett.129.120604}
  {\bibfield  {journal} {\bibinfo  {journal} {Phys. Rev. Lett.}\ }\textbf
  {\bibinfo {volume} {129}},\ \bibinfo {pages} {120604} (\bibinfo {year}
  {2022}{\natexlab{a}})}\BibitemShut {NoStop}%
\bibitem [{\citenamefont {Lavasani}\ \emph {et~al.}(2022)\citenamefont
  {Lavasani}, \citenamefont {Luo},\ and\ \citenamefont
  {Vijay}}]{lavasani_monitored_2022}%
  \BibitemOpen
  \bibfield  {author} {\bibinfo {author} {\bibfnamefont {A.}~\bibnamefont
  {Lavasani}}, \bibinfo {author} {\bibfnamefont {Z.-X.}\ \bibnamefont {Luo}},\
  and\ \bibinfo {author} {\bibfnamefont {S.}~\bibnamefont {Vijay}},\
  }\href@noop {} {\bibinfo {title} {Monitored quantum dynamics and the kitaev
  spin liquid}} (\bibinfo {year} {2022}),\ \Eprint
  {https://arxiv.org/abs/arXiv: 2207.02877} {arXiv: 2207.02877} \BibitemShut
  {NoStop}%
\bibitem [{\citenamefont {Zabalo}\ \emph {et~al.}(2023)\citenamefont {Zabalo},
  \citenamefont {Wilson}, \citenamefont {Gullans}, \citenamefont {Vasseur},
  \citenamefont {Gopalakrishnan}, \citenamefont {Huse},\ and\ \citenamefont
  {Pixley}}]{zabalo_infinite-randomness_2022}%
  \BibitemOpen
  \bibfield  {author} {\bibinfo {author} {\bibfnamefont {A.}~\bibnamefont
  {Zabalo}}, \bibinfo {author} {\bibfnamefont {J.~H.}\ \bibnamefont {Wilson}},
  \bibinfo {author} {\bibfnamefont {M.~J.}\ \bibnamefont {Gullans}}, \bibinfo
  {author} {\bibfnamefont {R.}~\bibnamefont {Vasseur}}, \bibinfo {author}
  {\bibfnamefont {S.}~\bibnamefont {Gopalakrishnan}}, \bibinfo {author}
  {\bibfnamefont {D.~A.}\ \bibnamefont {Huse}},\ and\ \bibinfo {author}
  {\bibfnamefont {J.~H.}\ \bibnamefont {Pixley}},\ }\bibfield  {title}
  {\bibinfo {title} {{I}nfinite-randomness criticality in monitored quantum
  dynamics with static disorder},\ }\href
  {https://doi.org/10.1103/PhysRevB.107.L220204} {\bibfield  {journal}
  {\bibinfo  {journal} {Phys. Rev. B}\ }\textbf {\bibinfo {volume} {107}},\
  \bibinfo {pages} {L220204} (\bibinfo {year} {2023})}\BibitemShut {NoStop}%
\bibitem [{\citenamefont {Zabalo}\ \emph {et~al.}(2022)\citenamefont {Zabalo},
  \citenamefont {Gullans}, \citenamefont {Wilson}, \citenamefont {Vasseur},
  \citenamefont {Ludwig}, \citenamefont {Gopalakrishnan}, \citenamefont
  {Huse},\ and\ \citenamefont {Pixley}}]{zabalo_operator_2022}%
  \BibitemOpen
  \bibfield  {author} {\bibinfo {author} {\bibfnamefont {A.}~\bibnamefont
  {Zabalo}}, \bibinfo {author} {\bibfnamefont {M.~J.}\ \bibnamefont {Gullans}},
  \bibinfo {author} {\bibfnamefont {J.~H.}\ \bibnamefont {Wilson}}, \bibinfo
  {author} {\bibfnamefont {R.}~\bibnamefont {Vasseur}}, \bibinfo {author}
  {\bibfnamefont {A.~W.~W.}\ \bibnamefont {Ludwig}}, \bibinfo {author}
  {\bibfnamefont {S.}~\bibnamefont {Gopalakrishnan}}, \bibinfo {author}
  {\bibfnamefont {D.~A.}\ \bibnamefont {Huse}},\ and\ \bibinfo {author}
  {\bibfnamefont {J.~H.}\ \bibnamefont {Pixley}},\ }\bibfield  {title}
  {\bibinfo {title} {Operator {Scaling} {Dimensions} and {Multifractality} at
  {Measurement}-{Induced} {Transitions}},\ }\href
  {https://doi.org/10.1103/PhysRevLett.128.050602} {\bibfield  {journal}
  {\bibinfo  {journal} {Phys. Rev. Lett.}\ }\textbf {\bibinfo {volume} {128}},\
  \bibinfo {pages} {050602} (\bibinfo {year} {2022})}\BibitemShut {NoStop}%
\bibitem [{\citenamefont {Ippoliti}\ and\ \citenamefont
  {Khemani}(2021)}]{ippoliti_postselection-free_2021}%
  \BibitemOpen
  \bibfield  {author} {\bibinfo {author} {\bibfnamefont {M.}~\bibnamefont
  {Ippoliti}}\ and\ \bibinfo {author} {\bibfnamefont {V.}~\bibnamefont
  {Khemani}},\ }\bibfield  {title} {\bibinfo {title} {Postselection-{Free}
  {Entanglement} {Dynamics} via {Spacetime} {Duality}},\ }\href
  {https://doi.org/10.1103/PhysRevLett.126.060501} {\bibfield  {journal}
  {\bibinfo  {journal} {Phys. Rev. Lett.}\ }\textbf {\bibinfo {volume} {126}},\
  \bibinfo {pages} {060501} (\bibinfo {year} {2021})}\BibitemShut {NoStop}%
\bibitem [{\citenamefont {Ippoliti}\ and\ \citenamefont
  {Ho}(2022)}]{ippoliti_dynamical_2022}%
  \BibitemOpen
  \bibfield  {author} {\bibinfo {author} {\bibfnamefont {M.}~\bibnamefont
  {Ippoliti}}\ and\ \bibinfo {author} {\bibfnamefont {W.~W.}\ \bibnamefont
  {Ho}},\ }\href@noop {} {\bibinfo {title} {Dynamical purification and the
  emergence of quantum state designs from the projected ensemble}} (\bibinfo
  {year} {2022}),\ \Eprint {https://arxiv.org/abs/arXiv: 2204.13657} {arXiv:
  2204.13657} \BibitemShut {NoStop}%
\bibitem [{\citenamefont {Szyniszewski}\ \emph {et~al.}(2022)\citenamefont
  {Szyniszewski}, \citenamefont {Lunt},\ and\ \citenamefont
  {Pal}}]{szyniszewski_disordered_2022}%
  \BibitemOpen
  \bibfield  {author} {\bibinfo {author} {\bibfnamefont {M.}~\bibnamefont
  {Szyniszewski}}, \bibinfo {author} {\bibfnamefont {O.}~\bibnamefont {Lunt}},\
  and\ \bibinfo {author} {\bibfnamefont {A.}~\bibnamefont {Pal}},\ }\href@noop
  {} {\bibinfo {title} {Disordered monitored free fermions}} (\bibinfo {year}
  {2022}),\ \Eprint {https://arxiv.org/abs/arXiv: 2211.02534} {arXiv:
  2211.02534} \BibitemShut {NoStop}%
\bibitem [{\citenamefont {Sriram}\ \emph {et~al.}(2022)\citenamefont {Sriram},
  \citenamefont {Rakovszky}, \citenamefont {Khemani},\ and\ \citenamefont
  {Ippoliti}}]{arxiv.2207.07096}%
  \BibitemOpen
  \bibfield  {author} {\bibinfo {author} {\bibfnamefont {A.}~\bibnamefont
  {Sriram}}, \bibinfo {author} {\bibfnamefont {T.}~\bibnamefont {Rakovszky}},
  \bibinfo {author} {\bibfnamefont {V.}~\bibnamefont {Khemani}},\ and\ \bibinfo
  {author} {\bibfnamefont {M.}~\bibnamefont {Ippoliti}},\ }\href@noop {}
  {\bibinfo {title} {Topology, criticality, and dynamically generated qubits in
  a stochastic measurement-only kitaev model}} (\bibinfo {year} {2022}),\
  \Eprint {https://arxiv.org/abs/arXiv: 2207.07096} {arXiv: 2207.07096}
  \BibitemShut {NoStop}%
\bibitem [{\citenamefont {Yu}\ and\ \citenamefont
  {Qi}(2022)}]{arxiv.2201.12704}%
  \BibitemOpen
  \bibfield  {author} {\bibinfo {author} {\bibfnamefont {X.}~\bibnamefont
  {Yu}}\ and\ \bibinfo {author} {\bibfnamefont {X.-L.}\ \bibnamefont {Qi}},\
  }\href@noop {} {\bibinfo {title} {Measurement-induced entanglement phase
  transition in random bilocal circuits}} (\bibinfo {year} {2022}),\ \Eprint
  {https://arxiv.org/abs/arXiv: 2201.12704} {arXiv: 2201.12704} \BibitemShut
  {NoStop}%
\bibitem [{\citenamefont {Turkeshi}\ \emph
  {et~al.}(2022{\natexlab{b}})\citenamefont {Turkeshi}, \citenamefont
  {Piroli},\ and\ \citenamefont {Schir\'o}}]{PhysRevB.106.024304}%
  \BibitemOpen
  \bibfield  {author} {\bibinfo {author} {\bibfnamefont {X.}~\bibnamefont
  {Turkeshi}}, \bibinfo {author} {\bibfnamefont {L.}~\bibnamefont {Piroli}},\
  and\ \bibinfo {author} {\bibfnamefont {M.}~\bibnamefont {Schir\'o}},\
  }\bibfield  {title} {\bibinfo {title} {Enhanced entanglement negativity in
  boundary-driven monitored fermionic chains},\ }\href
  {https://doi.org/10.1103/PhysRevB.106.024304} {\bibfield  {journal} {\bibinfo
   {journal} {Phys. Rev. B}\ }\textbf {\bibinfo {volume} {106}},\ \bibinfo
  {pages} {024304} (\bibinfo {year} {2022}{\natexlab{b}})}\BibitemShut
  {NoStop}%
\bibitem [{\citenamefont {Sierant}\ \emph {et~al.}(2022)\citenamefont
  {Sierant}, \citenamefont {Chiriac{\`{o}}}, \citenamefont {Surace},
  \citenamefont {Sharma}, \citenamefont {Turkeshi}, \citenamefont {Dalmonte},
  \citenamefont {Fazio},\ and\ \citenamefont
  {Pagano}}]{Sierant2022dissipativefloquet}%
  \BibitemOpen
  \bibfield  {author} {\bibinfo {author} {\bibfnamefont {P.}~\bibnamefont
  {Sierant}}, \bibinfo {author} {\bibfnamefont {G.}~\bibnamefont
  {Chiriac{\`{o}}}}, \bibinfo {author} {\bibfnamefont {F.~M.}\ \bibnamefont
  {Surace}}, \bibinfo {author} {\bibfnamefont {S.}~\bibnamefont {Sharma}},
  \bibinfo {author} {\bibfnamefont {X.}~\bibnamefont {Turkeshi}}, \bibinfo
  {author} {\bibfnamefont {M.}~\bibnamefont {Dalmonte}}, \bibinfo {author}
  {\bibfnamefont {R.}~\bibnamefont {Fazio}},\ and\ \bibinfo {author}
  {\bibfnamefont {G.}~\bibnamefont {Pagano}},\ }\bibfield  {title} {\bibinfo
  {title} {Dissipative {F}loquet {D}ynamics: from {S}teady {S}tate to
  {M}easurement {I}nduced {C}riticality in {T}rapped-ion {C}hains},\ }\href
  {https://doi.org/10.22331/q-2022-02-02-638} {\bibfield  {journal} {\bibinfo
  {journal} {{Quantum}}\ }\textbf {\bibinfo {volume} {6}},\ \bibinfo {pages}
  {638} (\bibinfo {year} {2022})}\BibitemShut {NoStop}%
\bibitem [{\citenamefont {Jian}\ \emph {et~al.}(2021)\citenamefont {Jian},
  \citenamefont {Liu}, \citenamefont {Chen}, \citenamefont {Swingle},\ and\
  \citenamefont {Zhang}}]{PhysRevLett.127.140601}%
  \BibitemOpen
  \bibfield  {author} {\bibinfo {author} {\bibfnamefont {S.-K.}\ \bibnamefont
  {Jian}}, \bibinfo {author} {\bibfnamefont {C.}~\bibnamefont {Liu}}, \bibinfo
  {author} {\bibfnamefont {X.}~\bibnamefont {Chen}}, \bibinfo {author}
  {\bibfnamefont {B.}~\bibnamefont {Swingle}},\ and\ \bibinfo {author}
  {\bibfnamefont {P.}~\bibnamefont {Zhang}},\ }\bibfield  {title} {\bibinfo
  {title} {Measurement-induced phase transition in the monitored
  sachdev-ye-kitaev model},\ }\href
  {https://doi.org/10.1103/PhysRevLett.127.140601} {\bibfield  {journal}
  {\bibinfo  {journal} {Phys. Rev. Lett.}\ }\textbf {\bibinfo {volume} {127}},\
  \bibinfo {pages} {140601} (\bibinfo {year} {2021})}\BibitemShut {NoStop}%
\bibitem [{\citenamefont {Koh}\ \emph {et~al.}(2022)\citenamefont {Koh},
  \citenamefont {Sun}, \citenamefont {Motta},\ and\ \citenamefont
  {Minnich}}]{koh2022experimental}%
  \BibitemOpen
  \bibfield  {author} {\bibinfo {author} {\bibfnamefont {J.~M.}\ \bibnamefont
  {Koh}}, \bibinfo {author} {\bibfnamefont {S.-N.}\ \bibnamefont {Sun}},
  \bibinfo {author} {\bibfnamefont {M.}~\bibnamefont {Motta}},\ and\ \bibinfo
  {author} {\bibfnamefont {A.~J.}\ \bibnamefont {Minnich}},\ }\href@noop {}
  {\bibinfo {title} {{E}xperimental realization of a measurement-induced
  entanglement phase transition on a superconducting quantum processor}}
  (\bibinfo {year} {2022}),\ \Eprint {https://arxiv.org/abs/arXiv: 2203.04338}
  {arXiv: 2203.04338} \BibitemShut {NoStop}%
\bibitem [{\citenamefont {Noel}\ \emph {et~al.}(2022)\citenamefont {Noel},
  \citenamefont {Niroula}, \citenamefont {Zhu}, \citenamefont {Risinger},
  \citenamefont {Egan}, \citenamefont {Biswas}, \citenamefont {Cetina},
  \citenamefont {Gorshkov}, \citenamefont {Gullans}, \citenamefont {Huse} \emph
  {et~al.}}]{noel2022}%
  \BibitemOpen
  \bibfield  {author} {\bibinfo {author} {\bibfnamefont {C.}~\bibnamefont
  {Noel}}, \bibinfo {author} {\bibfnamefont {P.}~\bibnamefont {Niroula}},
  \bibinfo {author} {\bibfnamefont {D.}~\bibnamefont {Zhu}}, \bibinfo {author}
  {\bibfnamefont {A.}~\bibnamefont {Risinger}}, \bibinfo {author}
  {\bibfnamefont {L.}~\bibnamefont {Egan}}, \bibinfo {author} {\bibfnamefont
  {D.}~\bibnamefont {Biswas}}, \bibinfo {author} {\bibfnamefont
  {M.}~\bibnamefont {Cetina}}, \bibinfo {author} {\bibfnamefont {A.~V.}\
  \bibnamefont {Gorshkov}}, \bibinfo {author} {\bibfnamefont {M.~J.}\
  \bibnamefont {Gullans}}, \bibinfo {author} {\bibfnamefont {D.~A.}\
  \bibnamefont {Huse}}, \emph {et~al.},\ }\bibfield  {title} {\bibinfo {title}
  {{M}easurement-induced quantum phases realized in a trapped-ion quantum
  computer},\ }\href {https://doi.org/10.1038/s41567-022-01619-7} {\bibfield
  {journal} {\bibinfo  {journal} {Nature Physics}\ }\textbf {\bibinfo {volume}
  {18}},\ \bibinfo {pages} {760} (\bibinfo {year} {2022})}\BibitemShut
  {NoStop}%
\bibitem [{\citenamefont {Hoke}\ \emph {et~al.}(2023)\citenamefont {Hoke},
  \citenamefont {Ippoliti}, \citenamefont {Abanin}, \citenamefont {Acharya},
  \citenamefont {Ansmann}, \citenamefont {Arute}, \citenamefont {Arya},
  \citenamefont {Asfaw}, \citenamefont {Atalaya}, \citenamefont {Bardin} \emph
  {et~al.}}]{hoke2023quantum}%
  \BibitemOpen
  \bibfield  {author} {\bibinfo {author} {\bibfnamefont {J.~C.}\ \bibnamefont
  {Hoke}}, \bibinfo {author} {\bibfnamefont {M.}~\bibnamefont {Ippoliti}},
  \bibinfo {author} {\bibfnamefont {D.}~\bibnamefont {Abanin}}, \bibinfo
  {author} {\bibfnamefont {R.}~\bibnamefont {Acharya}}, \bibinfo {author}
  {\bibfnamefont {M.}~\bibnamefont {Ansmann}}, \bibinfo {author} {\bibfnamefont
  {F.}~\bibnamefont {Arute}}, \bibinfo {author} {\bibfnamefont
  {K.}~\bibnamefont {Arya}}, \bibinfo {author} {\bibfnamefont {A.}~\bibnamefont
  {Asfaw}}, \bibinfo {author} {\bibfnamefont {J.}~\bibnamefont {Atalaya}},
  \bibinfo {author} {\bibfnamefont {J.~C.}\ \bibnamefont {Bardin}}, \emph
  {et~al.},\ }\href@noop {} {\bibinfo {title} {{Q}uantum information phases in
  space-time: measurement-induced entanglement and teleportation on a noisy
  quantum processor}} (\bibinfo {year} {2023}),\ \Eprint
  {https://arxiv.org/abs/arXiv: 2303.04792} {arXiv: 2303.04792} \BibitemShut
  {NoStop}%
\bibitem [{\citenamefont {Gullans}\ and\ \citenamefont
  {Huse}(2020{\natexlab{a}})}]{gullans2020scalable}%
  \BibitemOpen
  \bibfield  {author} {\bibinfo {author} {\bibfnamefont {M.~J.}\ \bibnamefont
  {Gullans}}\ and\ \bibinfo {author} {\bibfnamefont {D.~A.}\ \bibnamefont
  {Huse}},\ }\bibfield  {title} {\bibinfo {title} {{S}calable probes of
  measurement-induced criticality},\ }\href
  {https://doi.org/10.1103/PhysRevLett.125.070606} {\bibfield  {journal}
  {\bibinfo  {journal} {Phys. Rev. Lett.}\ }\textbf {\bibinfo {volume} {125}},\
  \bibinfo {pages} {070606} (\bibinfo {year} {2020}{\natexlab{a}})}\BibitemShut
  {NoStop}%
\bibitem [{\citenamefont {Li}\ \emph {et~al.}(2023{\natexlab{b}})\citenamefont
  {Li}, \citenamefont {Zou}, \citenamefont {Glorioso}, \citenamefont {Altman},\
  and\ \citenamefont {Fisher}}]{li2023cross}%
  \BibitemOpen
  \bibfield  {author} {\bibinfo {author} {\bibfnamefont {Y.}~\bibnamefont
  {Li}}, \bibinfo {author} {\bibfnamefont {Y.}~\bibnamefont {Zou}}, \bibinfo
  {author} {\bibfnamefont {P.}~\bibnamefont {Glorioso}}, \bibinfo {author}
  {\bibfnamefont {E.}~\bibnamefont {Altman}},\ and\ \bibinfo {author}
  {\bibfnamefont {M.~P.~A.}\ \bibnamefont {Fisher}},\ }\bibfield  {title}
  {\bibinfo {title} {Cross entropy benchmark for measurement-induced phase
  transitions},\ }\href {https://doi.org/10.1103/PhysRevLett.130.220404}
  {\bibfield  {journal} {\bibinfo  {journal} {Phys. Rev. Lett.}\ }\textbf
  {\bibinfo {volume} {130}},\ \bibinfo {pages} {220404} (\bibinfo {year}
  {2023}{\natexlab{b}})}\BibitemShut {NoStop}%
\bibitem [{\citenamefont {Weinstein}\ \emph
  {et~al.}(2022{\natexlab{b}})\citenamefont {Weinstein}, \citenamefont {Kelly},
  \citenamefont {Marino},\ and\ \citenamefont
  {Altman}}]{weinstein2022scrambling}%
  \BibitemOpen
  \bibfield  {author} {\bibinfo {author} {\bibfnamefont {Z.}~\bibnamefont
  {Weinstein}}, \bibinfo {author} {\bibfnamefont {S.~P.}\ \bibnamefont
  {Kelly}}, \bibinfo {author} {\bibfnamefont {J.}~\bibnamefont {Marino}},\ and\
  \bibinfo {author} {\bibfnamefont {E.}~\bibnamefont {Altman}},\ }\href@noop {}
  {\bibinfo {title} {Scrambling transition in a radiative random unitary
  circuit}} (\bibinfo {year} {2022}{\natexlab{b}}),\ \Eprint
  {https://arxiv.org/abs/arXiv: 2210.14242} {arXiv: 2210.14242} \BibitemShut
  {NoStop}%
\bibitem [{\citenamefont {Bao}\ \emph {et~al.}(2021{\natexlab{b}})\citenamefont
  {Bao}, \citenamefont {Choi},\ and\ \citenamefont {Altman}}]{bao2021}%
  \BibitemOpen
  \bibfield  {author} {\bibinfo {author} {\bibfnamefont {Y.}~\bibnamefont
  {Bao}}, \bibinfo {author} {\bibfnamefont {S.}~\bibnamefont {Choi}},\ and\
  \bibinfo {author} {\bibfnamefont {E.}~\bibnamefont {Altman}},\ }\bibfield
  {title} {\bibinfo {title} {{S}ymmetry enriched phases of quantum circuits},\
  }\href {https://doi.org/10.1016/j.aop.2021.168618.} {\bibfield  {journal}
  {\bibinfo  {journal} {Annals of Physics}\ }\textbf {\bibinfo {volume}
  {435}},\ \bibinfo {pages} {168618} (\bibinfo {year}
  {2021}{\natexlab{b}})}\BibitemShut {NoStop}%
\bibitem [{\citenamefont {Nahum}\ and\ \citenamefont
  {Skinner}(2020)}]{nahum2020entanglement}%
  \BibitemOpen
  \bibfield  {author} {\bibinfo {author} {\bibfnamefont {A.}~\bibnamefont
  {Nahum}}\ and\ \bibinfo {author} {\bibfnamefont {B.}~\bibnamefont
  {Skinner}},\ }\bibfield  {title} {\bibinfo {title} {{E}ntanglement and
  dynamics of diffusion-annihilation processes with {M}ajorana defects},\
  }\href {https://doi.org/10.1103/PhysRevResearch.2.023288} {\bibfield
  {journal} {\bibinfo  {journal} {Phys. Rev. Res.}\ }\textbf {\bibinfo {volume}
  {2}},\ \bibinfo {pages} {023288} (\bibinfo {year} {2020})}\BibitemShut
  {NoStop}%
\bibitem [{\citenamefont {Cardy}(1992)}]{cardy1992critical}%
  \BibitemOpen
  \bibfield  {author} {\bibinfo {author} {\bibfnamefont {J.~L.}\ \bibnamefont
  {Cardy}},\ }\bibfield  {title} {\bibinfo {title} {{C}ritical percolation in
  finite geometries},\ }\href
  {https://iopscience.iop.org/article/10.1088/0305-4470/25/4/009} {\bibfield
  {journal} {\bibinfo  {journal} {Journal of Physics A: Mathematical and
  General}\ }\textbf {\bibinfo {volume} {25}},\ \bibinfo {pages} {L201}
  (\bibinfo {year} {1992})}\BibitemShut {NoStop}%
\bibitem [{\citenamefont {Nienhuis}(1984)}]{nienhuis1984critical}%
  \BibitemOpen
  \bibfield  {author} {\bibinfo {author} {\bibfnamefont {B.}~\bibnamefont
  {Nienhuis}},\ }\bibfield  {title} {\bibinfo {title} {{C}ritical behavior of
  two-dimensional spin models and charge asymmetry in the {C}oulomb gas},\
  }\href {https://doi.org/10.1007/BF01009437} {\bibfield  {journal} {\bibinfo
  {journal} {J Stat Phys}\ }\textbf {\bibinfo {volume} {34}},\ \bibinfo {pages}
  {731} (\bibinfo {year} {1984})}\BibitemShut {NoStop}%
\bibitem [{\citenamefont {Saleur}\ and\ \citenamefont
  {Duplantier}(1987)}]{saleur1987exact}%
  \BibitemOpen
  \bibfield  {author} {\bibinfo {author} {\bibfnamefont {H.}~\bibnamefont
  {Saleur}}\ and\ \bibinfo {author} {\bibfnamefont {B.}~\bibnamefont
  {Duplantier}},\ }\bibfield  {title} {\bibinfo {title} {Exact determination of
  the percolation hull exponent in two dimensions},\ }\href
  {https://doi.org/10.1103/PhysRevLett.58.2325} {\bibfield  {journal} {\bibinfo
   {journal} {Phys. Rev. Lett.}\ }\textbf {\bibinfo {volume} {58}},\ \bibinfo
  {pages} {2325} (\bibinfo {year} {1987})}\BibitemShut {NoStop}%
\bibitem [{\citenamefont {Lang}\ and\ \citenamefont
  {B\"uchler}(2020)}]{lang2020entanglement}%
  \BibitemOpen
  \bibfield  {author} {\bibinfo {author} {\bibfnamefont {N.}~\bibnamefont
  {Lang}}\ and\ \bibinfo {author} {\bibfnamefont {H.~P.}\ \bibnamefont
  {B\"uchler}},\ }\bibfield  {title} {\bibinfo {title} {{E}ntanglement
  transition in the projective transverse field {I}sing model},\ }\href
  {https://doi.org/10.1103/PhysRevB.102.094204} {\bibfield  {journal} {\bibinfo
   {journal} {Phys. Rev. B}\ }\textbf {\bibinfo {volume} {102}},\ \bibinfo
  {pages} {094204} (\bibinfo {year} {2020})}\BibitemShut {NoStop}%
\bibitem [{\citenamefont {Li}\ and\ \citenamefont {Fisher}(2021)}]{Fisher2021}%
  \BibitemOpen
  \bibfield  {author} {\bibinfo {author} {\bibfnamefont {Y.}~\bibnamefont
  {Li}}\ and\ \bibinfo {author} {\bibfnamefont {M.~P.~A.}\ \bibnamefont
  {Fisher}},\ }\href@noop {} {\bibinfo {title} {{R}obust decoding in monitored
  dynamics of open quantum systems with $\mathbb{Z}_2$ symmetry}} (\bibinfo
  {year} {2021}),\ \Eprint {https://arxiv.org/abs/arXiv: 2201.12704} {arXiv:
  2201.12704} \BibitemShut {NoStop}%
\bibitem [{\citenamefont {Gullans}\ and\ \citenamefont
  {Huse}(2020{\natexlab{b}})}]{gullans2020dynamical}%
  \BibitemOpen
  \bibfield  {author} {\bibinfo {author} {\bibfnamefont {M.~J.}\ \bibnamefont
  {Gullans}}\ and\ \bibinfo {author} {\bibfnamefont {D.~A.}\ \bibnamefont
  {Huse}},\ }\bibfield  {title} {\bibinfo {title} {{D}ynamical purification
  phase transition induced by quantum measurements},\ }\href
  {https://doi.org/10.1103/PhysRevX.10.041020} {\bibfield  {journal} {\bibinfo
  {journal} {Phys. Rev. X}\ }\textbf {\bibinfo {volume} {10}},\ \bibinfo
  {pages} {041020} (\bibinfo {year} {2020}{\natexlab{b}})}\BibitemShut
  {NoStop}%
\bibitem [{\citenamefont {Kelly}\ \emph {et~al.}(2022)\citenamefont {Kelly},
  \citenamefont {Poschinger}, \citenamefont {Schmidt-Kaler}, \citenamefont
  {Fisher},\ and\ \citenamefont {Marino}}]{kelly2022coherence}%
  \BibitemOpen
  \bibfield  {author} {\bibinfo {author} {\bibfnamefont {S.~P.}\ \bibnamefont
  {Kelly}}, \bibinfo {author} {\bibfnamefont {U.}~\bibnamefont {Poschinger}},
  \bibinfo {author} {\bibfnamefont {F.}~\bibnamefont {Schmidt-Kaler}}, \bibinfo
  {author} {\bibfnamefont {M.}~\bibnamefont {Fisher}},\ and\ \bibinfo {author}
  {\bibfnamefont {J.}~\bibnamefont {Marino}},\ }\href@noop {} {\bibinfo {title}
  {Coherence requirements for quantum communication from hybrid circuit
  dynamics}} (\bibinfo {year} {2022}),\ \Eprint {https://arxiv.org/abs/arXiv:
  2210.11547} {arXiv: 2210.11547} \BibitemShut {NoStop}%
\bibitem [{\citenamefont {Agrawal}\ \emph {et~al.}(2022)\citenamefont
  {Agrawal}, \citenamefont {Zabalo}, \citenamefont {Chen}, \citenamefont
  {Wilson}, \citenamefont {Potter}, \citenamefont {Pixley}, \citenamefont
  {Gopalakrishnan},\ and\ \citenamefont {Vasseur}}]{agrawal2022entanglement}%
  \BibitemOpen
  \bibfield  {author} {\bibinfo {author} {\bibfnamefont {U.}~\bibnamefont
  {Agrawal}}, \bibinfo {author} {\bibfnamefont {A.}~\bibnamefont {Zabalo}},
  \bibinfo {author} {\bibfnamefont {K.}~\bibnamefont {Chen}}, \bibinfo {author}
  {\bibfnamefont {J.~H.}\ \bibnamefont {Wilson}}, \bibinfo {author}
  {\bibfnamefont {A.~C.}\ \bibnamefont {Potter}}, \bibinfo {author}
  {\bibfnamefont {J.~H.}\ \bibnamefont {Pixley}}, \bibinfo {author}
  {\bibfnamefont {S.}~\bibnamefont {Gopalakrishnan}},\ and\ \bibinfo {author}
  {\bibfnamefont {R.}~\bibnamefont {Vasseur}},\ }\bibfield  {title} {\bibinfo
  {title} {{E}ntanglement and charge-sharpening transitions in $u(1)$ symmetric
  monitored quantum circuits},\ }\href
  {https://doi.org/10.1103/PhysRevX.12.041002} {\bibfield  {journal} {\bibinfo
  {journal} {Phys. Rev. X}\ }\textbf {\bibinfo {volume} {12}},\ \bibinfo
  {pages} {041002} (\bibinfo {year} {2022})}\BibitemShut {NoStop}%
\bibitem [{\citenamefont {Barratt}\ \emph
  {et~al.}(2022{\natexlab{b}})\citenamefont {Barratt}, \citenamefont {Agrawal},
  \citenamefont {Gopalakrishnan}, \citenamefont {Huse}, \citenamefont
  {Vasseur},\ and\ \citenamefont {Potter}}]{barratt2022field}%
  \BibitemOpen
  \bibfield  {author} {\bibinfo {author} {\bibfnamefont {F.}~\bibnamefont
  {Barratt}}, \bibinfo {author} {\bibfnamefont {U.}~\bibnamefont {Agrawal}},
  \bibinfo {author} {\bibfnamefont {S.}~\bibnamefont {Gopalakrishnan}},
  \bibinfo {author} {\bibfnamefont {D.~A.}\ \bibnamefont {Huse}}, \bibinfo
  {author} {\bibfnamefont {R.}~\bibnamefont {Vasseur}},\ and\ \bibinfo {author}
  {\bibfnamefont {A.~C.}\ \bibnamefont {Potter}},\ }\bibfield  {title}
  {\bibinfo {title} {{F}ield theory of charge sharpening in symmetric monitored
  quantum circuits},\ }\href {https://doi.org/10.1103/PhysRevLett.129.120604}
  {\bibfield  {journal} {\bibinfo  {journal} {Phys. Rev. Lett.}\ }\textbf
  {\bibinfo {volume} {129}},\ \bibinfo {pages} {120604} (\bibinfo {year}
  {2022}{\natexlab{b}})}\BibitemShut {NoStop}%
\bibitem [{\citenamefont {Lin}\ \emph {et~al.}(2023)\citenamefont {Lin},
  \citenamefont {Ye}, \citenamefont {Zou}, \citenamefont {Sang},\ and\
  \citenamefont {Hsieh}}]{lin2023probing}%
  \BibitemOpen
  \bibfield  {author} {\bibinfo {author} {\bibfnamefont {C.-J.}\ \bibnamefont
  {Lin}}, \bibinfo {author} {\bibfnamefont {W.}~\bibnamefont {Ye}}, \bibinfo
  {author} {\bibfnamefont {Y.}~\bibnamefont {Zou}}, \bibinfo {author}
  {\bibfnamefont {S.}~\bibnamefont {Sang}},\ and\ \bibinfo {author}
  {\bibfnamefont {T.~H.}\ \bibnamefont {Hsieh}},\ }\bibfield  {title} {\bibinfo
  {title} {Probing sign structure using measurement-induced entanglement},\
  }\href {https://doi.org/10.22331/q-2023-02-02-910} {\bibfield  {journal}
  {\bibinfo  {journal} {Quantum}\ }\textbf {\bibinfo {volume} {7}},\ \bibinfo
  {pages} {910} (\bibinfo {year} {2023})}\BibitemShut {NoStop}%
\bibitem [{\citenamefont {Cardy}(1998)}]{cardy1998number}%
  \BibitemOpen
  \bibfield  {author} {\bibinfo {author} {\bibfnamefont {J.}~\bibnamefont
  {Cardy}},\ }\bibfield  {title} {\bibinfo {title} {{T}he number of incipient
  spanning clusters in two-dimensional percolation},\ }\href
  {https://iopscience.iop.org/article/10.1088/0305-4470/31/5/003} {\bibfield
  {journal} {\bibinfo  {journal} {Journal of Physics A: Mathematical and
  General}\ }\textbf {\bibinfo {volume} {31}},\ \bibinfo {pages} {L105}
  (\bibinfo {year} {1998})}\BibitemShut {NoStop}%
\bibitem [{\citenamefont {Smirnov}(2001)}]{smirnov2001critical}%
  \BibitemOpen
  \bibfield  {author} {\bibinfo {author} {\bibfnamefont {S.}~\bibnamefont
  {Smirnov}},\ }\bibfield  {title} {\bibinfo {title} {{C}ritical percolation in
  the plane: conformal invariance, {C}ardy's formula, scaling limits},\ }\href
  {https://doi.org/10.1016/S0764-4442(01)01991-7} {\bibfield  {journal}
  {\bibinfo  {journal} {Comptes Rendus de l'Acad{\'e}mie des Sciences-Series
  I-Mathematics}\ }\textbf {\bibinfo {volume} {333}},\ \bibinfo {pages} {239}
  (\bibinfo {year} {2001})}\BibitemShut {NoStop}%
\bibitem [{\citenamefont {Francesco}\ \emph {et~al.}(2012)\citenamefont
  {Francesco}, \citenamefont {Mathieu},\ and\ \citenamefont
  {S{\'e}n{\'e}chal}}]{francesco2012conformal}%
  \BibitemOpen
  \bibfield  {author} {\bibinfo {author} {\bibfnamefont {P.}~\bibnamefont
  {Francesco}}, \bibinfo {author} {\bibfnamefont {P.}~\bibnamefont {Mathieu}},\
  and\ \bibinfo {author} {\bibfnamefont {D.}~\bibnamefont {S{\'e}n{\'e}chal}},\
  }\href@noop {} {\emph {\bibinfo {title} {Conformal field theory}}}\ (\bibinfo
   {publisher} {Springer Science \& Business Media},\ \bibinfo {year}
  {2012})\BibitemShut {NoStop}%
\bibitem [{\citenamefont {Kondev}(1997)}]{kondev1997liouville}%
  \BibitemOpen
  \bibfield  {author} {\bibinfo {author} {\bibfnamefont {J.}~\bibnamefont
  {Kondev}},\ }\bibfield  {title} {\bibinfo {title} {Liouville field theory of
  fluctuating loops},\ }\href {https://doi.org/10.1103/PhysRevLett.78.4320}
  {\bibfield  {journal} {\bibinfo  {journal} {Phys. Rev. Lett.}\ }\textbf
  {\bibinfo {volume} {78}},\ \bibinfo {pages} {4320} (\bibinfo {year}
  {1997})}\BibitemShut {NoStop}%
\bibitem [{\citenamefont {Krastanov}(2019)}]{QuantumClifford}%
  \BibitemOpen
  \bibfield  {author} {\bibinfo {author} {\bibfnamefont {S.}~\bibnamefont
  {Krastanov}},\ }\href@noop {} {\bibinfo {title} {{QuantumClifford.jl}:
  {C}lifford circuits and other quantum {S}tabilizer formalism tools}},\
  \bibinfo {howpublished} {\url{https://juliapackages.com/p/quantumclifford}}
  (\bibinfo {year} {2019})\BibitemShut {NoStop}%
\bibitem [{\citenamefont {Stein}\ and\ \citenamefont
  {Shakarchi}(2010)}]{stein2010complex}%
  \BibitemOpen
  \bibfield  {author} {\bibinfo {author} {\bibfnamefont {E.~M.}\ \bibnamefont
  {Stein}}\ and\ \bibinfo {author} {\bibfnamefont {R.}~\bibnamefont
  {Shakarchi}},\ }\href@noop {} {\emph {\bibinfo {title} {Complex analysis}}},\
  Vol.~\bibinfo {volume} {2}\ (\bibinfo  {publisher} {Princeton University
  Press},\ \bibinfo {year} {2010})\BibitemShut {NoStop}%
\bibitem [{\citenamefont {Cardy}(2006)}]{cardy2006n}%
  \BibitemOpen
  \bibfield  {author} {\bibinfo {author} {\bibfnamefont {J.}~\bibnamefont
  {Cardy}},\ }\bibfield  {title} {\bibinfo {title} {{T}he ${O(n)}$ model on the
  annulus},\ }\href {https://doi.org/10.1007/s10955-006-9186-8} {\bibfield
  {journal} {\bibinfo  {journal} {Journal of statistical physics}\ }\textbf
  {\bibinfo {volume} {125}},\ \bibinfo {pages} {1} (\bibinfo {year}
  {2006})}\BibitemShut {NoStop}%
\bibitem [{\citenamefont {Bl\"ote}\ \emph {et~al.}(1986)\citenamefont
  {Bl\"ote}, \citenamefont {Cardy},\ and\ \citenamefont
  {Nightingale}}]{blote1986conformal}%
  \BibitemOpen
  \bibfield  {author} {\bibinfo {author} {\bibfnamefont {H.~W.~J.}\
  \bibnamefont {Bl\"ote}}, \bibinfo {author} {\bibfnamefont {J.~L.}\
  \bibnamefont {Cardy}},\ and\ \bibinfo {author} {\bibfnamefont {M.~P.}\
  \bibnamefont {Nightingale}},\ }\bibfield  {title} {\bibinfo {title}
  {Conformal invariance, the central charge, and universal finite-size
  amplitudes at criticality},\ }\href
  {https://doi.org/10.1103/PhysRevLett.56.742} {\bibfield  {journal} {\bibinfo
  {journal} {Phys. Rev. Lett.}\ }\textbf {\bibinfo {volume} {56}},\ \bibinfo
  {pages} {742} (\bibinfo {year} {1986})}\BibitemShut {NoStop}%
\end{thebibliography}%

\clearpage
\appendix

\section{Computing Cross entropy in Clifford circuits}
\label{app:XEB_clifford}
In this section, we explain how one can compute LXE $\chi(\rho,\sigma)$ when the circuit $C$ is Clifford and the  state $\sigma$  is a stabilizer state. We first note that LXE can be written as,
\begin{align}
    \chi(\rho,\sigma)=\frac{\sum_{\vec{m}}p_{\vec{m}}^{\rho}p_{\vec{m}}^{\sigma}}{\sum_{\vec{m}}(p_{\vec{m}}^{\sigma})^2}=\Big\langle \frac{p_{\vec{m}}^{\sigma}}{\sum_{\vec{m}'}(p_{\vec{m}'}^{\sigma})^2} \Big\rangle_{p_{\vec{m}}^{\rho}},
\end{align}
where $\big\langle \cdot \big\rangle_{p_{\vec{m}}^{\rho}}$ means average over $\vec{m}$ when $\vec{m}$ is sampled from the probability distribution $p_{\vec{m}}^{\rho}$. Sampling from the distribution $p_{\vec{m}}^{\rho}$ can be done by either running the circuit $C$ with initial state $\rho$ on a quantum computer or simulating that quantum circuit on a classical computer when it is possible and recording the measurement outcomes. Therefore, the task of computing $\chi$ boils down to computing $p_{\vec{m}}^{\sigma}/\sum_{\vec{m}'}(p_{\vec{m}'}^{\sigma})^2$ for a given set of measurement outcomes $\vec{m}$. In what follows, first we consider the case where the set of  measurement outcomes $\vec{m}$ includes the outcome of \textit{all} measurements in the circuit. Then we consider the more general case, where $\vec{m}$ includes only a subset of measurement outcomes in the circuit. 

\subsection{Including all measurement outcomes} 

In a given stabilizer circuit, i.e. a Clifford circuit with a stabilizer initial state, each measurement outcome is either deterministic or completely random. For a given circuit, there could be more than one measurement in each layer, and whether a measurement is deterministic or random could depend on the order in which different measurements in a layer are performed. Nevertheless, the total number of random measurements in a given layer (and hence in the circuit) is independent of the order in which the measurements are performed. Therefore, without loss of generality, we may assume a specific order is fixed for the circuit measurements (e.g. measurements are performed from left to right). Let $N_\text{rand}(C,\sigma)$ denote the total number of random measurements in circuit $C$ with the initial state $\sigma$. In general, for a given circuit $C$ and initial state $\sigma$, the outcome of a deterministic measurement could depend on the outcomes of previous random measurements in the circuit. Therefore, for a given list of measurement outcomes $\vec{m}$, two cases could happen: either the outcomes for all the deterministic measurements in $\vec{m}$ are compatible with the outcomes of random measurements in $\vec{m}$, in which case $p_{\vec{m}}^\sigma=2^{-N_\text{rand}(C,\sigma)}$, or there is at least one deterministic measurement outcome in $\vec{m}$ that is incompatible with the random measurement outcomes in $\vec{m}$, in which  case $p_{\vec{m}}^\sigma=0$. Noting that the total number of possible compatible measurement outcomes $\vec{m}$ is equal to $2^{N_\text{rand}(C,\sigma)}$, we find that
\begin{align}\label{eq_p_m_sigma}
    \frac{p_{\vec{m}}^{\sigma}}{\sum_{\vec{m}'}(p_{\vec{m}'}^{\sigma})^2}=
    \begin{cases}
        0 &  \text{ $\vec{m}$ is not compatible with $C$ and $\sigma$}\\
        1 &  \text{ $\vec{m}$ is compatible with $C$ and $\sigma$}
    \end{cases}
\end{align}
Checking whether a given set of measurement outcomes $\vec{m}$ is compatible with $C$ and $\sigma$ is straightforward in Clifford circuits. One simply simulates the circuit $C$, starting from the initial state $\sigma$. Whenever there is a measurement with a random outcome, one forces the outcome according to the corresponding value in $\vec{m}$.  When there is a deterministic measurement, one computes the outcome and compares it with the corresponding value in $\vec{m}$. If the two values do not agree, then $\vec{m}$ is incompatible with $C$ and $\sigma$, and one can halt the simulation. Otherwise, one proceeds with the simulation until the next measurement. If the simulation finishes without encountering any incompatible deterministic measurement, it means $\vec{m}$ is compatible with $C$ and $\sigma$.

\subsection{Including only a subset of measurement outcomes}
If $\vec{m}$ includes only a subset of measurement outcomes in the circuit, then the corresponding probability $p_{\vec{m}}^\sigma$ is obtained by summing over all possible measurement outcomes for the rest of the measurements which are not included in $\vec{m}$. The summation could be performed at the circuit level by replacing any measurement whose outcome is not included in $\vec{m}$ with a quantum channel with Kraus operators $\Pi_{\pm}$ where $\Pi_\pm$ is the projection operator into the $\pm$ subspace of the corresponding measured operator. For the Pauli measurements, the corresponding quantum channel would be a Clifford operation, mapping stabilizer density matrices to stabilizer density matrices. Therefore, the resulting quantum circuit $\widetilde{C}$ which is obtained from $C$ by replacing a subset of measurements with their corresponding quantum channel, is also a Clifford circuit. Importantly, $\vec{m}$ includes all the measurements in $\widetilde{C}$, so we may use the result of the previous section to compute $p_{\vec{m}}^\sigma$, after replacing $C$ with $\widetilde{C}$ in Eq.\eqref{eq_p_m_sigma}.

\section{Longer range measurement-only circuit}\label{ZIZ_XX_model}
\begin{figure*}
\center{\includegraphics[width=\textwidth]{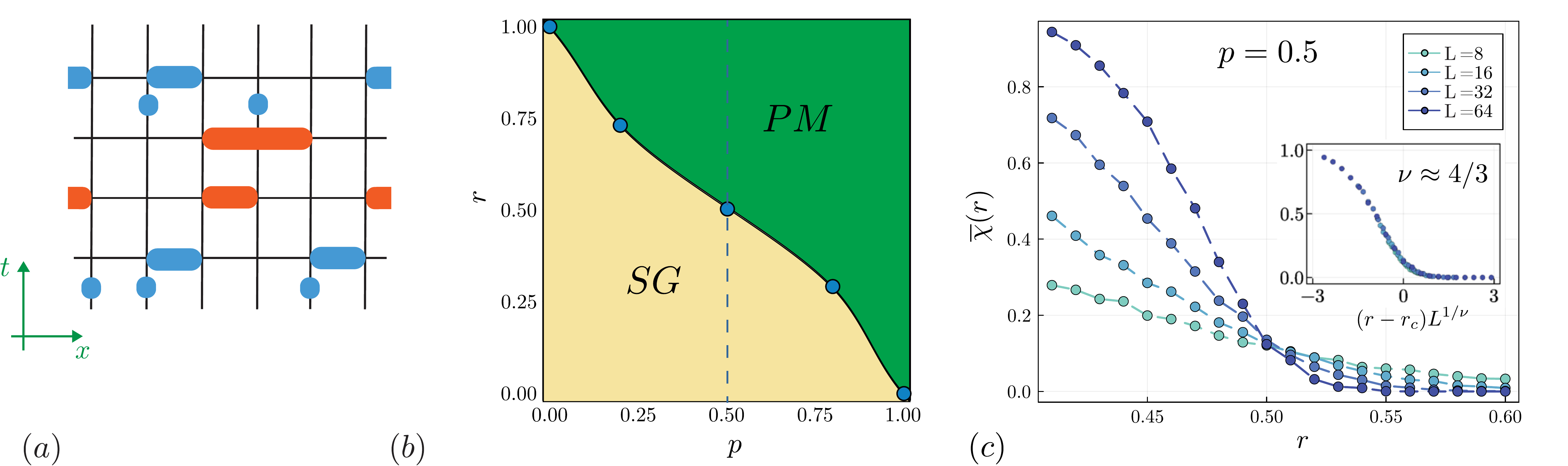}}
 \caption{Circuit architecture, phase diagram and the cross entropy behavior of a longer range symmetric circuit. (a) Circuit architecture with closed boundary conditions.  Blue operators correspond to two qubit $ZZ$ measurements and single qubit $X$  measurements. Orange operators are two qubit $XX$ (two qubit operator) measurements and two qubit $ZIZ$ (three qubit operator) measurements. (b) Approximate phase diagram of the model. We obtain several points at the phase transition line (shown in Fig.~\ref{fig:LR_cuts}). Number of performed interations is $N_{iter}=4000$. We initiate the circuit with $|GHZ\pm\rangle$ initial states. As described in the main text, for these initial states, the spin glass (SG) phase is characterized by the cross entropy reaching $\overline{\chi}=1$ in the thermodynamic limit. Paramagnetic phase (PM) is characterized by the cross entropy reaching $\overline{\chi}=0$ in the thermodynamic limit. The phase transition line (black solid line) is the set of crossing points for different system sizes. The purple vertical dashed line at $p=0.5$ corresponds to a plot in (c). (c) A vertical cut along the phase diagram (b) at the rate of $X$ measurements $p=0.5$. There is a clear signature of the phase transition at $r\approx0.5$. The scaling collapse is done for the value of the exponent $\nu\approx 4/3$. }
\label{fig:phase_diagram_LR}
\end{figure*}

\begin{figure}[h]
\center{\includegraphics[width=0.48\textwidth]{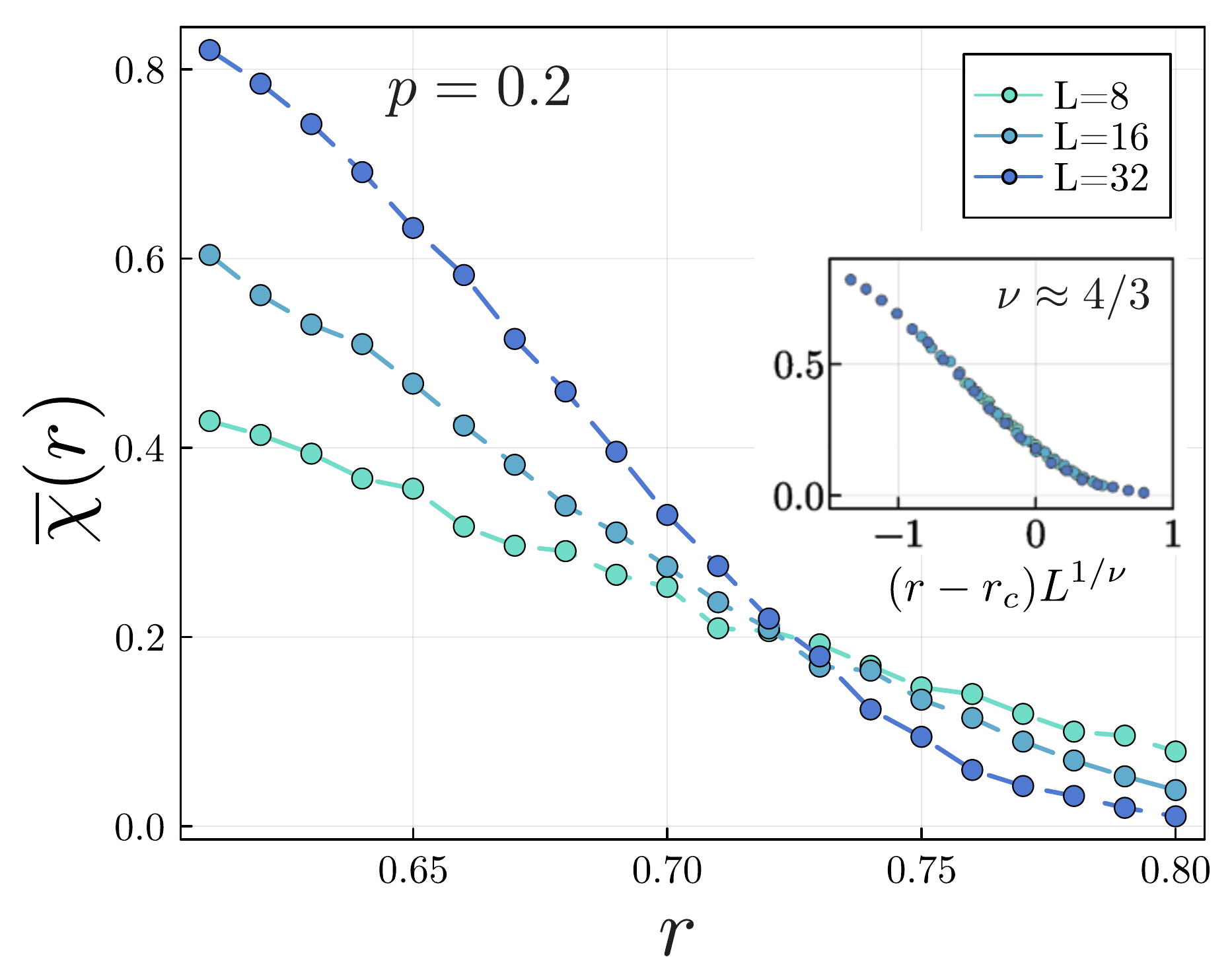}\ \ \ \ \ 
\includegraphics[width=0.48\textwidth]{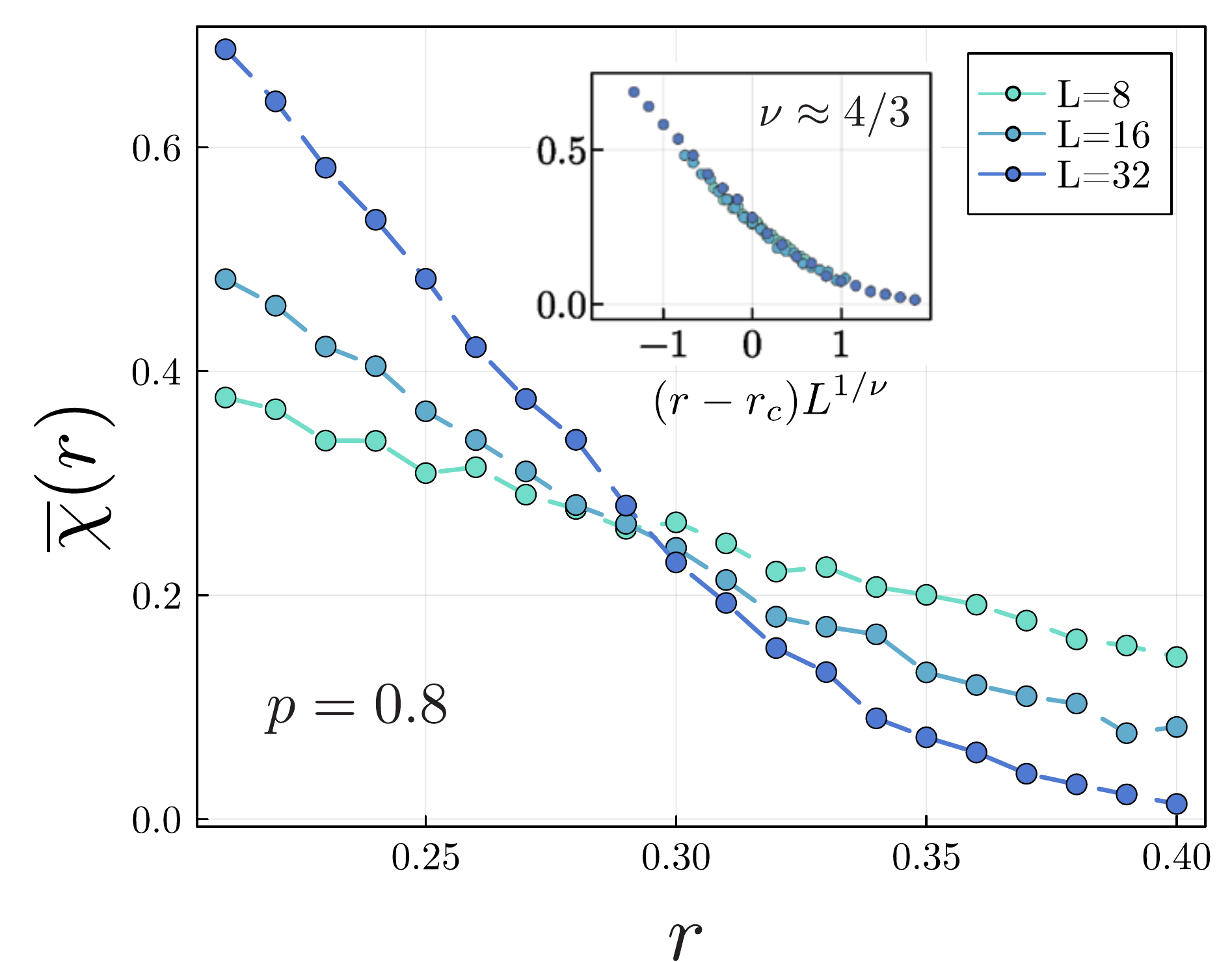}}
 \caption{Cuts on the phase diagrams Fig.~\ref{fig:phase_diagram_LR}(b) at $p=0.2$ (left) and $p=0.8$ (right). }
\label{fig:LR_cuts}
\end{figure}

In this section we consider a longer range measurement-only model where in addition to previously considered $ZZ$ and $X$ measurements we add two-qubit $ZIZ$ and $XX$ measurements as shown in Fig.~\ref{fig:phase_diagram_LR}(a). We measure $X$ with probability $p$ and $ZZ$ with probability $1-p$, and we assign $XX$ probability $r$ while $ZIZ$ is measured with probability $1-r$. The resulting phase diagram is shown in the Fig.~\ref{fig:phase_diagram_LR}(b). As an example of the phase transition between two area law phases, we show the behavior of LXE at $p=0.5$ in Fig.~\ref{fig:phase_diagram_LR}(c). The crossing point for different system sizes appears to be at $r=0.5$. Scaling collapse is performed at $\nu\approx4/3$. We show other cuts of the phase diagram in Fig.~\ref{fig:LR_cuts}. As we can see from the scaling collapse for finite system sizes, the critical exponents are close to percolation CFT exponents. 

\section{Probability of revealing non-local information by scrambling}
\label{app:scrambling}
\begin{figure}
    \centering
    \includegraphics[width=\columnwidth]{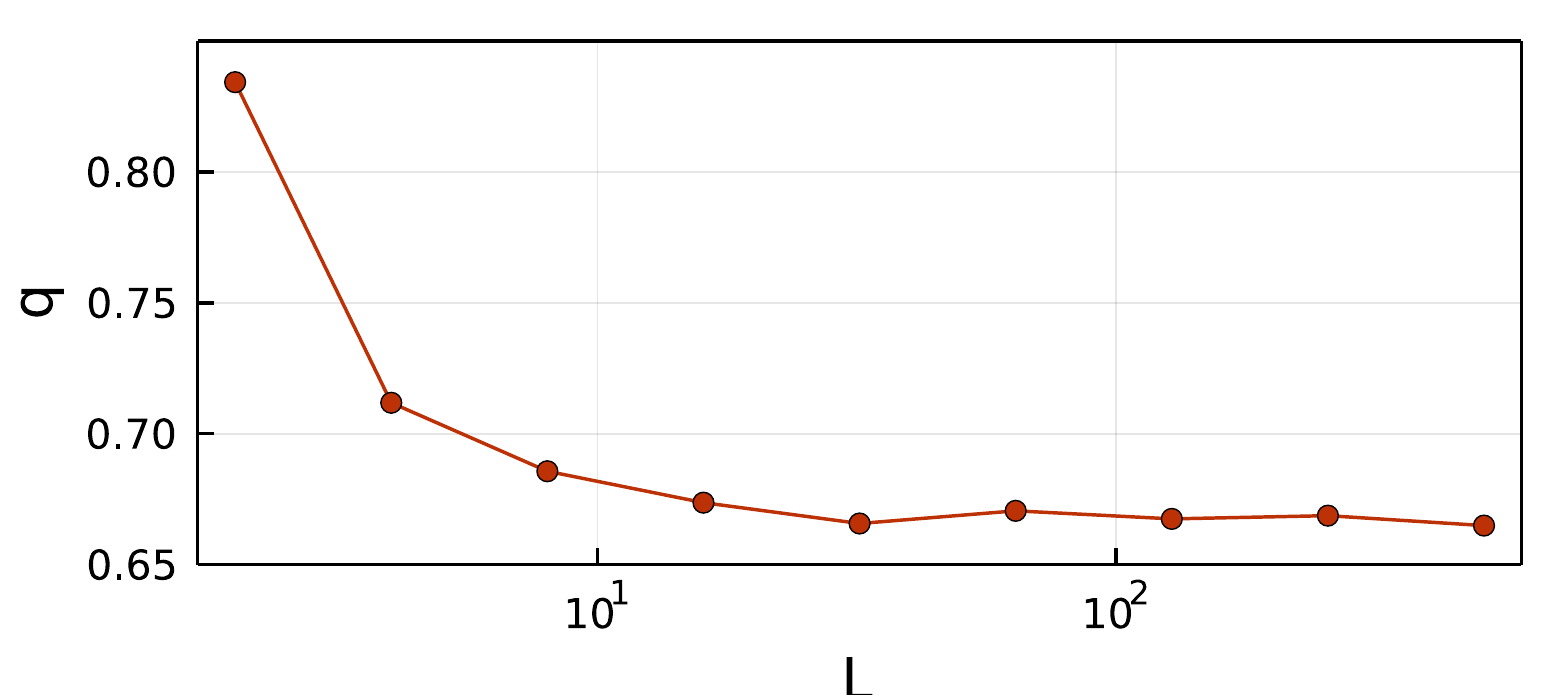}
    \caption{The probability that the scrambling of the $\ket{GHZ\pm}$ state exposes the non-local information to $Z$-type measurements. The $x$ axis is scaled logarithmically. }
    \label{fig:randomleak}
\end{figure}
Consider the stabilizer group of $\ket{GHZ\pm}$ states:
\begin{align}
    G_S=\langle Z_1Z_2,\cdots, Z_{L-1}Z_L, \pm X_1 X_2 \cdots X_N\rangle.
\end{align}
Let $G_Z=\langle Z_1 Z_2,\cdots, Z_{L-1}Z_L\rangle$ be the subgroup generated by the $Z$  type stabilizers. Let $U$ be a Clifford unitary that respects the $\mathbb{Z}_2$ symmetry, i.e $U \prod_{i=1}^L X_i U^\dagger  =\prod_{i=1}^L X_i$. Let $U G_Z U^\dagger\equiv\langle U Z_1 Z_2 U^\dagger,\cdots, U Z_{L-1}Z_L U^\dagger\rangle$ denote the image of $G_Z$ under $U$. If there exists an element $g\in U G_Z U^\dagger$ such that up to a phase,
\begin{align}
    g=\prod_{i=1}^L X_i ~\prod_{j=1}^L Z_j^{\alpha_j},
\end{align}
for some set of $\alpha_j=0,1$ values, then one can distinguish $U\ket{GHZ+}$ from $U\ket{GHZ-}$ by measuring only $Z$ type stabilizers. For an $L$ qubit system, let $q(L)$ be the probability that such a $g$ exists when $U$ is chosen randomly by a depth $L$ random local symmetric Clifford circuit with brickwork structure. While obtaining $q(L)$ analytically might be involved, it can be easily computed numerically. As is shown in Fig.~\ref{fig:randomleak}, $q$ approaches $0.66(1)$ for large $L$, which is consistent with the LXE result presented in Section \ref{sec:UE} of the main text.

\section{Conformal Transformation}\label{app:conf_transf}
Consider a $q$-state Potts model on the half-plane. The operator  $\phi_{f\rightarrow j}(x_{i})$ at a point $x_{i}$ on the real line changes the boundary conditions from a region where the Potts spins are free ($x<x_{i}$) to a region where the Potts spins are pinned in the $j\in\{1,\ldots,q\}$ state ($x>x_{i}$).  The four-point correlation function  $\langle\phi_{f\rightarrow1}\left({x}_{1}\right)\phi_{1\rightarrow f}\left({x}_{2}\right)\phi_{f\rightarrow2}\left({x}_{3}\right)\phi_{2\rightarrow f}\left({x}_{4}\right)\rangle$ with $x_{1}<x_{2}<x_{3}<x_{4}$ on the real-line is given in the limit $q\rightarrow 1$ by the expression
\begin{align}\label{eq:correlator_app}
C(\{x_{i}\}) = \frac{3\Gamma\left(\frac{2}{3}\right)}{\Gamma\left(\frac{1}{3}\right)^{2}}\,\,(1-\eta)^{1/3}\,\,_{2}F_{1}\left(\frac{1}{3},\frac{2}{3},\frac{4}{3};1-\eta\right)
\end{align}
where the cross-ratio $\eta \equiv (x_{12}x_{34})/(x_{14}x_{23})$ with $x_{ij} \equiv |{x}_{i} -{x}_{j}|$.  The scaling dimension of the boundary-condition changing operators $\phi_{f\rightarrow1}$, $\phi_{1\rightarrow f}$ is zero \cite{cardy1992critical} in the $q\rightarrow 1$ limit.  As a result, after a conformal transformation $w(z)$ this correlation function becomes $C(x_{1},x_{2},x_{3},x_{4}) = C(w_{1},w_{2},w_{3},w_{4})$ where $w_{i} \equiv w(x_{i})$\footnote{Recall that in a conformal field theory, the correlation function $\langle \prod_{j}\phi_{j}(w_{j})\rangle = \prod_{j}\left|\frac{dw}{dz}\right|_{w = w_{j}}^{-h_{j}}\langle \prod_{j}\phi_{j}(z_{j})\rangle$, where $h_{j}$ is the scaling dimension of $\phi_{j}$ \cite{francesco2012conformal}.}. 

We wish to compute the four-point correlation function of the bcc operators at the boundaries of a rectangular region, as shown in Fig.~ \ref{fig:XEB_Potts}b.  To do this, we choose the points $x_{4} = -x_{1} = y$ and $x_{3} = -x_{2} = x$ in (\ref{eq:correlator_app}), with $y>1>x>0$. and perform a conformal transformation to map the half-plane to a rectangle, so that these points map to the desired points on the boundaries of the rectangular region.  The Schwarz-Christoffel transformation \cite{stein2010complex}, is given by
\begin{align}\label{eq:y}
w(z) \equiv\frac{L}{2 K(1/y^{2})}\int_{0}^{z}\frac{dt}{\sqrt{(1-t^{2})[1- (t^{2}/y^{2})]}}
\end{align}
where
\begin{align}
K(x) \equiv \int_{0}^{1}\frac{dt}{\sqrt{(1-t^{2})(1- xt^{2})}}
\end{align}
transforms the points $\pm 1$, and $\pm y$ on the real line to the corners of a rectangle.  $\pm 1$ are mapped to $\pm L/2$, respectively. Requiring that $\pm y$ are mapped to $w(\pm y) = \pm (L/2) + iT$ fixes the position $y$ by the relation 
\begin{align}\label{eq:aspect_ratio}
\frac{L}{T} =\frac{2K(1/y^{2})}{K\left(1-(1/y^{2})\right)}.
\end{align}
Finally, we require that $\pm x$ are be mapped to $\pm r/2$.  This which fixes the point $x$ implicitly by the relation
$w(x) = r/2$.  This equation may be solved numerically to determine $x$.

 To conclude, we have shown that the four-point function (\ref{eq:four_pt_fn}) with cross-ratio
\begin{align}
\eta \equiv \frac{(x-y)^{2}}{(x+y)^{2}}
\end{align}
and with $y$ and $x$ determined implicitly by the above expressions, is equivalent to the desired four-point function on the boundaries of a rectangle with $\boldsymbol{x}_{1} = -(L/2)+iT$, $\boldsymbol{x}_{2} = -r/2$, $\boldsymbol{x}_{3} = r/2$, and $\boldsymbol{x}_{4} = (L/2)+iT$. When $r\ll L$, we find that
\begin{align}
&\underset{q\rightarrow 1}{\lim}\Big[\langle\phi_{f\rightarrow1}\left(\boldsymbol{x}_{1}\right)\phi_{1\rightarrow f}\left(\boldsymbol{x}_{2}\right)\phi_{f\rightarrow2}\left(\boldsymbol{x}_{3}\right)\phi_{2\rightarrow f}\left(\boldsymbol{x}_{4}\right)\rangle\Big]\nonumber\\
&= \frac{3\Gamma\left(\frac{2}{3}\right)}{\Gamma\left(\frac{1}{3}\right)^{2}}\left(\frac{4 K(1/y^{2})}{3y}\frac{r}{L}\right)^{1/3} + O((r/L)^{2/3})
\end{align}

\section{Height Field Representation of Critical  Percolation}\label{app:field_theory_percolation}
We review a continuum field-theoretic description of critical bond percolation in two dimensions which we then use to determine the behavior of the LXE with periodic boundary conditions.  The derivation of this continuum field theory has been extensively discussed (see, for example Ref.~\cite{nienhuis1984critical, cardy2006n, kondev1997liouville}).  Here, we will provide a heuristic derivation that reproduces the known results which have been more formally and carefully derived in the literature, namely we will argue that critical percolation on a cylinder with compact direction $(x)$ of width $L$ and of length $(\tau)$ $T$, is described by the following action for a continuum field $\varphi(x,\tau)$ 
\begin{align}
    S[\varphi] = \frac{g}{4\pi}\int dx \,d\tau\,(\nabla \varphi)^{2} + i\delta\,[\varphi(0,T) - \varphi(0,0)]
\end{align}
which describes the Gaussian fluctuations of a height field $\varphi$.  Here the parameters 
\begin{align}
    g = \frac{2}{3} \hspace{.25in} \delta = \frac{1}{3}.
\end{align}

Two-dimensional bond percolation, the boundaries (hulls) of percolating regions may be thought of as enclosing regions of constant ``height". To consistently define the height for a given configuration of percolating bonds, we must orient the hulls, so that the height jumps by a positive or negative increment depending on the local orientation of the hull.  Since this assigned orientation is arbitrary, it is natural to choose to sum over both orientations of each hull when re-casting percolation as the statistical mechanics of a fluctuating height variable \cite{saleur1987exact}.  The weights for this height variable are chosen as follows.  An infinitesimal patch of an oriented hull receives a weight $\exp\left[{\pm i \theta_{0}\,d\theta/2\pi}\right]$ with the sign depending on the assigned orientation, and we choose the constant $\theta_{0} = \pi/3$ so that the weight for each closed, contractible hull after summing over both orientations is $2\cos\theta_{0} = 1$ as is required of critical percolation.  

From this microscopic description, in which local weights are assigned to a given height-field configuration to reproduce the partition sum for bond percolation, it is natural to postulate that the field theory for critical percolation is described by the Gaussian fluctuations of a coarse-grained height field $\varphi$, which is 
described by the action
\begin{align}\label{eq:S1}
S_{1}[\varphi] = \frac{g}{4\pi}\int dx\,d\tau\,(\nabla \varphi)^{2}
\end{align}
This description is incomplete.  First, the central charge of this free boson $c = 1$ does not match the known central charge ($c=0$) of critical percolation (the fact that percolation has zero central charge follows trivially from the fact that the partition function for percolation is $Z = 1$, independent of the percolation probability). Second, on a compact manifold, it is possible to have percolating hulls which wrap around non-contractible cycles, and these will be weighted incorrectly.  According to the previous microscopic description, each non-contractible loop will have weight $2$ after summing over both orientations, since the total winding angle of such a loop is zero. 

Both of these issues may be rectified by introducing background charges in the continuum field theory. Consider critical percolation on a cylinder  with compact direction of length $L$ and finite length $T$. On this manifold, we may add an additional term to the action  
\begin{align}\label{eq:bkgnd_charge}
S_{2}[\varphi] = i\delta\,[\varphi(0,T) - \varphi(0,0)]
\end{align}
This term does not alter the weights of closed, contractible percolating hulls.  If there is a single oriented loop wrapping around the cylinder, however, the height difference $\varphi(0,T) - \varphi(0,0) = \pm\pi$ and so we take 
\begin{align}
\delta = \frac{1}{3}
\end{align}
so that the non-contractible loop appears with the correct weight $2\cos\pi\delta = 1$ after summing over both orientations.  

The constant $g$ in Eq. (\ref{eq:S1}) may be fixed by requiring that this insertion of a background charge shifts the central charge to the correct value, $c = 0$, for percolation.  Let $Z_{\delta}$ be the partition function for the height field in the presence of the $\pm\delta$ background charges, as described by Eq. (\ref{eq:bkgnd_charge}). We note that on the cylinder \cite{francesco2012conformal}
\begin{align}\label{eq:ratio}
\frac{Z_{\delta}}{Z_{0}} &= \langle e^{i\delta\,\varphi(0,T)} e^{-i\delta\,\varphi(0,0)}\rangle_{0}\\
&= \left(\frac{2\pi}{L}\right)^{2\Delta_{0}}\left[2\cosh\left(\frac{2\pi T}{L}\right) - 2\right]^{-\Delta_{0}}
\end{align}
Here, the expectation value $\langle\cdots\rangle_{0}$ is taken with respect to $Z_{0}$, and $\Delta_{0}/2 = \delta^{2}/4g$ is the scaling dimension of the operator $e^{i\delta}$.  When $T\gg L$ this reduces to
\begin{align}
\frac{Z_{\delta}}{Z_{0}} \sim e^{-2\pi \Delta_{0} T/L}
\end{align}
Because of this, the free energy of the system with background charges $\pm \delta$, per unit length of the cylinder 
\begin{align}
f_{\delta} \equiv -T^{-1}\ln Z_{\delta}
\end{align}
is given by
\begin{align}
f_{\delta} = f_{0} + \frac{1}{L}\frac{\pi\,\delta^{2}}{g} + \cdots
\end{align}
where the ellipsis denotes corrections which vanish as $T\rightarrow\infty$. For a conformal field theory with central charge $c$, the free energy per unit length of the cylinder with compact direction $L$ is given by $f(L) = f(\infty) - (\pi c/6L)$ \cite{blote1986conformal}.  Since the central charge of a compact boson is $1$, the central charge of the new theory in the presence of background charges is then 
\begin{align}
c = 1 - \frac{6 \delta^{2}}{g}
\end{align}
As a result, we must choose $g = 6\delta^{2} = 2/3$ in order for the theory to have the desired central charge $c = 0$.

With these preliminary results in hand, we may now determine the fraction of configurations in critical percolation for which there are no non-contractible loops on the cylinder.  To forbid non-contractible loops entirely, we must insert a background charge $\delta' = 1/2$ so that non-contractible loops receive zero weight.  Then, using Eq. (\ref{eq:ratio}) it is easy to see that 
\begin{align}
\frac{Z_{\delta'}}{Z_{\delta}} = \left(\frac{2\pi}{L}\right)^{2\Delta'}\left[2\cosh\left(\frac{2\pi T}{L}\right) - 2\right]^{-\Delta'}
\end{align}
where $\Delta' = [(\delta')^{2} - \delta^{2}]/2g = 5/48$.

\newpage
\pagebreak
\widetext
\section{Additional plots}\label{apx:additional_plots}

\begin{figure}[h]
    \includegraphics[width=0.48\textwidth]{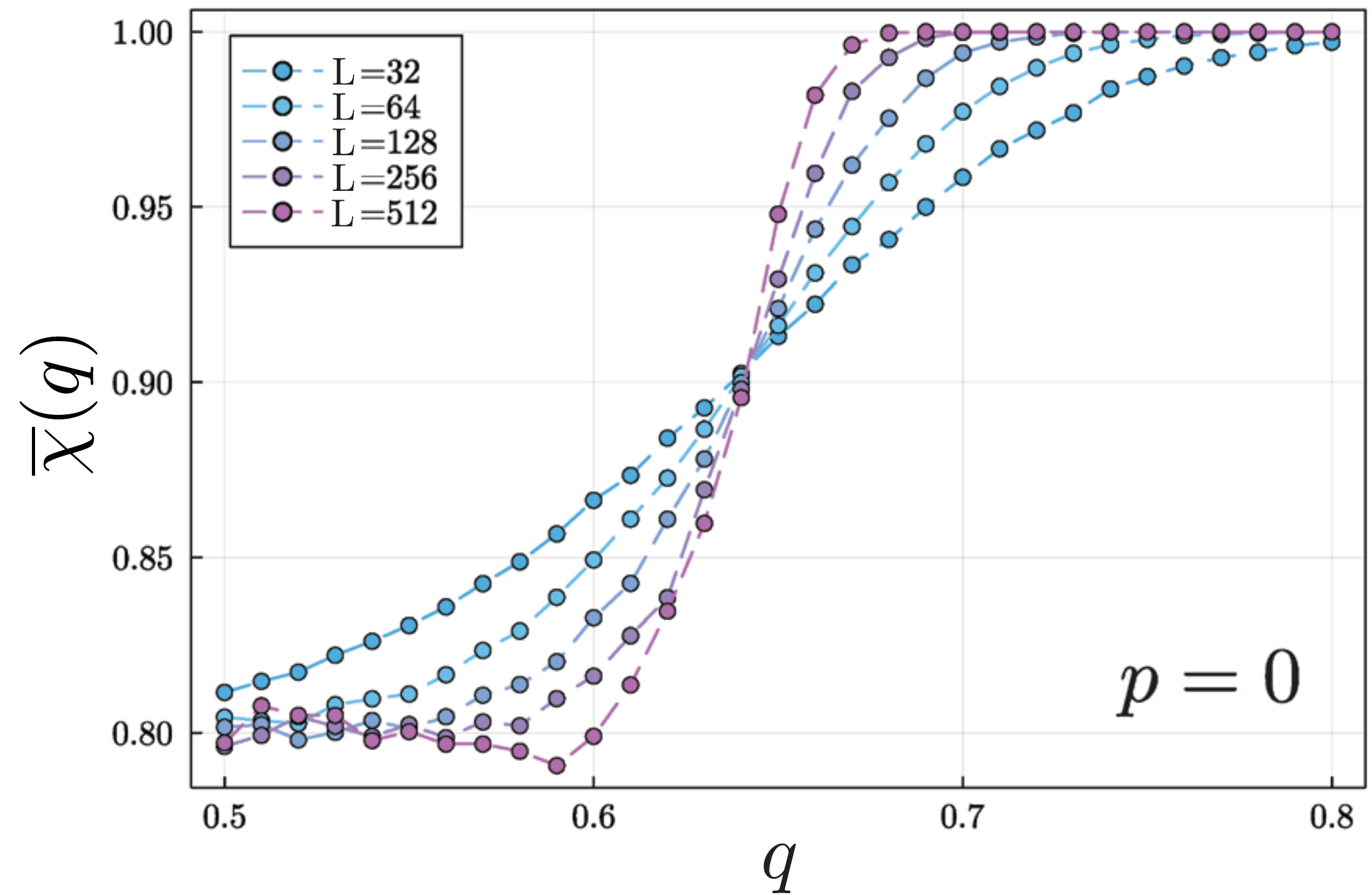}
    \includegraphics[width=0.48\textwidth]{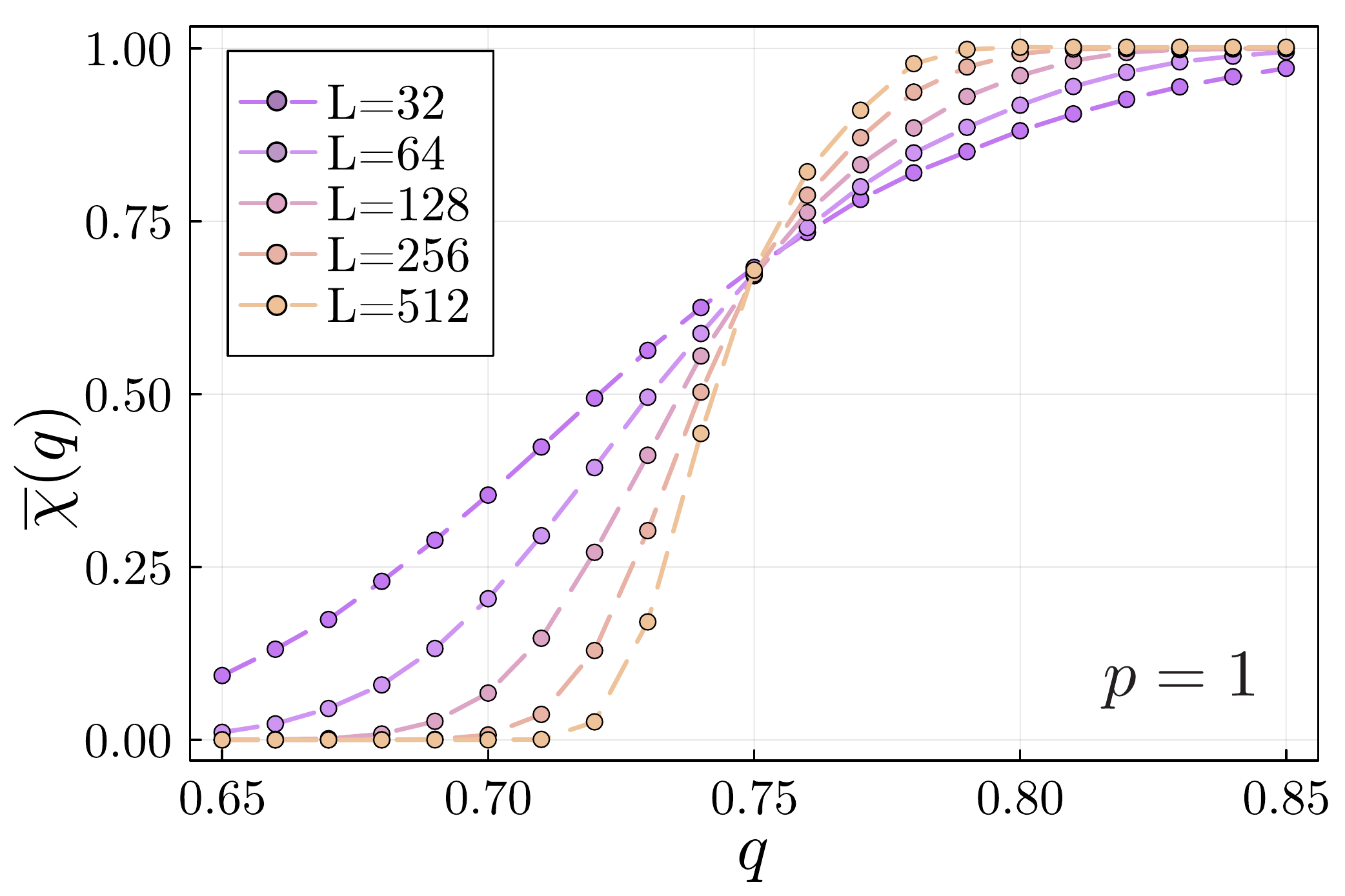}
    \caption{Vertical cuts on the phase diagram (Fig.~\ref{fig:phase_diagram}(b)) Left: phase transition between spin glass and volume law phases ($p=0$). The phase transition is observed at $q\approx0.64$. Initial states are $|GHZ_\pm\rangle$ after scrambling with $\mathbb{Z}_2$ symmetric unitaries for $t_{scr}=L$.  Right: phase transition between paramagnet and volume law phases ($p=1$). The phase transition is observed at $q\approx0.75$. Initial states are $|GHZ_\pm\rangle$.}
    \label{fig:add_plots1}
\end{figure}

\begin{figure}[h]
    \includegraphics[width=0.48\textwidth]{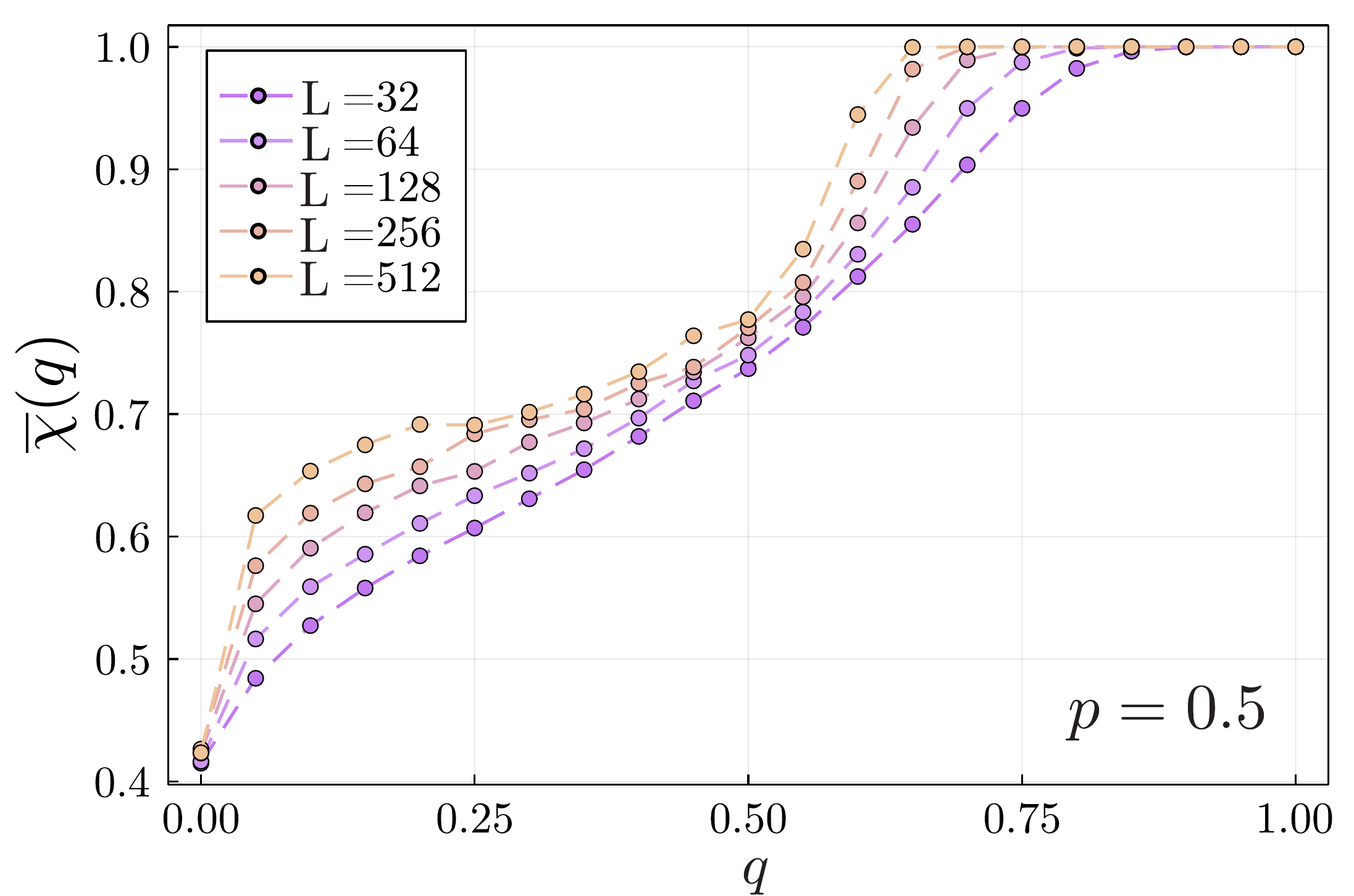}
    \includegraphics[width=0.48\textwidth]{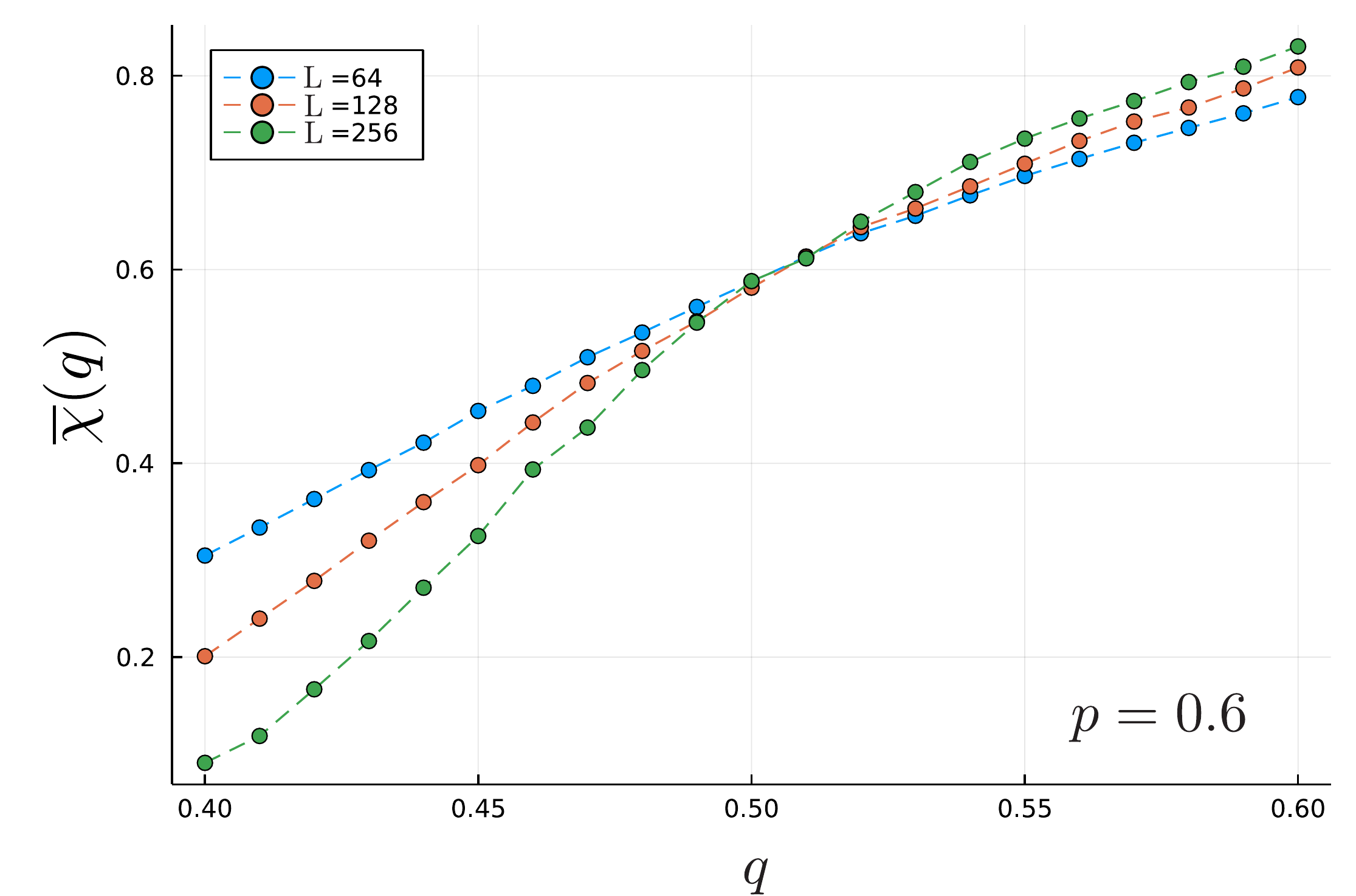}
    \caption{Vertical cuts on the phase diagram (Fig.~\ref{fig:phase_diagram}(b)). Left: behavior of the cross entropy at $p=0.5$. Initial states are $|GHZ_\pm\rangle$.  Right: phase transition between paramagnet and volume law phases ($p=0.6$). The phase transition is observed at $q\approx0.5$. Initial states are $|GHZ_\pm\rangle$.}
    \label{fig:add_plots2}
\end{figure}

\end{document}